\documentclass{article}
\usepackage[utf8]{inputenc}
\usepackage[round]{natbib}

\usepackage{algorithm}

  \usepackage{setspace}
\usepackage{verbatim}
\usepackage{amsmath,graphicx,centernot, caption, subcaption}
\usepackage{amssymb}
\allowdisplaybreaks

\newcommand{\bigCI}{\mathrel{\text{\scalebox{1.07}{$\perp\mkern-10mu\perp$}}}}
\newcommand{\nbigCI}{\centernot{\bigCI}}

\usepackage{tikz}
\usepackage{graphicx}
\tikzstyle{ov}=[shape=rectangle,
                draw=black!50,
                thick,
                minimum width=0.7cm,
                minimum height=0.7cm]
\tikzstyle{av}=[shape=rectangle,
                draw=black!50,
                fill=black!10,
                thick,
                minimum width=0.7cm,
                minimum height=0.7cm]
\tikzstyle{lv}=[shape=circle,draw=black!50,thick]
\usetikzlibrary{shapes,calendar,matrix,backgrounds,folding,snakes}

\usepackage[english]{babel}
\usepackage{authblk}

\title{Principled Selection of Baseline Covariates to Account for Censoring in Randomized Trials with a Survival Endpoint}
\author[1]{Kelly Van Lancker}
\author[1]{Oliver Dukes}
\author[1,2]{Stijn Vansteelandt}

\affil[1]{Department of Applied Mathematics, Computer Science and Statistics, Ghent University, Ghent, Belgium}
\affil[2]{Department of Medical Statistics, London School of Hygiene and Tropical Medicine, London, United Kingdom}

\usepackage{natbib}
\usepackage{graphicx}

\providecommand{\keywords}[1]
{
  \small	
  \textbf{\textit{Keywords---}} #1
}

\begin{document}

\maketitle

\begin{abstract}
The analysis of randomized trials with time-to-event endpoints is nearly always plagued by the problem of censoring. As the censoring mechanism is usually unknown, analyses typically employ the assumption of non-informative censoring. While this assumption usually becomes more plausible as more baseline covariates are being adjusted for, such adjustment also raises concerns. Pre-specification of which covariates will be adjusted for (and how) is difficult, thus prompting the use of data-driven variable selection procedures, which may impede valid inferences to be drawn. The adjustment for covariates moreover adds concerns about model misspecification, and the fact that each change in adjustment set, also changes the censoring assumption and the treatment effect estimand. In this paper, we discuss these concerns and propose a simple variable selection strategy that aims to produce a valid test of the null in large samples. The proposal can be implemented using off-the-shelf software for (penalized) Cox regression, and is empirically found to work well in simulation studies and real data analyses. 
\end{abstract}


\section{Introduction}
Randomized trials on the effect of treatment on time-to-event endpoints are typically plagued by dropout or intercurrent events, resulting in censored event times. 
The logrank test constitutes the gold standard for the design (in determining sample size) and analysis of such data \citep[e.g.][]{Lin2017, halabi2019} under the assumption that censoring is non-informative in each treatment arm.
This assumption is strong but can usually be relaxed via adjustment for prognostic baseline covariates, which moreover tend to yield increased power in randomized controlled trials \citep{HERNANDEZ2006}.
Even so, unadjusted analyses still dominate the evaluation of primary endpoints in practice \citep{Austin2010, yu2010}. 
The reason is that covariate adjustment demands (proportional hazards) modelling, which raises concerns about model misspecification bias and the difficulty in pre-specifying which covariates to adjust for and in which functional form. 
 
Interestingly, model misspecification appears not to be a major concern for testing the null hypothesis of no treatment effect in randomized experiments. In particular, standard tests obtained under a Cox proportional-hazards model are valid in spite of misspecification, so long as a sandwich estimator of the variance is used \citep[e.g.][]{Lin1989, Kong1997, DiRienzo2001}. 
However, this robustness is attained only when censoring is either independent of the treatment conditional on the covariates, or independent of covariates in each treatment arm \citep[e.g.][]{Kong1997, Lagakos1984, DiRienzo2001}. These assumptions are stronger than one is typically willing to make in practice.
Variable selection procedures (eg, based on hypothesis tests, AIC or the Lasso) can help in choosing a model and thereby temper concerns about misspecification. 
Data-driven selection is however known to typically inflate the Type I error of the test of no treatment effect, as we will also demonstrate in this article.
Although several methods have been proposed to deal with this problem of inference after variable selection \citep{ning2017, fang2017, chai2019}, little work has specifically focused on Type I error rate control in randomized experiments with censored data. In particular, to the best of our knowledge, little or no attention has been given to the fact that each change in adjustment set, also changes the censoring assumptions as well as the treatment effect estimand.

In view of this, we will here propose a simple variable selection method which controls the Type I error of the test of no treatment effect and evaluate the impact of selection on estimating marginal survival curves. 


\section{The Impact of Variable Selection}\label{sec:impact}
Let $T$ denote the time from randomization to event and $C$ the censoring time. We observe $\left(T^{*}, \delta\right),$ where $T^{*}=\min (T, C)$ is the
observed portion of $T$ and $\delta=I(T \leqslant C)$ is a censoring indicator. Let the binary random variable $A$ denote the treatment group and let $X$ denote a covariate that for the moment is scalar.  
Throughout this paper we assume that $A$ is independent of $X$ by virtue of the randomization and that censoring is non-informative conditional on $A$ and $X$ in the sense that $T \bigCI C |(X, A)$. We are interested in testing the null hypothesis that the event time distribution does not depend on the treatment group. 

To develop insight into the aforementioned concerns about variable selection, suppose that the following Cox model
\begin{align}\label{eq:unrestr_model}
 \lambda\left(t\mid A, X\right)=\lambda_0(t)e^{\alpha A+\beta X},
\end{align}
is correctly specified, and that censoring is non-informative conditional on $A$ and $X$. Here, $\lambda_0(t)$ is the unspecified baseline hazard function, and $\alpha$ encodes the treatment effect of interest. Likewise, suppose that the model for censoring obeys
\begin{align}\label{eq:cens_model}
 \lambda^C\left(t\mid A, X\right)=\lambda_0^C(t)e^{\gamma_1A+\gamma_2X}, 
\end{align}
where $\lambda_0^C(t)$ is the baseline hazard function.

In practice, it is difficult to know a priori whether adjustment for $X$ in model (\ref{eq:unrestr_model}) is required in order to do inference for $\alpha$.
We therefore typically need to decide between fitting the unrestricted model (\ref{eq:unrestr_model}), or the restricted model:
\begin{align}\label{eq:restr_model}
 \lambda\left(t\mid A\right)=\lambda_0^r(t)e^{\alpha_0 A},
\end{align}
where $\lambda_0^r(t)$ is the baseline hazard function.

Routine practice to decide whether to adjust for $X$, is often based on data-adaptive covariate selection strategies.
These employ the data to make a binary decision $D$ on whether to report the $p$-value $p_\alpha$ (when $D=1$), corresponding with the test statistic (denoted by $Z_\alpha$) for the null hypothesis that $\alpha=0$ in the unrestricted model (\ref{eq:unrestr_model}) or to report the $p$-value $p_{\alpha_0}$ (when $D=0$), corresponding with the test statistic (denoted by $Z_{\alpha_0}$) for the null hypothesis that $\alpha_0=0$ in the restricted model (\ref{eq:restr_model}). For instance, it is common to adjust for $X$ if and only if it is significantly (eg, at the $5\%$ level) associated with the survival endpoint conditional on the treatment $A$. In this case, $D=I(p_\beta<0.05)$, where $p_\beta$ is the $p$-value corresponding with the null hypothesis that $\beta=0$.
Thus, for scalar $X$, routine practice is based on reporting the following, so-called post-model-selection $p$-value for the parameter of interest $\alpha$
\begin{align}\label{eq:pvalue}
\Tilde{p}_\alpha = \left\{ \begin{matrix} p_\alpha & \mbox{if } D=1 \\ p_{\alpha_0} & \mbox{if }D= 0. \end{matrix}\right.
\end{align}
Note that these $p$-values are obtained under different censoring assumptions. 
In particular, under model (\ref{eq:restr_model}) we implicitly make the stronger assumption that censoring is independent of the event time in each treatment arm.
As $T\bigCI C\mid A, X$ does not imply $T\bigCI C\mid A$, failing to control for the baseline covariate $X$ by prioritising the unadjusted analysis, as is common, or as a result of variable selection errors, may induce informative censoring.
Selecting model (\ref{eq:restr_model}) may therefore lead to bias in the corresponding test statistic $Z_{\alpha_0}$.
As a consequence, the test statistic corresponding with the $p$-value in (\ref{eq:pvalue})
$$
\Tilde{Z}_\alpha = \left\{ \begin{matrix} Z_\alpha & \mbox{if } D=1 \\ Z_{\alpha_0} & \mbox{if } D= 0, \end{matrix}\right.
$$
generally has a complex, bimodal distribution that may sharply deviates from the standard normal distribution. 
This is the result of the test statistic jumping back and forth between the test statistics $Z_\alpha$ and $Z_{\alpha_0}$, where the distribution of the latter might not be centered at zero (ie, is biased) as a result of informative censoring when not adjusting for $X$ (model (\ref{eq:restr_model})). 


For the more technical reader, it is shown in Appendix A.1 that, when failing to select the covariate $X$, the bias of the (score) test statistic under the null hypothesis of no treatment effect (See \eqref{eq:cbias_score} in Appendix A.1) has a leading term that is proportional to $\beta \gamma_2$. 
This bias is problematic as it may inflate the Type I error of the corresponding score test \citep{leeb2006}. This happens primarily if it has the same order of magnitude as the standard error of the mean score as is the case when $\beta\gamma_2$ has order of magnitude equal to or larger than $a/\sqrt{n}$ for some constant $a$. For example, when $\beta=b/\sqrt{n}$ for some constant $b$ and $\gamma_2$ is large, or vice versa. 
This is illustrated in Figure \ref{fig:LRT_bias} of Appendix B, which shows the distribution of the score test statistic when failing to control for $X$ (which corresponds to the logrank test statistic) for different values of $\beta$, $\gamma_1$ and $\gamma_2$. The problem of bias is most pronounced when $\beta$, $\gamma_1$ and $\gamma_2$ are large and disappears when either at least one of the coefficients is zero or both $\beta$ and $\gamma_2$ are small.


\begin{figure}[htp]
\centering
\includegraphics[width=.49\textwidth]{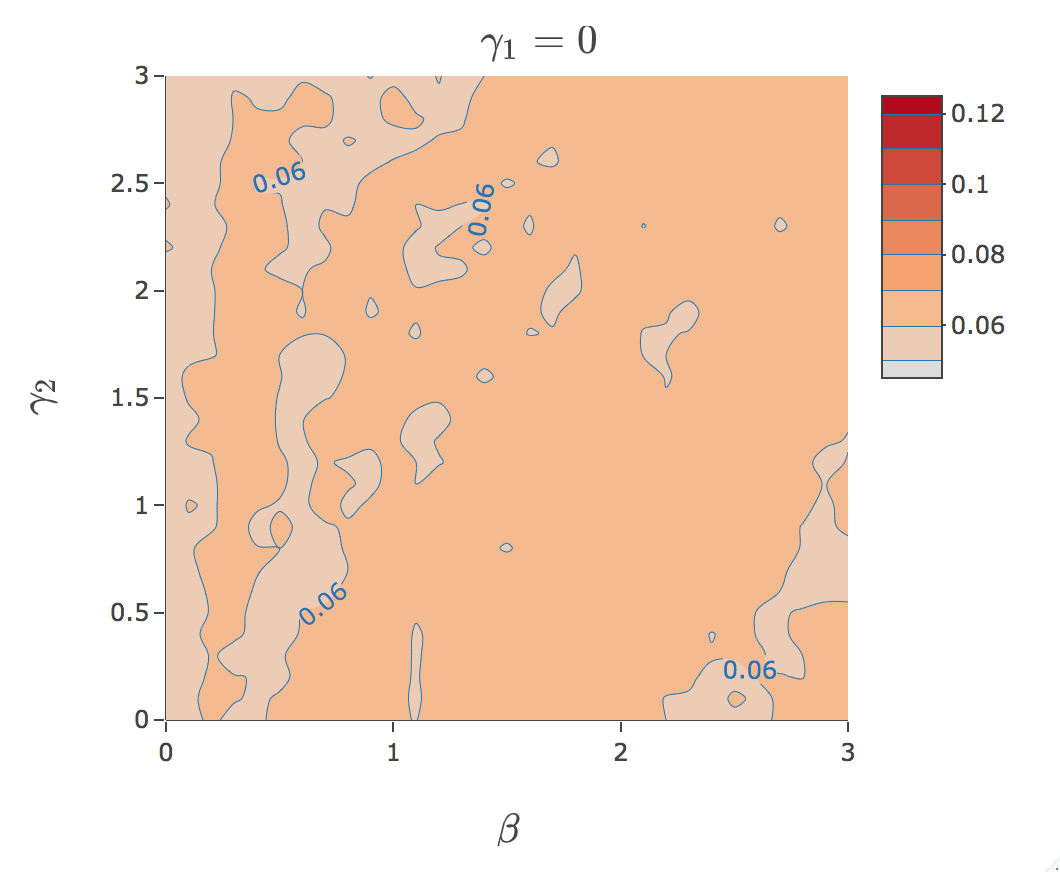}
\includegraphics[width=.49\textwidth]{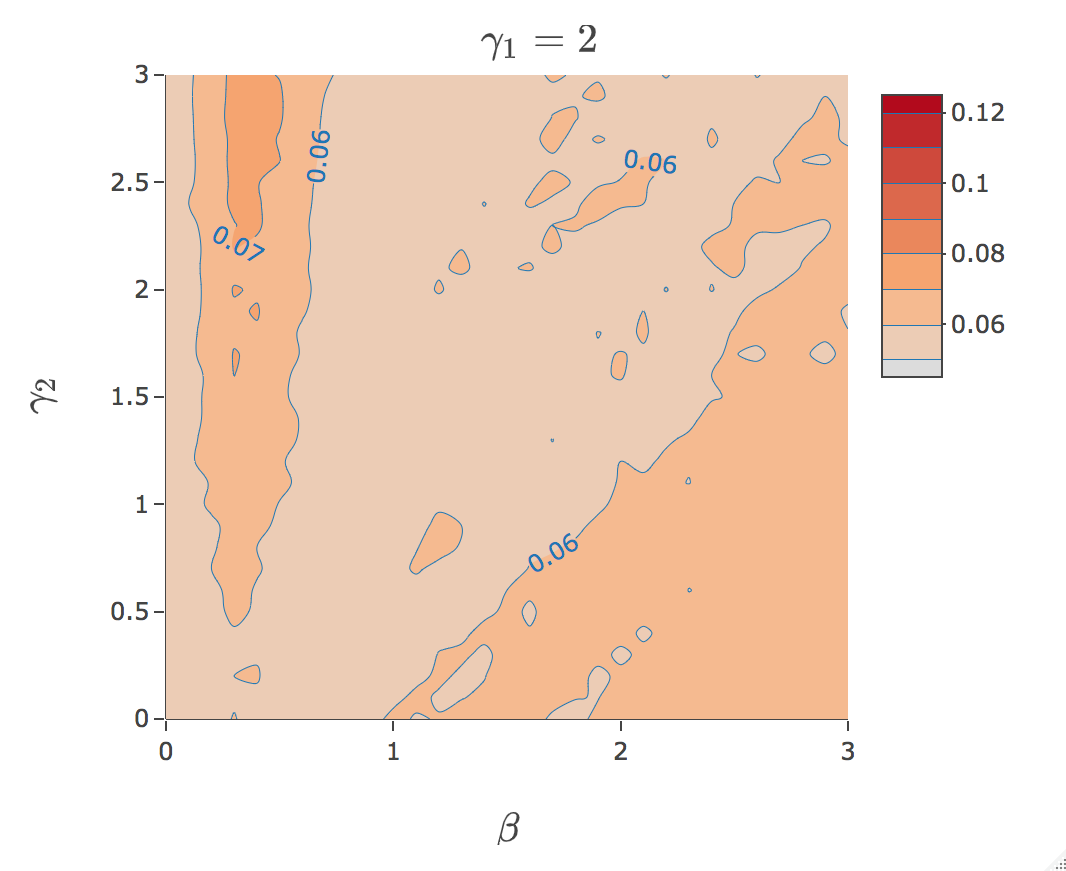}
\medskip
\includegraphics[width=.49\textwidth]{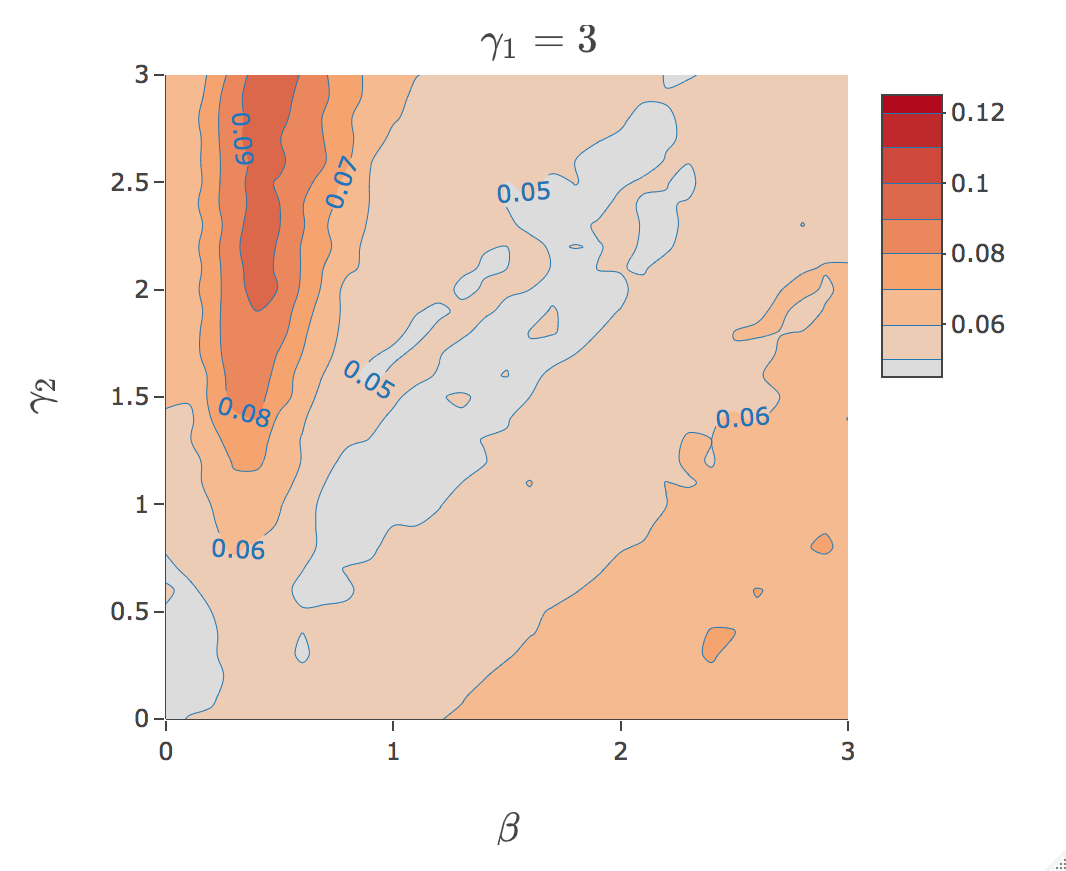}
\includegraphics[width=.49\textwidth]{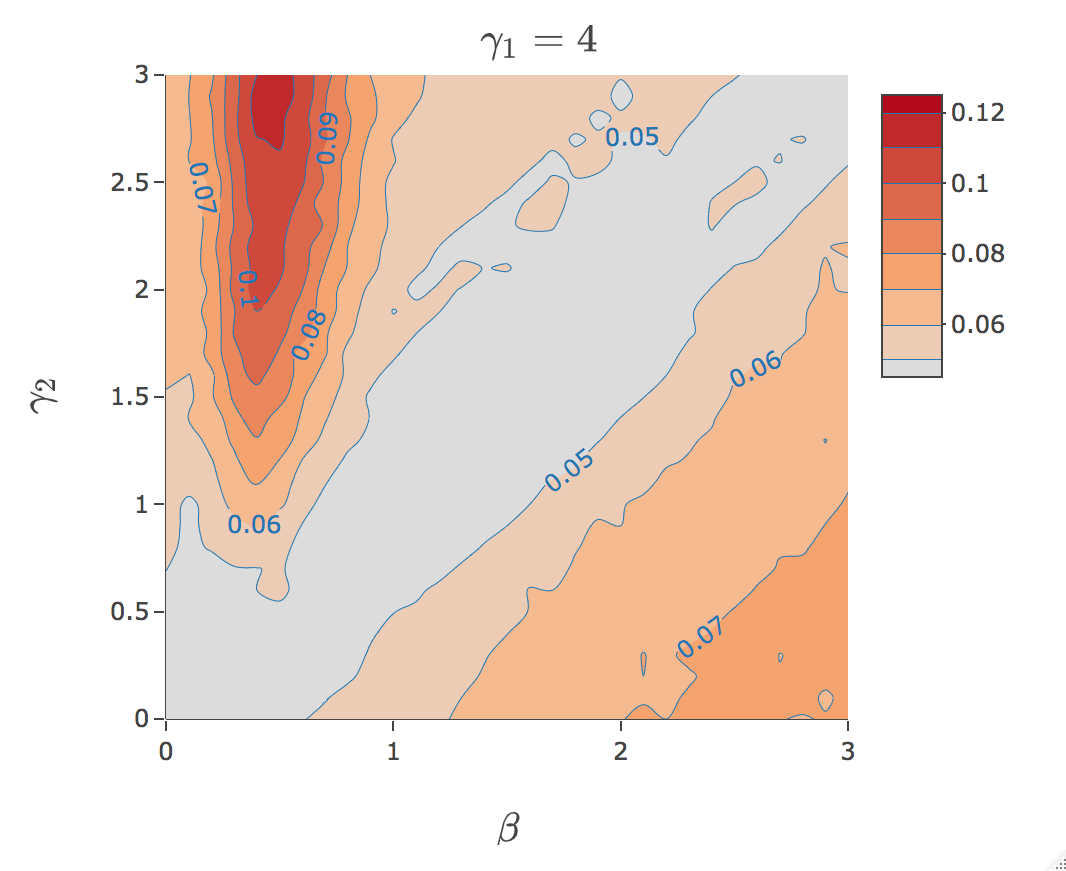}
\caption{Empirical Type I error rate at the $5\%$ significance level under covariate selection at the $2.5\%$ significance level in function of $\beta$, $\gamma_1$ and $\gamma_2$.  Results are based on $5000$ simulations, $n=100$, $X\sim N(0,1)$, $\lambda_0(t)=e$ and $\lambda_0^C(t)=e^{-1}$.}
\label{fig:contour}
\end{figure}
To appreciate the severity of the problem, Figure \ref{fig:contour} summarizes the Type I error under hypothesis-based covariate selection at the $2.5\%$ significance level for different values of $\beta$, $\gamma_1$ and $\gamma_2$. First, note that the nominal Type I error rate is maintained when $A$ is independent of $C$ conditional on $X$ (top-left panel). When $A$ and $C$ are associated conditional on $X$, the problem of inflated Type I error is most pronounced when $X$ has a strong effect on $C$ (high $\gamma_2$) but a weak-to-moderate effect on $T$ (low $\beta$). This is because the variable selection procedure likely fails to pick up $X$ when it is only weakly predictive of the survival time, as a consequence of the information loss induced by censoring as well as the fact that censoring may even reduce the variation in $X$ in the risk set. This in turn induces informative censoring, which may lead to bias in the corresponding test statistic and hence to inflated Type I errors. This problem becomes more severe for larger values of $\gamma_1$ (aka when $A$ is highly predictive of censoring). 

Surprisingly, these standard variable selection procedures do not affect tests of the null hypothesis of no treatment effect in generalised linear models for uncensored outcomes \citep{Wager2016,Dukes2020}, but they here arise due to the change in censoring assumption with each change in adjustment set.
This is the case regardless of whether selection is done through hypothesis testing (eg, stepwise regression) or related approaches such as the Lasso. 
Although the extent of Type I error rate inflation is less severe with increasing sample size, at any sample size there may be values of $\beta$ that are small but different from zero so that they will be missed by the variable selection procedure. As we don't know the underlying data-generating mechanism and as the problem may potentially get worse with multivariate $X$, it will be difficult to know in advance whether we have enough data to guarantee selection of the right variables in order to control for informative censoring and hence Type I error inflation. In the next section, we therefore propose a simple method that has a better chance of selecting the right variables and thereby controlling the Type I error.


\section{Proposal for Variable Selection}
In order to prevent inflated Type I errors, we recommend using a ``double selection'' approach \citep{belloni2014} based on the Lasso.
Double selection was developed in the context of selection of possible confounders for inclusion in the analysis of observational studies. It uses two steps to identify covariates for inclusion, first selecting variables that predict the outcome of interest and then those that predict the exposure/treatment.
Using similar ideas, our proposal
will rely on the selection of variables associated with either survival or censoring (or both).  

In particular, let $X$ now be the $p$-dimensional vector of baseline covariates $(X_1, \dots, X_p)'$, adjustment for which we assume to be sufficient to render censoring independent. For the remainder of the paper, we will focus on the Lasso to select covariates since it is known to perform well with a large number of covariates and it can now be implemented easily in many statistical software packages. However, our proposal can allow for more general variable selection procedures.
Building on the idea of double selection, we perform a two stage selection procedure followed by a final
estimation step as follows:
\begin{enumerate}

    \item Fit a Cox model for survival time $T$ on the treatment $A$ and baseline covariates $X$ using Lasso (penalising all coefficients in the model), and select the covariates with non-zero estimated coefficients:
    $$\lambda\left(t\mid A, X\right)=\lambda_0(t)e^{\alpha A+\beta'X}$$
    \item Fit a Cox model for censoring $C$ on the treatment $A$ and baseline covariates $X$ using Lasso (penalising all coefficients in the model), and select the covariates with non-zero estimated coefficients:
    $$\lambda^C\left(t\mid A, X\right)=\lambda_0^C(t)e^{\gamma_1 A+\gamma_2'X}$$
    \item  Fit a Cox model for the survival time on the treatment $A$ and all covariates selected in either one of the first two steps:
    $$\lambda\left(t\mid A, X_U\right)=\lambda_0^*(t)e^{\alpha^* A+\beta^{*'}X_U},$$
    with $X_U$ the covariates estimated to have non-zero coefficients in Steps $1$ and/or $2$. 
    Inference on the treatment effect $\alpha^*$ may then be performed using conventional methods, provided that a robust standard error is used. 
    
\end{enumerate}
Note that the last regression may also include additional variables that were not selected in the first two steps, but that were identified a priori as being important.

Our proposal differs from standard methods that rely on a single selection step by introducing this extra (second) step. This step helps to identify variables that are weakly-to-moderately predictive of survival but strongly related to censoring. As shown in Section \ref{sec:impact}, eliminating these variables may induce informative censoring, even if all common causes
of survival and censoring are measured, and this may in turn inflate Type I errors. Our procedure will only reject covariates that are weakly associated with both survival and censoring; asymptotically, this is not problematic because the omission of such covariates induces such weak degrees of informative censoring that the resulting bias in the test statistic for the null will be small enough that inference is not jeopardised (see Section \ref{sec:impact}).

The above proposal is closely linked to a more rigorous method proposed by \cite{fang2017} for observational studies. 
These authors construct a decorrelated score function by applying a projection of the score function of the parameter of interest $\alpha$ under model (\ref{eq:unrestr_model}) orthogonal to the nuisance tangent space. In addition to Step $1$ of our proposal, this involves fitting a weighted linear model for the treatment in the risk set across all time points. 
In this approach, the predictors of censoring are indirectly picked up via the treatment model as such predictors become associated with treatment in the risk sets over time.
Our proposal is motivated by the fact that the treatment is randomised (at baseline) and that score tests under a Cox model for censoring have more power to pick up predictors of censoring. For more details we refer the interested reader to Appendix A.3 and \cite{fang2017}.

\section{Monte Carlo Simulations}\label{sec:montecarlo}
In this section, we compare the finite-sample Type I error rate of the logrank test statistic, the test statistic based on the proposed double selection strategy and that based on the post-Lasso approach, which refits the Cox model to the variables selected by the first-step penalized variable selection method (ie, Lasso).
The last two considered approaches require the selection of penalty parameters and base inference for the treatment effect parameter on the robust standard error. 
In our simulation study, we use a $20$-fold cross validation technique with the negative cross-validated penalized (partial) log-likelihood as the loss function.
We obtain the two default penalty parameters $\lambda_\text{min}$, which gives minimum mean cross-validated error, and $\lambda_\text{1se}$, which is the largest value of the penalty parameter such that the cross-validated error is within $1$ standard error of the minimum, using the function \texttt{cv.glmnet()} in the \texttt{R} package \texttt{glmnet}.

In the simulation studies, we generate $n$ mutually independent vectors $\left\{\left(T_{i}, C_{i}, A_{i}, X_{i}'\right)'\right\}, i=1, \ldots, n .$ Here, $X_{i}=\left(X_{i, 1}, \ldots, X_{i, p}\right)$ is a mean zero multivariate normal covariate with covariance matrix $\mathrm{I}_{p \times p}$. The binary treatment $A_i$ is Bernoulli distributed with probability $0.5$ and the $i$th survival and censoring time are based on the following distributions:
\begin{align*}
T_i&\overset{d}{=}\exp(\lambda_{T,i}), \text{ with } \lambda_{T,i}=\exp(\beta_0+\beta' X_i)\\
C_i&\overset{d}{=}\exp(\lambda_{C,i}), \text{ with } \lambda_{C,i}=\exp(\gamma_0+\gamma_1 A_i+\gamma_2'X_i),
\end{align*}
where $\beta_0$, $\gamma_0$ and $\gamma_1$ are scalar parameters, and $\beta=b\cdot\nu_T$ and $\gamma_2=g\cdot\nu_C$ are $p$-dimensional parameters with $b$ and $g$ scalar and $\nu_T$ and $\nu_C$ $p$-dimensional parameters. In our study, we consider the following coefficient vectors $\nu_T$ and $\nu_C$
\begin{enumerate}
    \item $\nu_T=(1, 1/2, \dots,1/9, 1/10, 0_{11}, \dots, 0_p)',$\\
    $\nu_C=(1, 1/2, 1/3, 1/4, 1/5, 1, 1/2,1/3, 1/4, 1/5, 0_{11}, \dots, 0_p)',$
     \item $\nu_T=(1, 1/2, 1/3, 1/4, 1/5, 1, 1/2, 1/3, 1/4, 1/5, 0_{11}, \dots, 0_p)',$\\
   $\nu_C=(1, 1/2, \dots, 1/9, 1/10, 0_{11}, \dots, 0_p)',$
    \item $\nu_T=(1, 1/2, 1/3, 1/4, 1/5, 1, 1/2, 1/3, 1/4, 1/5, 0_{11}, \dots, 0_p)',$\\
   $\nu_C=(1/5,1/4, 1/3,1/2, 1, 1/5,1/4, 1/3, 1/2, 1, 0_{11}, \dots, 0_p)',$
\end{enumerate}
where the subscripts indicate the index (i.e., position) of $0$ in the vector. 
The coefficients $\nu_T$ and $\nu_C$ are used to control the association between the covariates and respectively the survival and censoring times. The coefficients $\beta_0$ and $\gamma_0$ are set equal to $0$, corresponding with $\lambda_0(t)=1$.
For the data generating mechanisms described
above, we perform $1,000$ Monte Carlo runs for $n = 400$ and $p = 30$.
In these set-ups, the coefficients
feature declining patterns, with the smallest coefficients being hard to distinguish from zero at the given
sample size. Therefore, we would expect the single step model selection procedure to mistakenly remove variables with smaller coefficients.

\begin{figure}[htp]
\centering

        \includegraphics[width=0.32\textwidth]{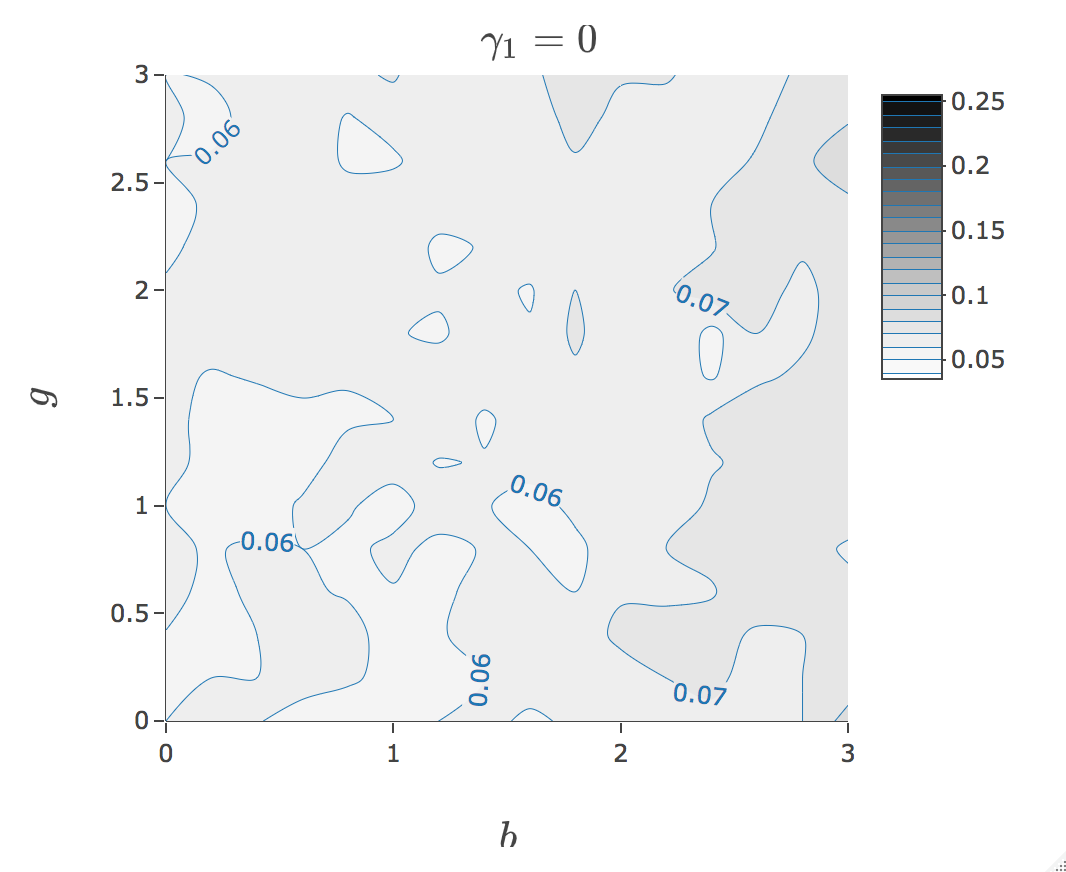}
        \includegraphics[width=0.32\textwidth]{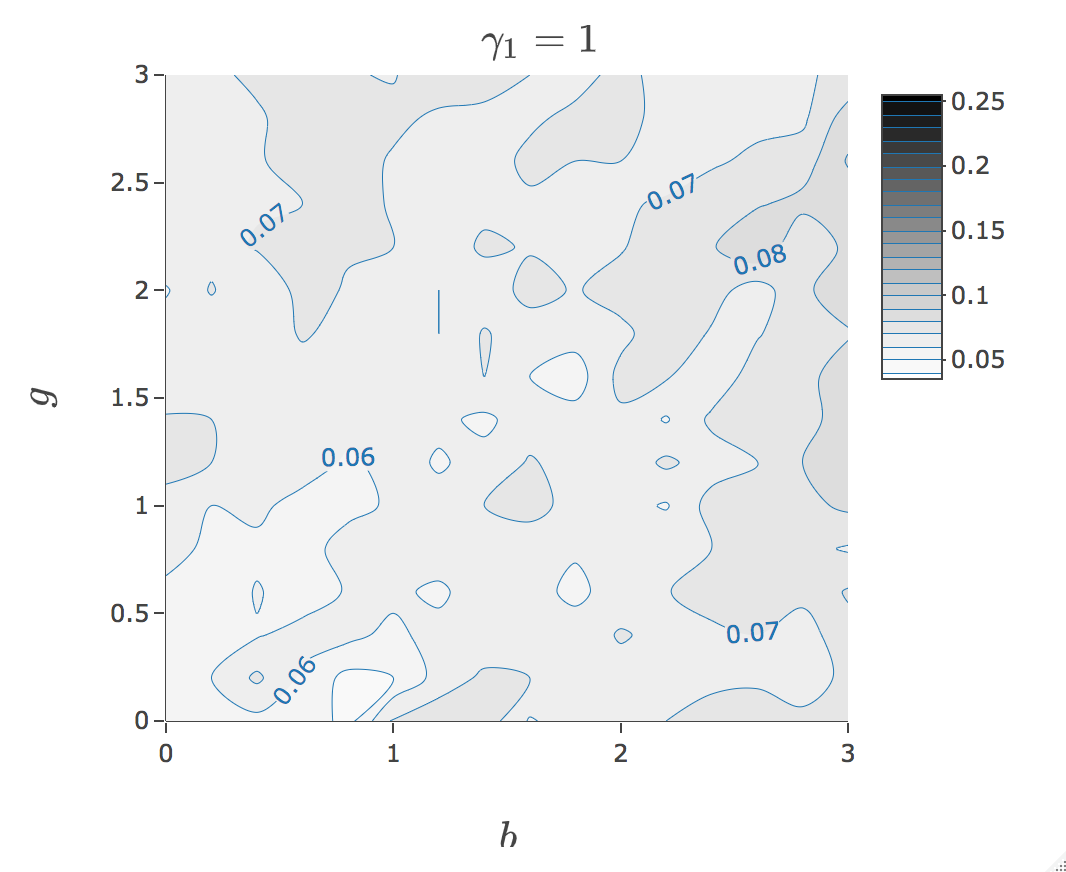}
        \includegraphics[width=0.32\textwidth]{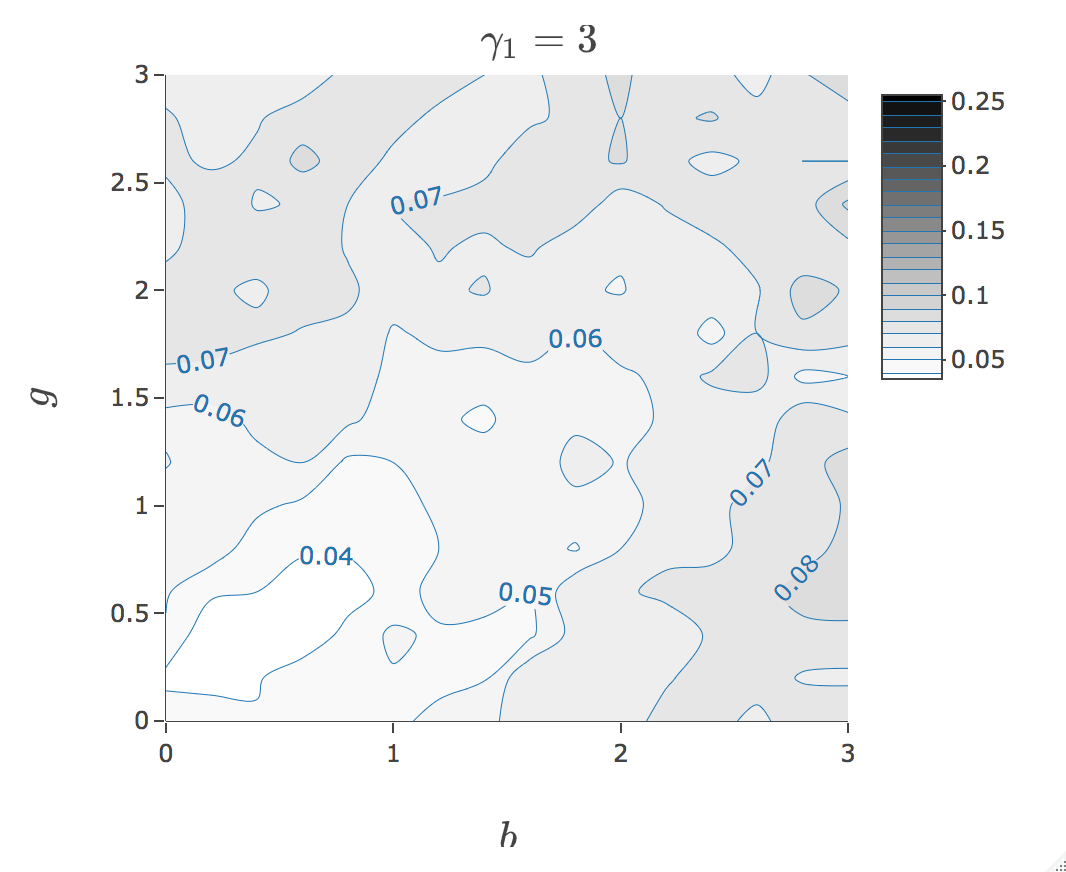}
\medskip

        \includegraphics[width=0.32\textwidth]{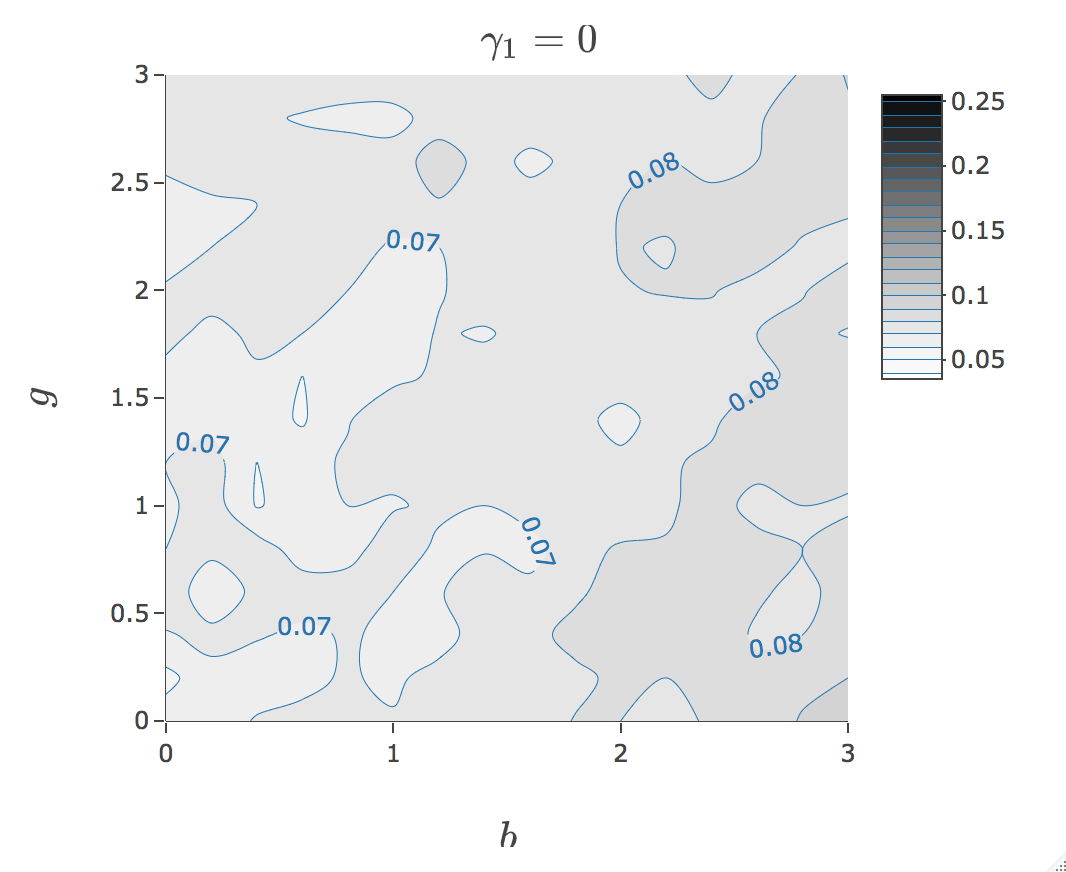}
        \includegraphics[width=0.32\textwidth]{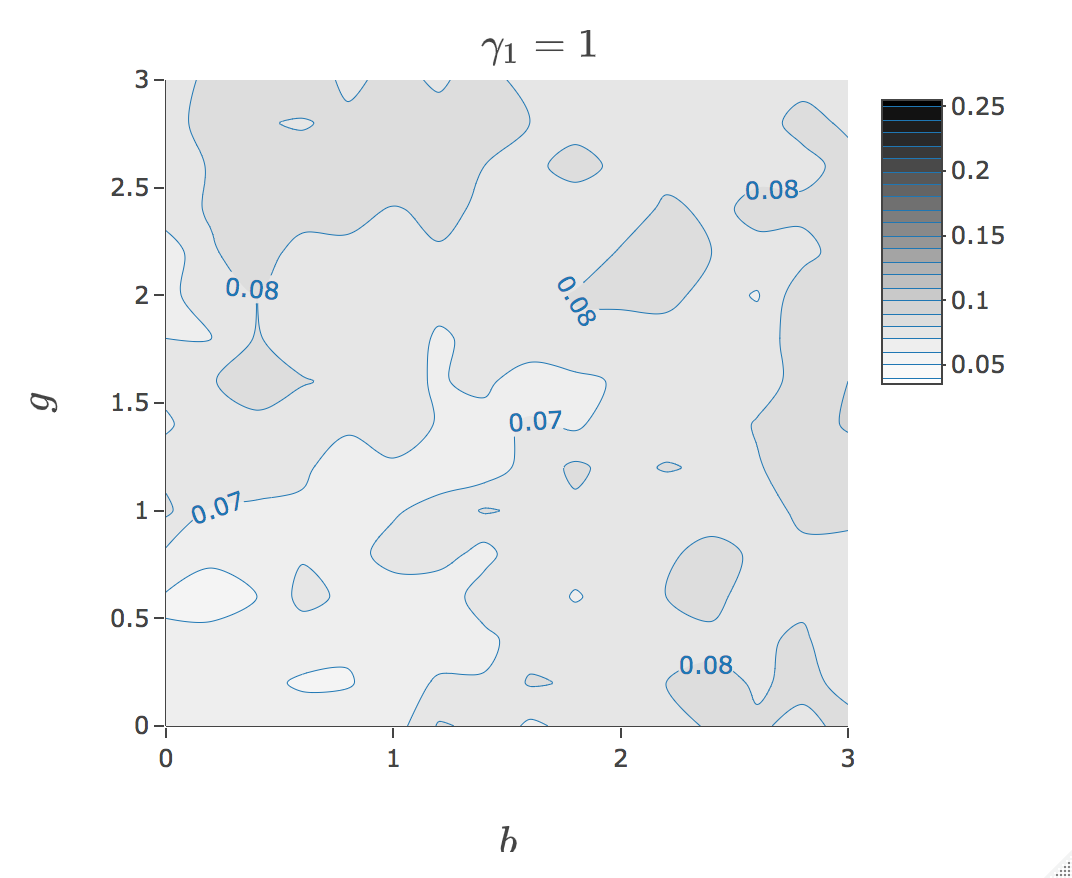}
        \includegraphics[width=0.32\textwidth]{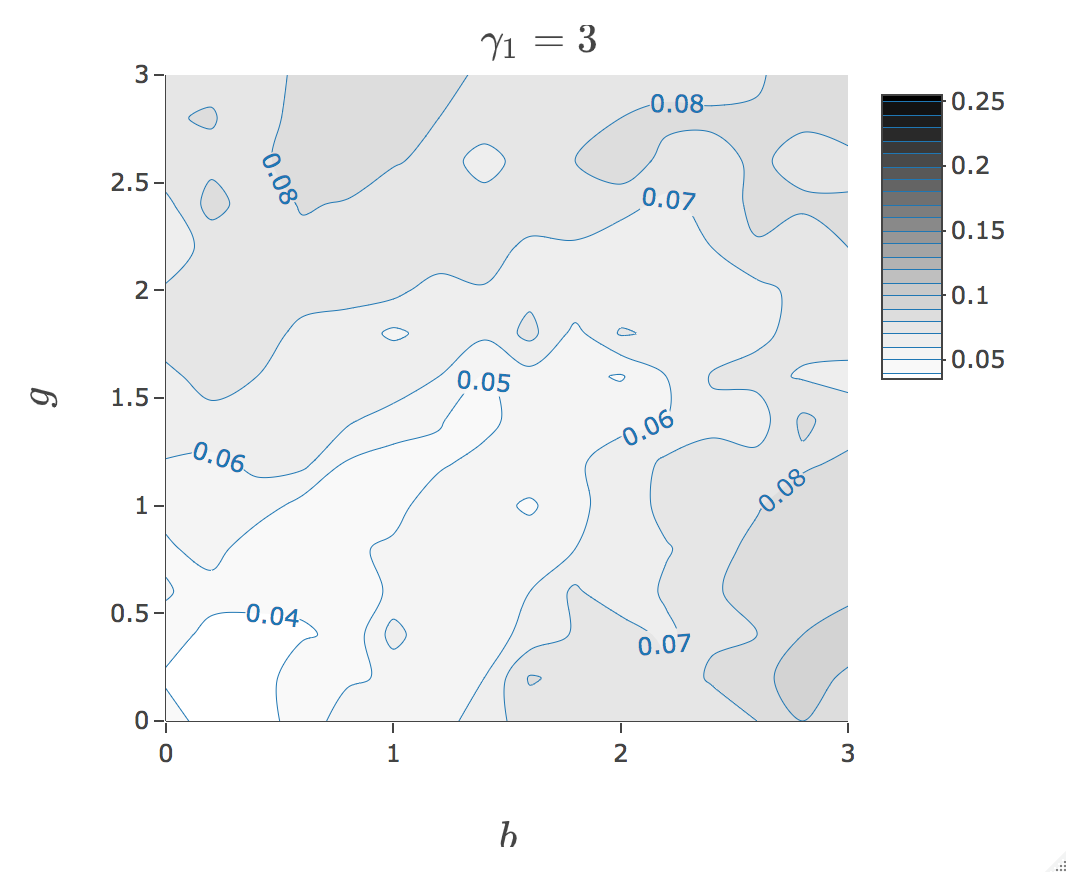}
\medskip

        \includegraphics[width=0.32\textwidth]{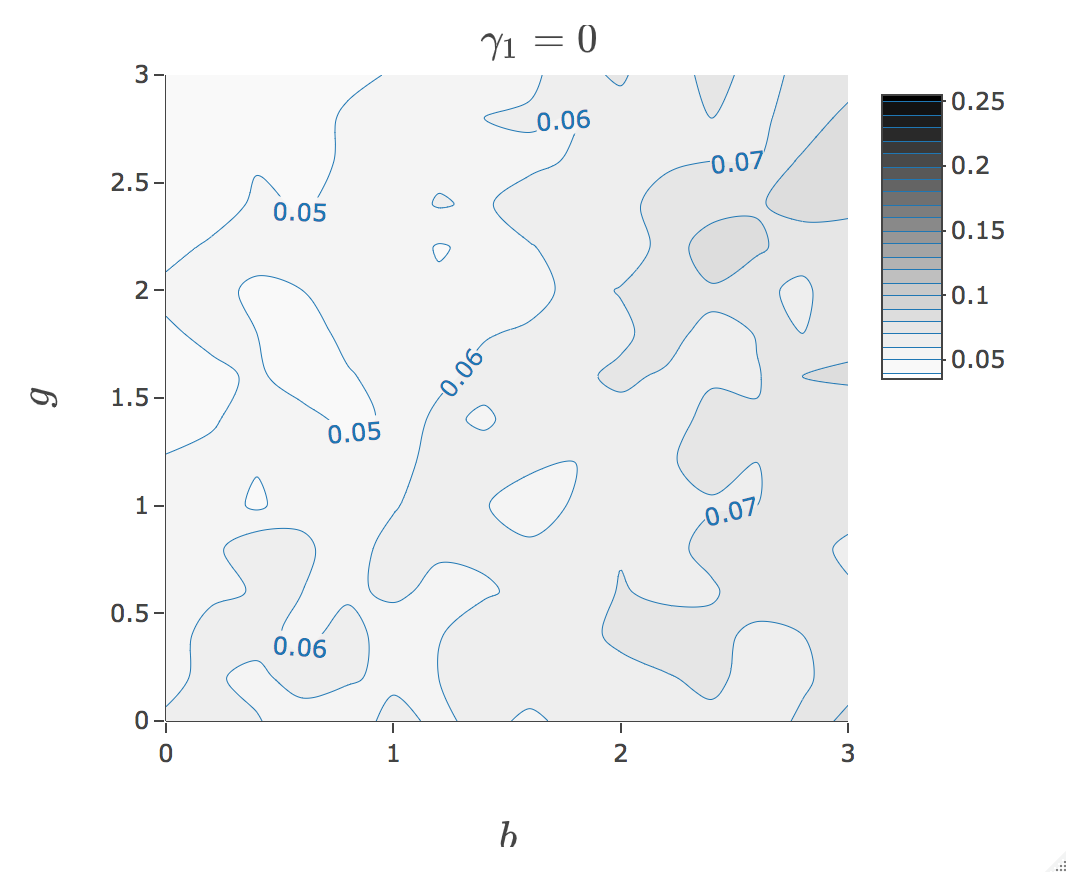}
        \includegraphics[width=0.32\textwidth]{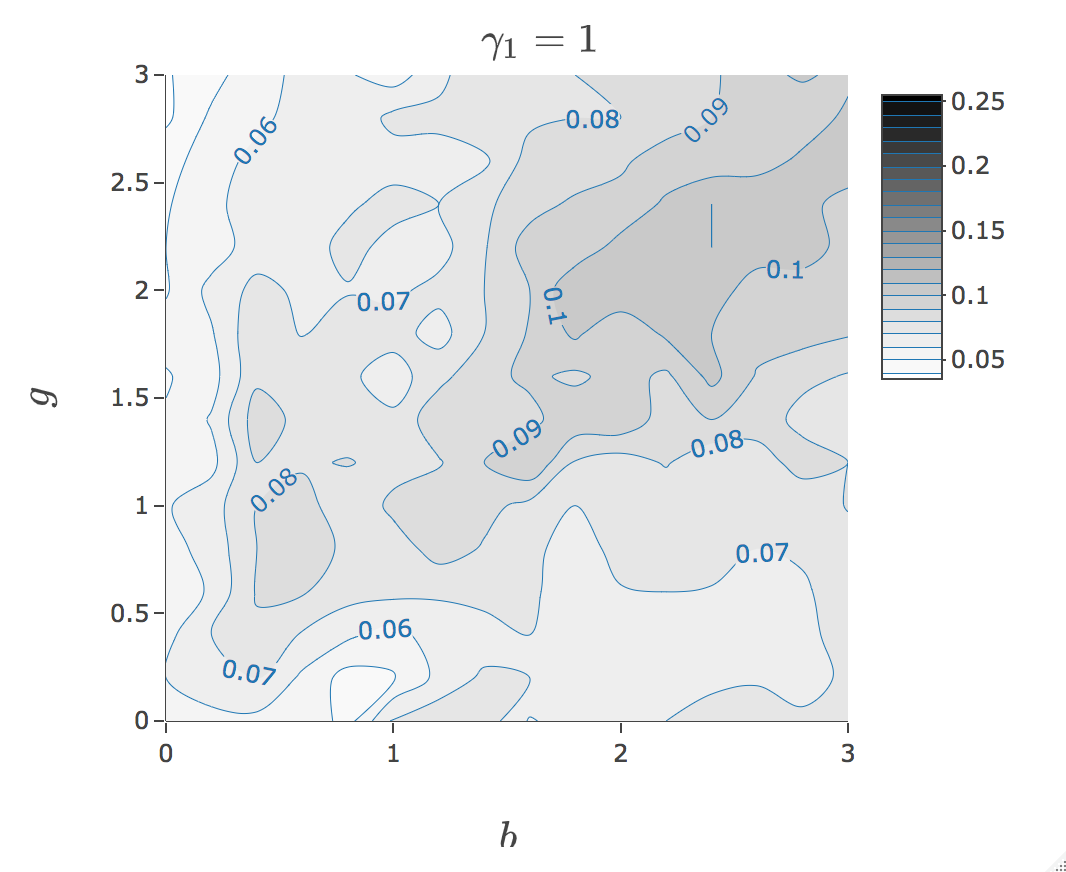}
        \includegraphics[width=0.32\textwidth]{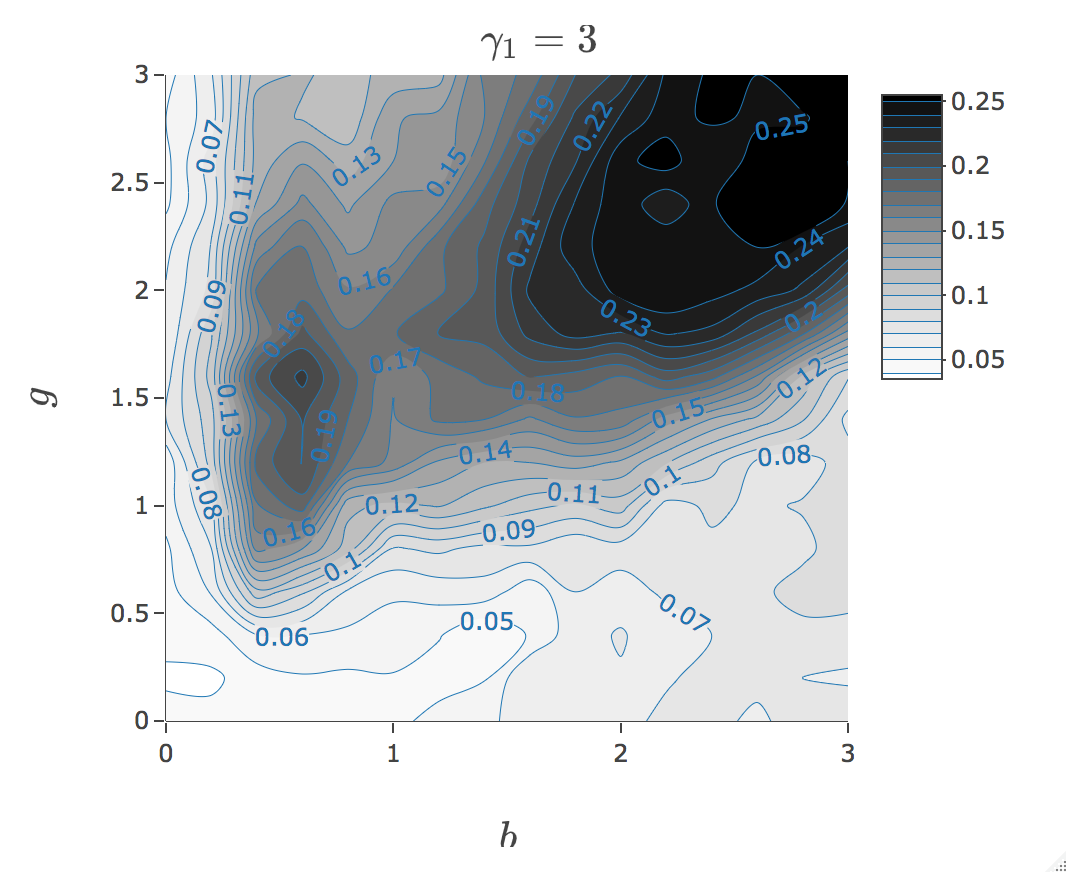}
\medskip

        \includegraphics[width=0.32\textwidth]{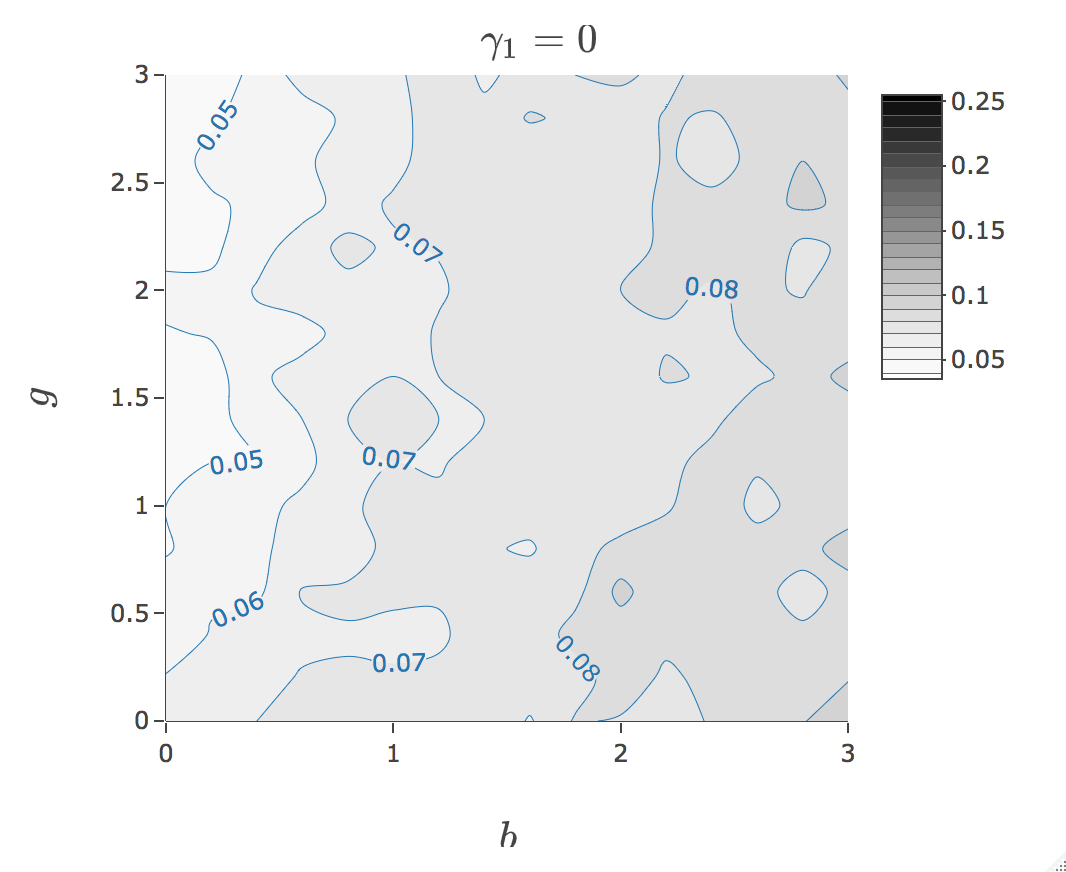}
        \includegraphics[width=0.32\textwidth]{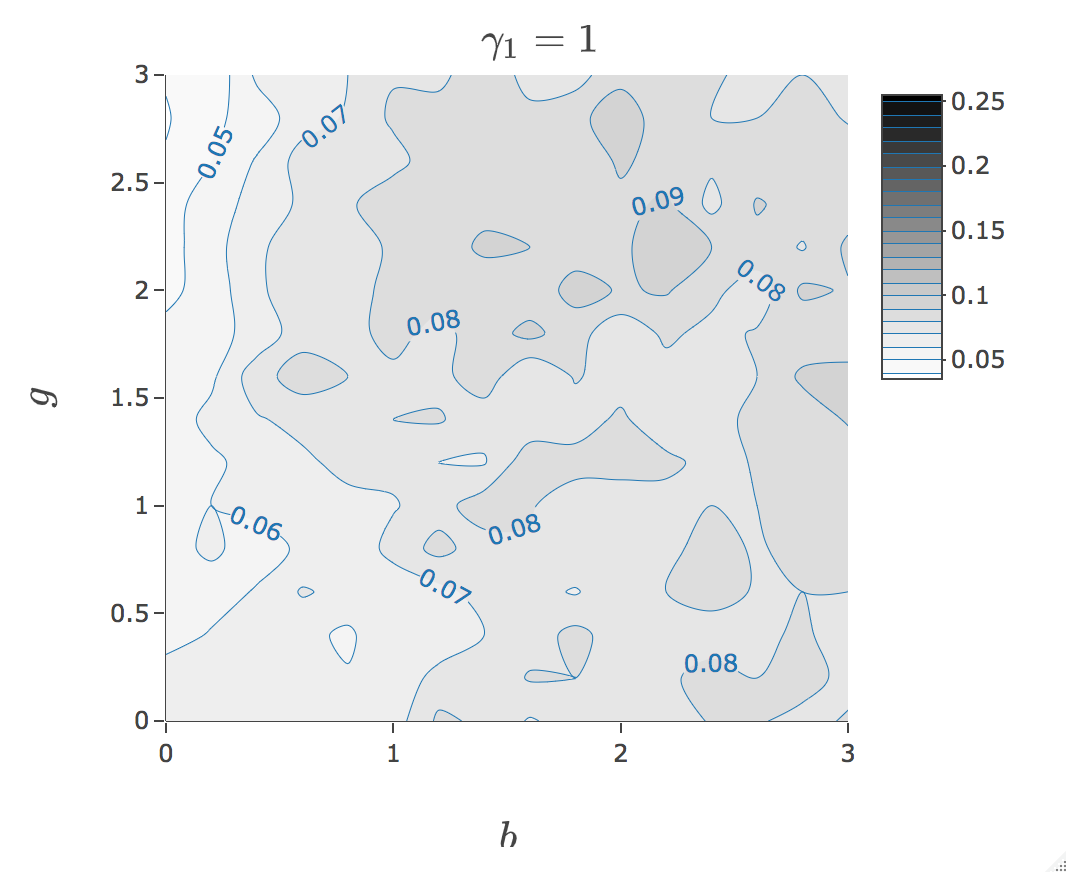}
        \includegraphics[width=0.32\textwidth]{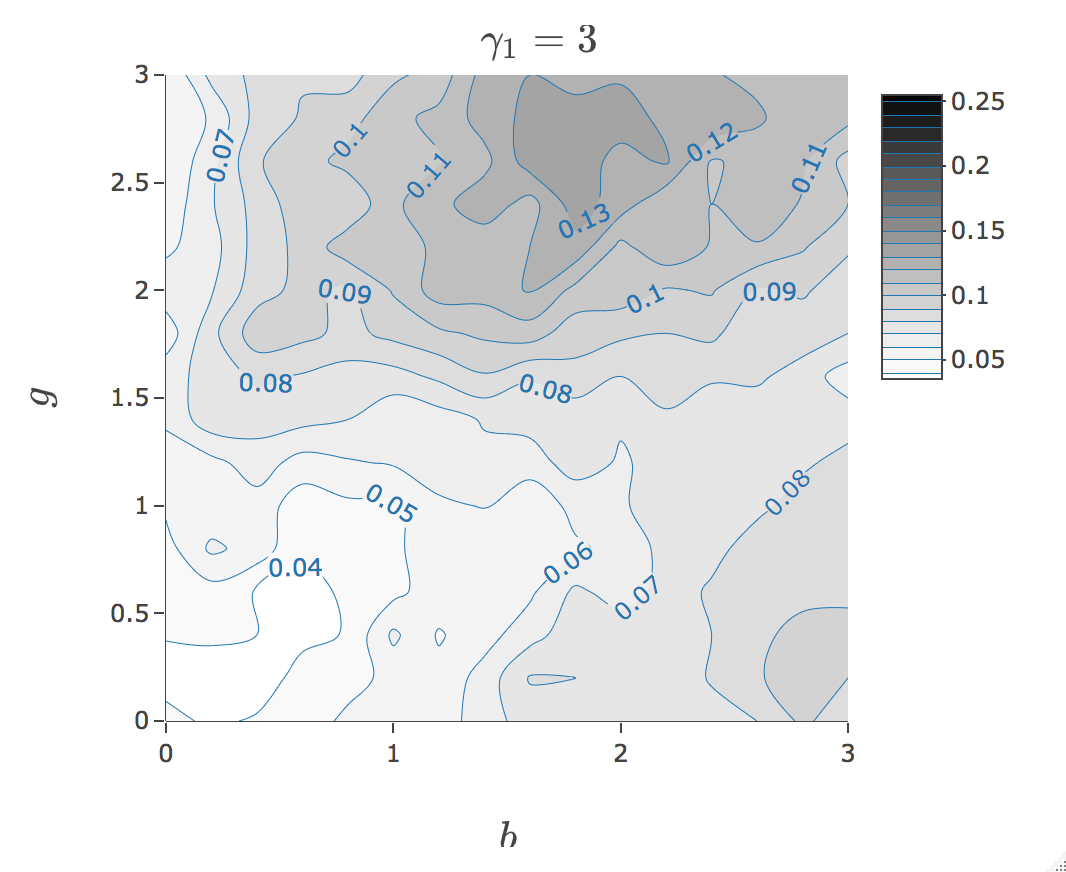}

\caption{Empirical Type I error rate at the $5\%$ significance level of the post-Lasso and the proposed double selection method under Setting $1$. Row $1$: double selection method with penalty parameter $\lambda_{1se}$; Row $2$:  double selection method with penalty parameter $\lambda_{min}$; Row $3$: Post-Lasso with penalty parameter $\lambda_{1se}$ ; Row $4$: Post-Lasso with penalty parameter $\lambda_{min}$. Left: $\gamma_1=0$; Middle: $\gamma_1=1$; Right: $\gamma_1=3$.
Results are based on $1,000$ simulations, $n=400$, $p=30$, $\lambda_0(t)=1$ ($\beta_0=0$) and $\lambda_0^C(t)=1$ ($\gamma_0=0$).
}
\label{fig:Sim_results1}
\end{figure}

\begin{figure}[htp]
\centering

        \includegraphics[width=0.32\textwidth]{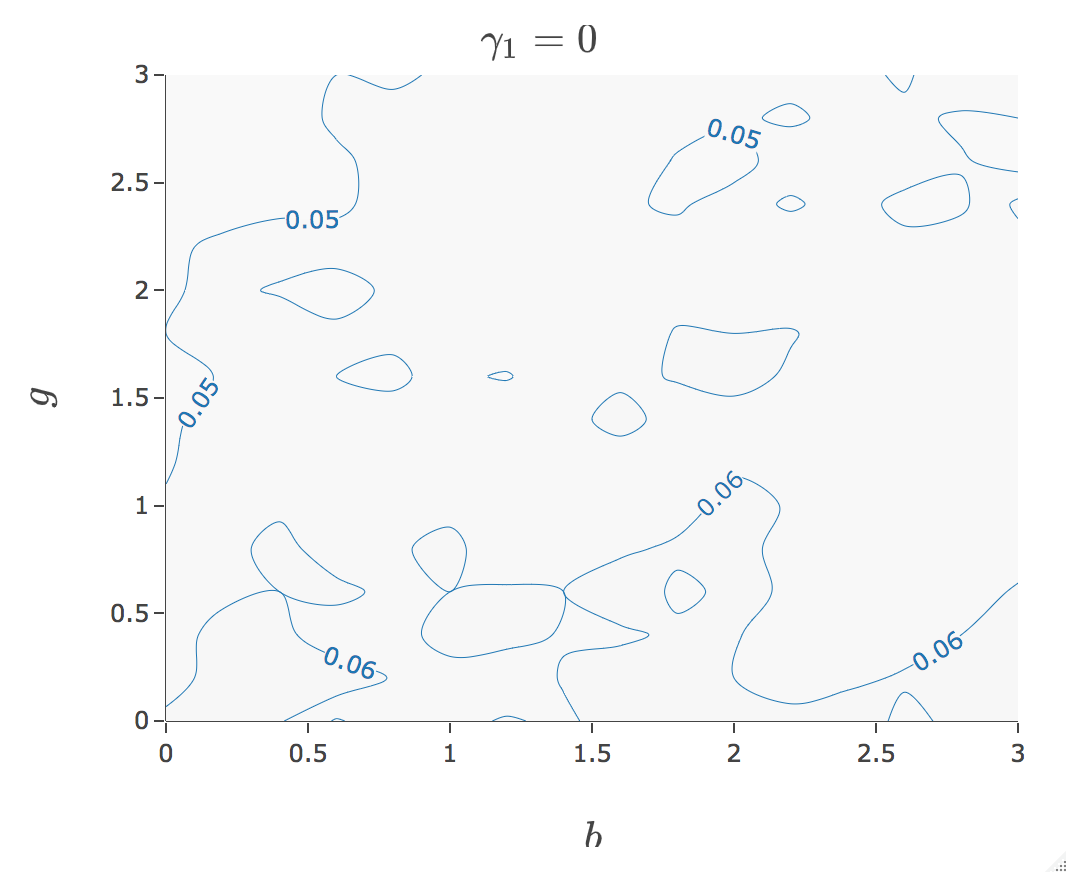}
        \includegraphics[width=0.32\textwidth]{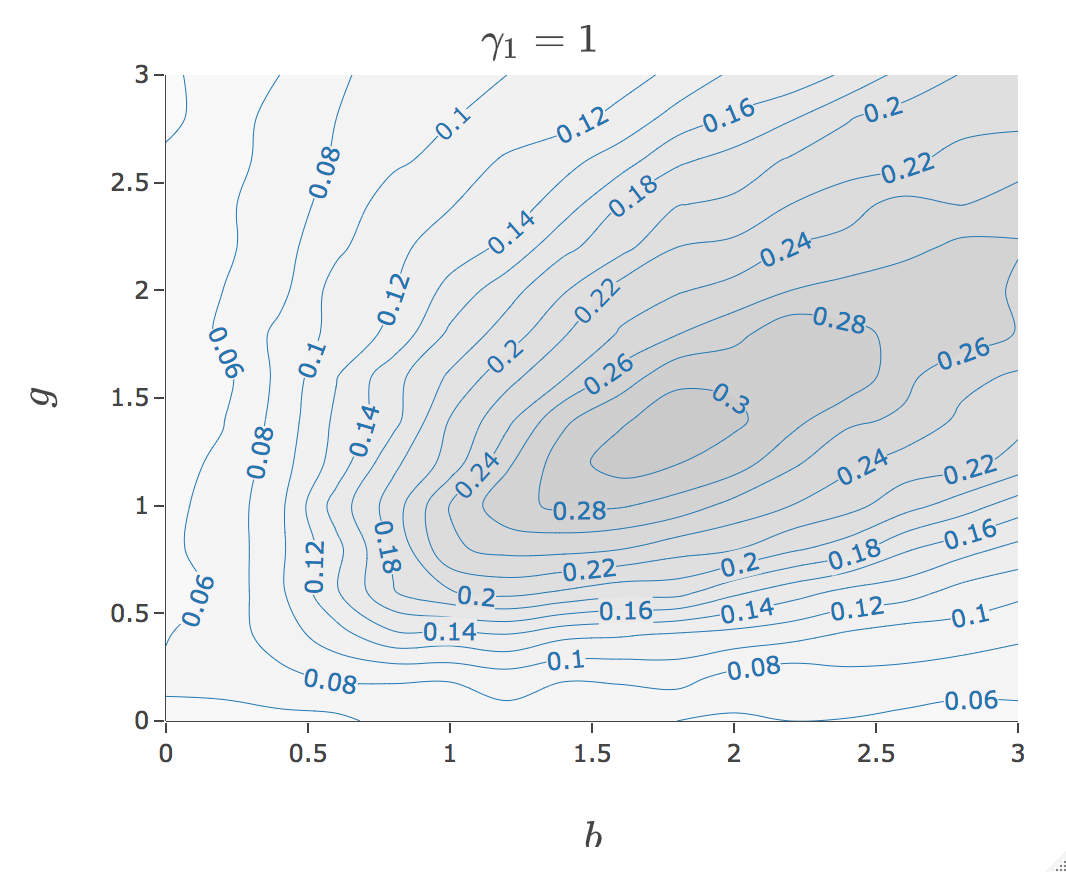}
        \includegraphics[width=0.32\textwidth]{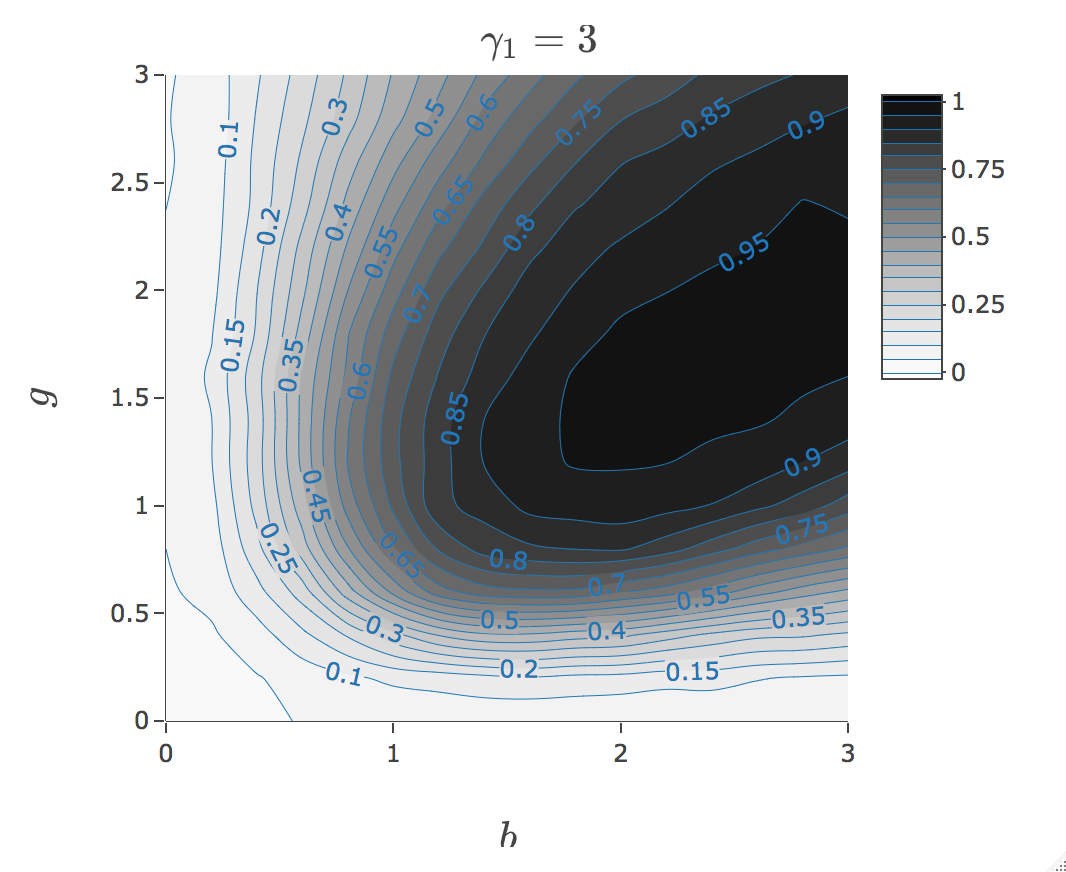}

\caption{Empirical Type I error rate at the $5\%$ significance level of the logrank test under Setting 1. Left: $\gamma_1=0$; Middle: $\gamma_1=1$; Right: $\gamma_1=3$.
Results are based on $1,000$ simulations, $n=400$, $p=30$, $\lambda_0(t)=1$ ($\beta_0=0$) and $\lambda_0^C(t)=1$ ($\gamma_0=0$).
}
\label{fig:Sim_results1_LRT}
\end{figure}

The empirical type I errors for Setting $1$ are summarized in Figure \ref{fig:Sim_results1} and Figure \ref{fig:Sim_results1_LRT}. 
First, note that the Type I error is maintained for all considered tests when no baseline covariate is simultaneously associated with both $C$ and $T$ (ie, $b=0$ or $g=0$; see Section \ref{sec:impact}). 
Moreover, the empirical
Type I errors of the different tests are close to the desired significance level of $5\%$ for $\gamma_1=0$.
For values of $\gamma_1$ deviating from zero, the Type I error rate of the logrank test is highly inflated throughout large parts of the parameter space (see Figure \ref{fig:Sim_results1_LRT}). This subspace as well as the inflation itself, become larger with stronger strength of association between $A$ and $C$ (ie, absolute value of $\gamma_1$). 
Surprisingly, the na\"ive test based on the post-Lasso still performs reasonably well for $\gamma_1$ close to zero.
However, better results are achieved using our proposal (see column $2$ in Figure \ref{fig:Sim_results1}).
When $\gamma_1=3$, the Type I error rates based on the post-Lasso selection deviate
strongly from the desired significance level of $5\%$ throughout large parts of the parameter space (Rows $3$ and $4$ in Figure \ref{fig:Sim_results1}). This is a result of eliminating too many covariates from the outcome model (Rows $3$ and $4$). Although using the value of the penalty parameter that gives minimum mean cross-validated error (ie, $\lambda_\text{min}$) tends to improve the results (Row $4$), it does not resolve the problem entirely.
In contrast, the empirical
Type I errors of both tests based on double selection (Row $1$ and $2$) are substantially closer to the desired significance level of $5\%$, which confirms our expectations. This holds across the different parameter values considered. In addition, although the proposal seems to have best performance with $\lambda_\text{1se}$, the simulation results for the proposal are less sensitive to the choice of the penalty parameter (ie, $\lambda_\text{min}$ or $\lambda_\text{1se}$) compared to the post-Lasso approach. 
We refer the readers to Appendix B for the detailed
results under Setting $2$ and $3$.


\section{Understanding the Selection Bias due to Censoring}
While the double selection proposal performs well across the different data generating mechanisms considered, somewhat surprisingly the Type I error of the na\"ive Lasso-based test is almost as accurate in several settings.
To develop insight into this, we will use causal diagrams. For pedagogical purposes, we will focus on settings with a single covariate.

\begin{figure}[h!]
\begin{subfigure}{.5\linewidth}
\centering
\begin{tikzpicture}[scale=1]

\node[] (a) at (0, 0)   {$A$};
\node[] (t) at (4, 0)   {$T$};
\node[] (c) at (2, 0)   {$C$};
\node[] (x) at (3,2) {$X$};
\path[->] (a) edge node {} (c);
\path[->] (x) edge node {} (c);

\end{tikzpicture}
\caption{Non-informative censoring and $A\nbigCI C$} \label{fig:DAG_LRT.2a}
\end{subfigure}\hfill
\begin{subfigure}{.5\linewidth}
\centering
\begin{tikzpicture}[scale=1]

\node[] (a) at (0, 0)   {$A$};
\node[] (t) at (4, 0)   {$T$};
\node[] (c) at (2, 0)   {$C$}; 
5

\node[] (x) at (3,2) {$X$};
\path[->] (a) edge node {} (c);
\path[->] (x) edge node {} (t);

\end{tikzpicture}
\caption{Non-informative censoring and $A\nbigCI C$} \label{fig:DAG_LRT.2}
\end{subfigure}
\begin{subfigure}{.5\linewidth}
\centering
\begin{tikzpicture}[scale=1]
\node[] (a) at (0, 0)   {$A$};
\node[] (t) at (4, 0)   {$T$};
\node[] (c) at (2, 0)   {$C$};
\node[] (x) at (3,2) {$X$};
\path[->] (x) edge node {} (c);
\path[->] (x) edge node {} (t);

\end{tikzpicture}
\caption{Informative censoring and $A\bigCI C$} \label{fig:DAG_LRT.1}
\end{subfigure}\hfill
\centering
\begin{subfigure}{.5\linewidth}
\centering
\begin{tikzpicture}[scale=1]

\node[] (a) at (0, 0)   {$A$};
\node[] (t) at (4, 0)   {$T$};
\node[] (c) at (2, 0)   {$C$};
\node[] (x) at (3,2) {$X$};
\path[->] (a) edge node {} (c);
\path[->] (x) edge node {} (t);
\path[->] (x) edge node {} (c);

\end{tikzpicture}
\caption{Informative censoring and $A\nbigCI C$} \label{fig:DAG_LRT.3}
\end{subfigure}
\caption{DAGs under different censoring mechanisms.} \label{fig:DAG_LRT}
\end{figure}
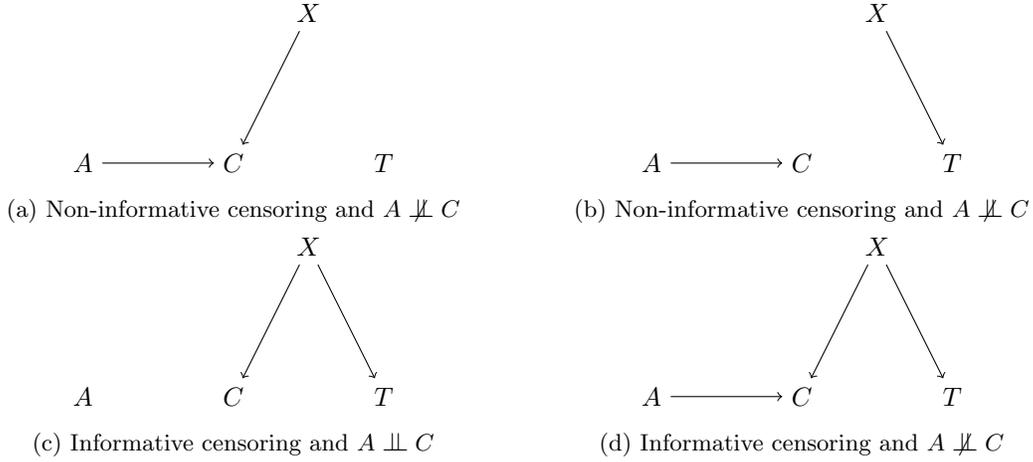

Surprisingly the validity of the three tests of the null hypothesis of no treatment effect is maintained even if censoring does depend on survival time, as long as censoring is independent of treatment. 
To see why, in a Directed Acyclic Graph (DAG), if indeed $A$ has no causal effect on $T$ and $C$ (which is the case since there is no arrow going from $A$ to $T$ or from $A$ to $C$), then, even if censoring depends on the survival time through the backdoor pathway $C\leftarrow X \rightarrow T$, there is no path connecting $A$ and $T$ in the risk set at each time point (see Figure \ref{fig:DAG_LRT.1}). This means that, even if censoring is strongly informative, no systematic association between $A$ and $T$ could be observed, and hence a valid test of the null will be obtained. 
Analyses unconditional on $X$ are valid under the null hypothesis if censoring is (statistically) independent of survival time (see Figures \ref{fig:DAG_LRT.2a} and \ref{fig:DAG_LRT.2}).
This validity, however, breaks down when censoring does depend on $A$ and $T$. To see why, note that the hazard, by definition, conditions on being in the risk set. In particular, once censoring is involved, then by conditioning on the risk set, one is also conditioning on patients being uncensored. If we consider the causal diagram in Figure \ref{fig:DAG_LRT.3}, it is clear that we thereby condition on a `collider', thus opening a spurious association between treatment $A$ and $X$ in the risk set at each time point.
This phenomenon, where a distorted association between treatment $A$ and outcome $T$ is induced when in fact none exists, is better known as collider stratification bias \citep{greenland1999}. This can lead to inflated Type I error rates. 

This can also be seen in the expression for (the bias in) the score (Equation \ref{eq:score_LRT}) in Appendix A.1, which equals zero if either $C$ is independent of $T$ (corresponding with $\beta=0$ or  $\gamma_2=0$) or $C$ is independent of $A$ (corresponding with $\gamma_1=0$). This is also shown more rigorously in Appendix A.2. 

To summarize, theory suggests that failing to adjust for $X$ is only problematic in settings where censoring is informative (due to $X$) and depends on treatment. In particular, the logrank test is a valid test of the null if censoring is either independent of treatment or independent of the survival time in each treatment arm.
Correspondingly, in settings where censoring depends on treatment and survival, variable selection approaches may induce an inflated Type I error if they fail to pick up a common cause of $C$ and $T$.
As biases from collider-stratification often tend to be much smaller than confounding bias \citep{Greenland2003}, the inflation in Type I error tends to be more limited than that due to selecting confounders in observational studies. 
This explains why the post-Lasso is performing reasonably well in several settings where one would expect it to fail; ie, in settings where (some) variables are weakly associated with $T$ and strongly associated with $C$ (see Section \ref{sec:montecarlo}).
In particular, its inflation in Type I error is only severe when censoring is strongly associated with treatment and informative.
Even so, the proposed double selection approach, by making it more likely to pick up the right variables, controls the Type I error better, even in these more extreme yet plausible settings.

The previous results suggest that we can obtain valid tests of the null hypothesis of no treatment effect under different data-generating mechanisms, even if we fail to adjust for (certain) baseline covariates. As we will see, this does not imply that Kaplan-Meier curves or marginal survival curves obtained via the Cox model after na\"ive (eg, post-Lasso) selection are unbiased. This is important as investigators are increasingly encouraged to also present marginal survival curves rather than just testing the null hypothesis \citep{austin2014}. In the following section, we therefore discuss the implications for estimated survival curves and give a proposal to reduce bias in estimated survival curves after variable selection.

\section{Implications for Estimated Survival Curves}
\begin{figure}[h!]
    \centering
    \begin{subfigure}[b]{0.45\textwidth}
        \includegraphics[width=\textwidth]{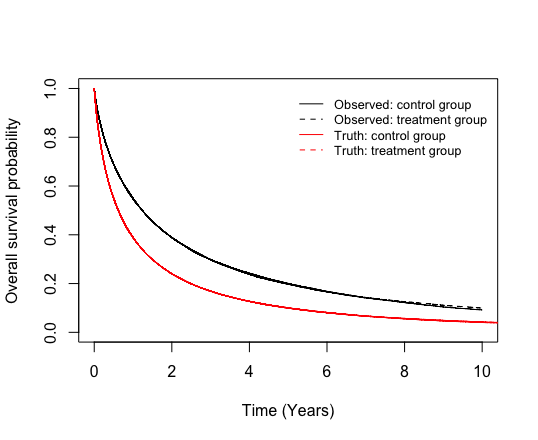}
        \caption{Censoring independent of treatment ($b=1$, $\gamma_1=0$ and $g=2$).}
        \label{fig:KM_LRT.1}
    \end{subfigure}
    \begin{subfigure}[b]{0.45\textwidth}
        \includegraphics[width=\textwidth]{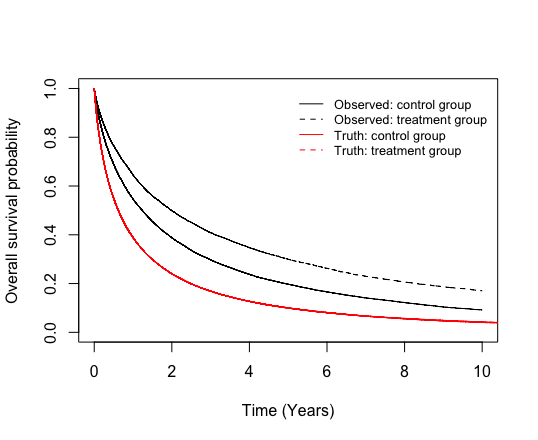}
        \caption{Censoring dependent on treatment ($b=1$, $\gamma_1=2$ and $g=2$).}
        \label{fig:KM_LRT.2}
    \end{subfigure}
    \caption{True and estimated Kaplan-Meier surival curves under an informative-censoring setting. Data are simulated under Setting 1 (see Section \ref{sec:montecarlo}). Results are based on one dataset with $n=100,000$, $p=30$, $\lambda_0(t)=1$ ($\beta_0=0$) and $\lambda_0^C(t)=1$ ($\gamma_0=0$).}
\label{fig:KM_LRT}
\end{figure}

Unbiasedness of the Kaplan–Meier estimator relies on the assumption that there is no dependence between time to event $T$ and censoring $C$. 
To illustrate the possible bias under dependent censoring, Figure \ref{fig:KM_LRT} shows the traditional Kaplan-Meier estimator and the true survival curves for a simulated dataset in which covariates are associated with both time to event and time to censoring. 
Note that Kaplan-Meier survival curves are biased if treatment has no effect on both censoring and survival, but in the same way across treatment groups.
In particular, there is then no difference between the observed curves in both treatment arms (in large samples), regardless of whether survival and censoring are independent (see Figure \ref{fig:KM_LRT.1}). As shown in Figure \ref{fig:KM_LRT.2}, the validity of the logrank test is no longer maintained if censoring depends on both treatment and survival. In that case the two observed curves no longer overlap.

It follows from the above that correction for dependent censoring is more important to obtain valid/interpretable survival curves than it is to test equality of survival curves in two treatment arms. 
To avoid bias, regression standardization -also referred to as 'directly adjusted survival curves' \citep{Gail1986}- based on a model that includes all variables necessary to render $T$ independent of $C$ can then be employed \citep{Rothman2008}. 
This method uses the regression model to predict the risk of survival at each time, for treated and untreated/controls separately, at every observed level of the measured baseline covariates.  These predictions are then averaged over the baseline covariate distribution observed in the sample to produce the survival function under treatment and control.

\begin{figure}[h!]
    \centering
    \begin{subfigure}[b]{0.45\textwidth}
        \includegraphics[width=\textwidth]{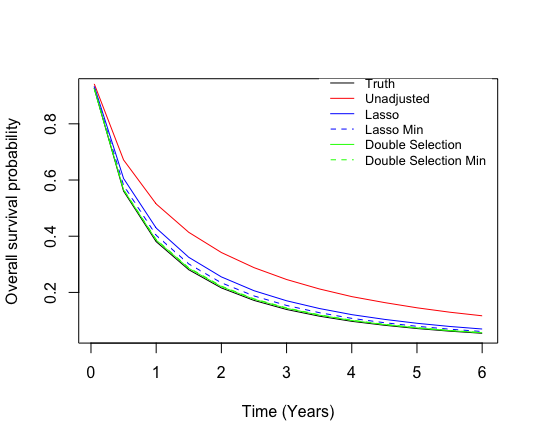}
        \caption{Censoring independent of treatment ($b=0.8$, $\gamma_1=0$ and $g=1.6$).}
    \end{subfigure}
    \begin{subfigure}[b]{0.45\textwidth}
        \includegraphics[width=\textwidth]{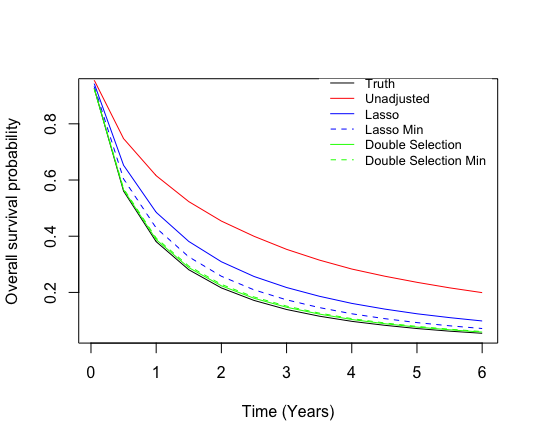}
        \caption{Censoring dependent on treatment ($b=0.8$, $\gamma_1=2$ and $g=1.6$).}
    \end{subfigure}
    \caption{True survival curves, unadjusted (Kaplan-Meier based) survival curves and adjusted/standardized survival curves based on the post-Lasso and double selection models evaluated at $13$ timepoints ($0, 0.5, \dots, 5.5, 6$).
    Results are shown for $A=1$ based on $1,000$ simulations under Setting 1 (see Section \ref{sec:montecarlo}), $n=400$, $p=30$, $\lambda_0(t)=1$ ($\beta_0=0$) and $\lambda_0^C(t)=1$ ($\gamma_0=0$). 
    Truth: True survival curves; Unadjusted: unadjusted/Kaplan-Meier survival curves; Double Selection: standardized survival curves based on double selection model with penalty parameter $\lambda_{1se}$; Double Selection Min: standardized survival curves based on double selection model with penalty parameter $\lambda_{min}$; Lasso: standardized survival curves based on Post-Lasso model penalty parameter $\lambda_{1se}$ ; Lasso Min: standardized survival curves based on Post-Lasso model with penalty parameter $\lambda_{min}$.
    }
\label{fig:standardized}
\end{figure}

We performed a limited simulation study to evaluate the performance of standardized survival curves based on post-Lasso and double selection. 
Figure \ref{fig:standardized} compares standardized survival curves to the true and Kaplan-Meier survival curves evaluated at $13$ timepoints ($0, 0.5, \dots, 5.5, 6$) under treatment. 
We see that standardization reduces bias in the estimated survival curves, with the ones based on the proposed double selection approach closest to the real survival curve in all settings.  
This improved performance is due to the higher chance of selecting the right variables; ie, the variables that are associated with both survival and censoring. This in turn reduces the impact of selection bias.
We repeated these simulations for several different data generating mechansims. Since the results were largely unchanged, the additional results are shown in Figures \ref{fig:curves0} and \ref{fig:curves1} in Appendix B.

Although the combination of double selection and regression standardisation performed well in simulations, it lacks a rigorous justification.
An alternative with better theoretical justification would be to use augmented inverse probability of censoring weighting (AIPCW) \citep{Robins1992} with separately fitted Cox models for the time to event and censoring. 
The resulting estimators, by virtue of being double robust, are not sensitive to the use of standard variable selection procedures, provided that both Cox regression models are correctly specified, for similar reasons as in \cite{Farrell2015}.

Whilst the AIPCW methodology, combined with data-adaptive estimation of nuisance parameters, will likely play an important role in future methodological developments, the focus of this article is how to improve upon standard practice in the analysis of trials with time-to-event endpoints. We have given attention to simple methods that can be implemented using tools familiar to trial statisticians; the use of weighting-based approaches (as well as machine learning for time-to-event endpoints) is currently not widespread. Moreover, in settings where covariates and treatment are strongly predictive of censoring, our proposal may yield better finite sample performance compared with AIPCW estimators by avoiding the use of weights (which can become extreme), despite the lack of theoretical justication. 

\section{Data Analysis}
We illustrate the proposal on data from the PBC-3 trial \citep{Lombard1993}, a multicenter randomized clinical trial conducted in six European hospitals with patient accrual between January 1983 and January $1987$. In this period, 349 patients with the liver disease primary biliary cirrhosis (PBC) were randomized to either treatment with Cyclosporin A (CyA, 176 patients) or placebo (173 patients).
The purpose of the study was to investigate the effect of treatment with CyA (compared to placebo) on the survival time. An increased use of liver transplantation for patients with this disease made the investigators redefine the main response variable to be time to ``failure of medical treatment'' defined as either death or liver transplantation.
At entry a number of possible prognostic factors were measured:
histological stage (1-2-3-4), previous gastrointestinal bleeding (yes/no), creatinine (micromoles/L), serum bilirubin (micromoles/L), serum albumin (g/L), alkaline phosphatase (IU/L), aspartate transaminase (IU/L), body weight (kg), age (years) and sex.

An unadjusted analysis comparing survival between the two arms (where patients receiving transplant were censored) showed no statistical differences between the two arms (Logrank with robust SE $p=0.88$). The same was true for an additional test for the combined outcome of progression to death or transplant (Logrank with robust SE $p=0.78$). 
In the original study, the possibility of a chance imbalance between the arms at entry with respect to important prognostic factors motivated \cite{Lombard1993} to conduct a multivariate analysis. Variables for the Cox proportional hazards regression model were identified by backward elimination procedures. Cox multivariate analysis showed time from entry to death or transplantation was significantly prolonged in the CyA-treated group compared to placebo. On the other hand, for the Cox model with death as the sole endpoint (transplants censored), no significant effect of treatment was found.

A multivariate analysis can take these imbalances into account, but is recommended even more generally. This because the assumption of non-informative censoring is more plausible conditional on covariates; hence a log rank test may not be valid. 


A complication of multivariate analysis is that data may be missing for certain patients on one or more baseline covariates. 
This was the case here: the percentage of patients with missing data on histological stage was 16.6 (58 our of 349).
In view of this, we first selected variables using the 275 complete cases (141 in CyA arm and 134 in placebo arm), and then performed multiple imputation before conducting all regression analyses. While this approach may not be optimal \citep{wood2008}, it improves upon fitting adjusted Cox models using complete cases only. Five different imputed datasets were created for each endpoint separately. 
The imputed values were obtained using the function \texttt{smcfcs} in the eponymous package, which imputes missing values of covariates using Substantive Model Compatible Fully Conditional Specification \citep{Bartlett2015}. On each imputed dataset, we then fitted a Cox model for time to the considered endpoint (ie, all deaths or the combined endpoint) on treatment and the variables selected either by i) the Lasso and ii) the double selection approach. 
After analyzing each imputed dataset, the five sets of results were pooled using Rubin's Rules \citep{rubin1987}. At the selection stage, we used the grouped LASSO offered by the \texttt{grpreg} package in order to deal with categorical predictors in LASSO regression. 

Considering main effects only, performing post-Lasso with penalty parameter $\lambda_\text{1se}$ (selected via leave-one-out cross-validation) for the Cox model with death as the
sole endpoint (transplants censored), resulted in a model adjusted for histological stage, previous gastrointestinal bleeding, serum bilirubin, serum albumin, weight, age and sex. 
No significant treatment effect was found ($p=0.23$). 
The double selection method with penalty parameter $\lambda_\text{1se}$ additionally included the variables hospital ($7$ levels) and creatinine, which also did not render the treatment effect on time to death significant ($p=0.13$).
In the model for progression to death or transplantation, only histological stage, serum bilirubin, serum albumin and sex were selected by post-Lasso and a significant treatment effect was found ($p=0.02$). The double selection method (with penaly parameter $\lambda_\text{1se}$) additionally included the variables hospital and creatinine, and also resulted in a significant treatment effect ($p=0.02$).

Figure \ref{fig:DataAnalysis} in Appendix B compares the standardized survival curves based on the post-Lasso and double selection to the unadjusted survival curves, under CyA and placebo and for both endpoints. 
The standardized survival curves showed an attenuated difference in survival compared to the survival curves without accounting for any covariates. This difference was most pronounced under treatment with CyA. Under placebo the standardized survival curves based on post-Lasso and double selection were nearly identical, while a (rather small) difference between the standardized survival curves was observed under treatment with CyA. This was expected based on the simulation results.

\section{Discussion}
In this paper, we aimed to improve on the standard practice of conducting tests of the null hypothesis of no treatment effects in randomized controlled trials with time-to-event endpoints. 
The next step is to obtain valid estimates of treatment effect along with valid confidence intervals. The method we have proposed in this paper is readily applicable for this.
In particular, the treatment effect is directly obtained from the final Cox model for survival $T$ on treatment $A$ and the union of the sets of variables selected in the two variable selection steps. 
A theoretically justified but more involved procedure is the method proposed by \cite{fang2017}.

The results in this paper have implications for protocol writing. 
While one may be concerned that the covariates that will be adjusted for are difficult to prespecify in a protocol, one need not to decide in advance which variables to adjust for, but rather how variable selection will be done. 
Specifically, as long as the variable selection approach is pre-specified and works along the lines of our proposal, the analysis is pre-specified and there is no risk of inflating the Type I error. 
We recommend using the Lasso in combination with cross-validation in the two selection steps. 
Although Lasso is less common in survival analyses in randomized trials, software for penalized maximum likelihood estimation has become increasingly available to statisticians.
It is hereby important that the choice of the penalty parameter, which is based on cross-validation, is also discussed in the protocol. 
As our proposal seems to perform better with the largest value of the penalty parameter such that the error is within $1$ standard error of the minimum (ie, $\lambda_\text{1se}$), we suggest using this choice of penalty parameter.
That this choice of penalty is preferable is not surprising as we only need the very strong predictors of censoring for collider bias to be impactful. 
We remind the reader that the proposal extends to variable selection procedures other than the Lasso.


In our work, we so far did not consider the fact that the need for modelling also brings an increased risk of model misspecification. 
Model misspecification turns out not to be a concern for testing the null hypothesis of no treatment effect when censoring is either independent of the treatment conditional on the covariates or independent of predictors of survival given the treatment group \citep[e.g.][]{Kong1997, Lagakos1984, DiRienzo2001}.
However, our procedure needs the Cox model to be correctly specified when censoring depends on both $A$ and the predictors of survival. 
Although one might see this as a limitation, some degree of modelling is unavoidable when adjusting makes for non-informative censoring.
Moreover, we conjecture that tests of the null hypothesis of no treatment effect will continue to approach the nominal Type I error rate under model misspecification, though this remains to be studied. 

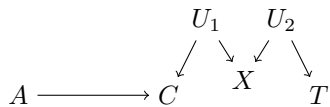
\begin{figure}[h!]
\centering
\begin{tikzpicture}
\node[] (a) at (0, 0)   {$A$};
\node[] (t) at (4, 0)   {$T$};
\node[] (c) at (2, 0)   {$C$};
\node[] (x) at (3,0.2) {$X$};
\node[] (u1) at (2.5,1) {$U_1$};
\node[] (u2) at (3.5,1) {$U_2$};
\path[->] (a) edge node {} (c);
\path[->] (u1) edge node {} (c);
\path[->] (u2) edge node {} (t);
\path[->] (u1) edge node {} (x);
\path[->] (u2) edge node {} (x);
\end{tikzpicture}
\caption{Example of M-bias. $U_1$ and $U_2$ are two unmeasured variables; one a cause of $C$, the other a cause of $T$.} \label{fig:dag_not}
\end{figure}

Note that, although adjustment generally makes the censoring assumption more plausible and thereby leads to a higher chance of obtaining a valid test, there are settings where adjustment may make things worse. This is so in the causal diagram in Figure \ref{fig:dag_not}, where $X$ is caused by two other (unmeasured) variables, one a cause of $C$, the other a cause of $T$. Because $X$ is a collider on a path from $C$ to $T$ , adjusting for it may introduce selection bias, referred to as M-bias \citep{hernan2020}.
In our data analysis, we assumed the absence of such variables. Also the selection of instruments (variables solely predictive of the censoring mechanism in this case) is well known to be detrimental for (finite-sample) bias and efficiency \citep{Brookhart2006} and, where possible, is best avoided by eliminating them prior to the data analysis based on subject-matter knowledge. 


In this paper, we only considered adjustment for baseline covariates. 
In practice, there may also be time dependent variables influencing survival and censoring. Remaining work is needed to adapt the procedure, as time varying covariates should not be directly included in the Cox model for the event time of interest. Augmented inverse probability of censoring weighting (AIPCW) \citep{Robins1992, Scharfstein1999}, which is valid under selection of variables \citep{Farrell2015}, may then be preferable.


\section*{Appendix}
\subsection*{Appendix A.1: Bias in Unadjusted Score Function}
In this section we derive an expression for (the bias in) the score test statistic for the treatment effect in a Cox proportional hazards model comparing two groups, without adjusting for any baseline covariate while in fact one should.
To develop further insight, suppose that
\begin{align*}
 \lambda\left(t\mid X\right)=\lambda_0(t)e^{\beta X}.
\end{align*}
Here, $\lambda_0(t)$ is the true baseline hazard function and $\beta$ is the true coefficient corresponding with the scalar $X$. Likewise, suppose that the true model for censoring obeys
\begin{align*}
 \lambda^C\left(t\mid A, X\right)=\lambda_0^C(t)e^{\gamma_1A+\gamma_2X},  
\end{align*}
where $\lambda_0^C(t)$ is the true baseline hazard function and $\gamma_1$ and $\gamma_2$ are the true coefficients corresponding with $A$ and $X$ respectively.
Without adjustment for $X$, the partial score function under the null hypothesis of no treatment effect is given by
\begin{align}\label{eq:score_LRT}
    \int_0^\infty\left(A-\frac{E(A\cdot R(t))}{E(R(t))}\right)\left(dN(t)-R(t)\lambda_0^r(t)dt\right),
\end{align}
where $R(t)$ is the at risk indicator at time $t$, $dN(t)$ the increment in the counting process $N(t)$ at time $t$ and $\lambda_0^r(t)$ the unconditional baseline hazard at time $t$.
The expected value of the score in Expression (\ref{eq:score_LRT}) is then given by
\begin{align}\label{eq:cbias_score1}
\begin{split}
&\int_0^\infty E\left(A I(T\geq t)e^{-\Lambda_0^C(t)e^{\gamma_1A+\gamma_2X}}\lambda_0(t)e^{\beta X}\right)dt\\
-&\int_0^\infty\frac{E\left(A I(T\geq t)e^{-\Lambda_0^C(t)e^{\gamma_1A+\gamma_2X}}\right)}{E\left(I(T\geq t)e^{-\Lambda_0^C(t)e^{\gamma_1A+\gamma_2X}}\right)}E\left(I(T\geq t)e^{-\Lambda_0^C(t)e^{\gamma_1A+\gamma_2X}}\lambda_0(t)e^{\beta X}\right)dt\\
-&\int_0^\infty E\left(AI(T\geq t)e^{-\Lambda_0^C(t)e^{\gamma_1A+\gamma_2X}}\lambda_0^r(t)dt\right)\\
+&\int_0^\infty\frac{E(A\cdot I(T\geq t)e^{-\Lambda_0^C(t)e^{\gamma_1A+\gamma_2X}})}{E(I(T\geq t)e^{-\Lambda_0^C(t)e^{\gamma_1A+\gamma_2X}})}E\left(I(T\geq t)e^{-\Lambda_0^C(t)e^{\gamma_1A+\gamma_2X}}\lambda_0^r(t)dt\right),
\end{split}
\end{align}
where $\Lambda_0^C(t)=\int_0^t\lambda_0^C(s)ds$ is the cumulative baseline hazard for censoring and $\lambda_0(t)e^{\beta X}$ the true baseline hazard for survival. As the latter two terms in Expression (\ref{eq:cbias_score1}) both equal 
\begin{align*}
  \int_0^\infty E\left(AI(T\geq t)e^{-\Lambda_0^C(t)e^{\gamma_1A+\gamma_2X}}\right)\lambda_0^r(t)dt, 
\end{align*}
Expression (\ref{eq:cbias_score1}) reduces to 
\begin{align}\label{eq:cbias_score}
\begin{split}
&\int_0^\infty E\left(A I(T\geq t)e^{-\Lambda_0^C(t)e^{\gamma_1A+\gamma_2X}}\lambda_0(t)e^{\beta X}\right)dt\\
-&\int_0^\infty\frac{E\left(A I(T\geq t)e^{-\Lambda_0^C(t)e^{\gamma_1A+\gamma_2X}}\right)}{E\left(I(T\geq t)e^{-\Lambda_0^C(t)e^{\gamma_1A+\gamma_2X}}\right)}E\left(I(T\geq t)e^{-\Lambda_0^C(t)e^{\gamma_1A+\gamma_2X}}\lambda_0(t)e^{\beta X}\right)dt.
\end{split}
\end{align}
By using the fact that $A\bigCI(T,X)$ -which is guaranteed to hold in a randomized trial under the null hypothesis, it can be easily seen that (the bias in) the score is zero if either $\beta=0$, $\gamma_1=0$ or $\gamma_2=0$. 
Using a Taylor expansion, Expression (\ref{eq:cbias_score}) can be rewritten as
\begin{align*}
&\int_0^\infty\left\{ E\left(A I(T\geq t)e^{-\Lambda_0^C(t)e^{\gamma_1A}}\lambda_0(t)\right)-\frac{E\left(A I(T\geq t)e^{-\Lambda_0^C(t)e^{\gamma_1A}}\right)}{E\left(I(T\geq t)e^{-\Lambda_0^C(t)e^{\gamma_1A}}\right)}E\left(I(T\geq t)e^{-\Lambda_0^C(t)e^{\gamma_1A}}\lambda_0(t)\right)\right\}dt\\
+&\int_0^\infty\left\{ E\left(A I(T\geq t)e^{-\Lambda_0^C(t)e^{\gamma_1A}}\lambda_0(t)X\right)-\frac{E\left(A I(T\geq t)e^{-\Lambda_0^C(t)e^{\gamma_1A}}\right)}{E\left(I(T\geq t)e^{-\Lambda_0^C(t)e^{\gamma_1A}}\right)}E\left(I(T\geq t)e^{-\Lambda_0^C(t)e^{\gamma_1A}}\lambda_0(t)X\right)\right\}dt\beta\\
+&\int_0^\infty\left\{\vphantom{\frac{E\left(A I(T\geq t)e^{-\Lambda_0^C(t)e^{\gamma_1A}}\right)}{E\left(I(T\geq t)e^{-\Lambda_0^C(t)e^{\gamma_1A}}\right)^2}} E\left(A I(T\geq t)e^{-\Lambda_0^C(t)e^{\gamma_1A}}\lambda_0(t)(-\Lambda_0^C(t)e^{\gamma_1A})X\right)\right.\\
+&\frac{E\left(A I(T\geq t)e^{-\Lambda_0^C(t)e^{\gamma_1A}}\right)}{E\left(I(T\geq t)e^{-\Lambda_0^C(t)e^{\gamma_1A}}\right)^2}E\left(I(T\geq t)e^{-\Lambda_0^C(t)e^{\gamma_1A}}(-\Lambda_0^C(t)e^{\gamma_1A})X\right)E\left(I(T\geq t)e^{-\Lambda_0^C(t)e^{\gamma_1A}}\lambda_0(t)\right)\\
-&\frac{E\left(A I(T\geq t)e^{-\Lambda_0^C(t)e^{\gamma_1A}}(-\Lambda_0^C(t)e^{\gamma_1A})X\right)}{E\left(I(T\geq t)e^{-\Lambda_0^C(t)e^{\gamma_1A}}\right)}E\left(I(T\geq t)e^{-\Lambda_0^C(t)e^{\gamma_1A}}\lambda_0(t)\right)\\
-&\left.\frac{E\left(A I(T\geq t)e^{-\Lambda_0^C(t)e^{\gamma_1A}}\right)}{E\left(I(T\geq t)e^{-\Lambda_0^C(t)e^{\gamma_1A}}\right)}E\left(I(T\geq t)e^{-\Lambda_0^C(t)e^{\gamma_1A}}\lambda_0(t)(-\Lambda_0^C(t)e^{\gamma_1A})X\right)\right\}dt\gamma_2\\
    &+\int_0^\infty\left\{\vphantom{\frac{E\left(A I(T\geq t)e^{-\Lambda_0^C(t)e^{\gamma_1A+\gamma_2X}}\right)}{E\left(I(T\geq t)e^{-\Lambda_0^C(t)e^{\gamma_1A+\gamma_2X}}\right)}} \frac{\partial}{\partial \gamma_2}\frac{\partial}{\partial \beta}E\left(A I(T\geq t)e^{-\Lambda_0^C(t)e^{\gamma_1A+\gamma_2X}}\lambda_0(t)e^{\beta X}\right)\Bigr|_{\substack{\beta=0, \gamma_2=0}}\right.\\
    &\left.-\frac{\partial}{\partial \gamma_2}\frac{\partial}{\partial \beta}\left(\frac{E\left(A I(T\geq t)e^{-\Lambda_0^C(t)e^{\gamma_1A+\gamma_2X}}\right)}{E\left(I(T\geq t)e^{-\Lambda_0^C(t)e^{\gamma_1A+\gamma_2X}}\right)}E\left(I(T\geq t)e^{-\Lambda_0^C(t)e^{\gamma_1A+\gamma_2X}}\lambda_0(t)e^{\beta X}\right)\right)\Bigr|_{\substack{\beta=0, \gamma_2=0}}\right\}dt\beta\gamma_2\\
    &+\text{higher order terms}
\end{align*}
It is straightforward to show that the integrand of the first integral is zero for each $t$. It can be shown that the second and third integral are zero when $A\bigCI(T,X)$. Similarly, it can be shown that the last integral equals
\begin{align*}
 &-\Lambda_0^C(t)  \lambda_0(t) \left\{E\left(Ae^{-\Lambda_0^C(t)e^{\gamma_1A}}e^{\gamma_1A}\right)E\left(e^{-\Lambda_0^C(t)e^{\gamma_1A}}\right)-E\left(Ae^{-\Lambda_0^C(t)e^{\gamma_1A}}\right)E\left(e^{-\Lambda_0^C(t)e^{\gamma_1A}}e^{\gamma_1A}\right)\right\}\cdot\\
 &\left\{E\left(I(T\geq t)L^2\right)E\left(I(T\geq t)\right)-E\left(I(T\geq t)L\right)^2\right\}\left/\left\{E\left(I(T\geq t)\right)E\left(e^{-\Lambda_0^C(t)e^{\gamma_1A}}\right)\right\}\right.,
\end{align*}
so that (the bias in) the score is proportional to $\beta\gamma_2$ up to higher order terms.

\subsection*{Appendix A.2: Censoring Assumptions}
Here, we give a more formal proof of the validity of the unadjusted score test, without making assumptions about the underlying models for survival and censoring. In particular, we show that this score test (which corresponds to the logrank test) is valid if censoring is either independent of treatment or survival (conditional on treatment).
The expectation of the partial score function in Equation (\ref{eq:score_LRT}) is
\[
\int_0^\infty E\left\{\left(A-\frac{E(A\cdot R(t))}{E(R(t))}\right)R(t)E\left(dN(t)\mid \mathcal{F}_{T}(t), \mathcal{F}_{C}(t), A\right)\right\},
\]
with $\mathcal{F}_{T}(t)$ the history spanned by the counting process $N(t)$ and $\mathcal{F}_{C}(t)$ the history spanned by the counting process $N_C(t)$ for censoring. First, we prove that the expectation has mean zero and hence provides a valid test of the null hypothesis when censoring is independent of treatment. Note that in a randomized trial, $C\bigCI A$ and $T\bigCI A$ implies that $A\bigCI(C,T)$. 
We will assume that $A$ is independent of $(C,T)$, which usually holds (in a randomized trial) under the null when $A$ is independent of censoring.
This, however, wouldn't hold in observational studies as there might be common causes of $A$ and $T$.
Therefore,
\begin{align*}
&\int_0^\infty E\left\{\left(A-\frac{E(A\cdot R(t))}{E(R(t))}\right)R(t)E\left(dN(t)\mid \mathcal{F}_{T}(t), \mathcal{F}_{C}(t), A\right)\right\}\\
=&\int_0^\infty E\left\{\left(A-\frac{E(A\cdot I(T\geq t)I(C\geq t))}{E(I(T\geq t)I(C\geq t))}\right)I(T\geq t)I(C\geq t)E\left(dN(t)\mid \mathcal{F}_{T}(t), \mathcal{F}_{C}(t)\right)\right\}\\
=&\int_0^\infty E\left\{\left(A-\frac{E(A)E(I(T\geq t)I(C\geq t))}{E(I(T\geq t)I(C\geq t))}\right)I(T\geq t)I(C\geq t)E\left(dN(t)\mid \mathcal{F}_{T}(t), \mathcal{F}_{C}(t)\right)\right\}\\
=&0,
\end{align*}
where the second and third equality follow from $A\bigCI (C,T)$.

Similarly, we prove that the expectation has mean zero and hence provides a valid test of the null hypothesis when censoring is independent of survival in each treatment arm. Under the null hypothesis,
\begin{align*}
&\int_0^\infty E\left\{\left(A-\frac{E(A\cdot R(t))}{E(R(t))}\right)R(t)E\left(dN(t)\mid \mathcal{F}_{T}(t), \mathcal{F}_{C}(t), A\right)\right\}\\
=&\int_0^\infty E\left\{\left(A-\frac{E(A\cdot I(T\geq t)I(C\geq t))}{E(I(T\geq t)I(C\geq t))}\right)I(T\geq t)I(C\geq t)E\left(dN(t)\mid \mathcal{F}_{T}(t)\right)\right\}\\
=&\int_0^\infty E\left\{\left(A-\frac{E(A\cdot I(C\geq t))}{E(I(C\geq t))}\right)I(T\geq t)I(C\geq t)E\left(dN(t)\mid \mathcal{F}_{T}(t)\right)\right\}\\
=&\int_0^\infty \left(E(A I(C\geq t))-\frac{E(A\cdot I(C\geq t))E(I(C\geq t))}{E(I(C\geq t))}\right)E(I(T\geq t)E\left(dN(t)\mid \mathcal{F}_{T}(t)\right))\\
&=0.
\end{align*}
Here, the second equality follows from $E(A\cdot I(T\geq t)I(C\geq t))=E(A\cdot I(C\geq t))E(I(T\geq t))$ by the assumption that $T\bigCI C\mid A$ and $E(I(T\geq t)I(C\geq t))=E(I(C\geq t))E(I(T\geq t))$ by the fact that assumption $T\bigCI C\mid A$ also implies that $T\bigCI C$ under the null hypothesis in a randomized trial. The third equality follows from the same assumption(s).

Thus, as commonly known the log-rank test is valid under the null hypothesis if censoring is non-informative - namely, if censoring is (statistically) independent of survival time. It is more surprising that its validity under the null is maintained even if censoring does depend on survival time, as long as $A$ is jointly independent of $C$ and $T$. 

\subsection*{Appendix A.3: Method by \cite{fang2017}}
\cite{fang2017} propose a decorrelated score test for $H_{0}: \alpha=0$, as follows. First, they estimate $\beta$ as $\hat{\beta}$ by fitting a Cox model for survival time $T$ on treatment $A$ and baseline covariates $X$ using Lasso. 
Next, they linearly regress
$$\int_0^\tau \left(A_i-\frac{\sum_{i=1}^nR_i(t)A_ie^{\hat{\beta}' L_i}}{\sum_{i=1}^nR_i(t)e^{\hat{\beta}' L_i}}\right)dN_i(t)$$
	on
$$\int_0^\tau \left(L_i-\frac{\sum_{i=1}^nR_i(t)L_ie^{\hat{\beta}' L_i}}{\sum_{i=1}^nR_i(t)e^{\hat{\beta}' L_i}}\right)dN_i(t),
$$
using Lasso and with $\hat{\beta}$ the Lasso estimates obtained in the previous step. Here, $\tau$ denotes the end-of-study time. Denoting the $p$-dimensional (estimated) linear regression coefficient from the previous step by $\hat{w}$, they propose the following decorrelated score function for $\alpha$
\[
\hat{U}(\alpha, \hat{\beta})=-\frac{1}{n} \sum_{i=1}^{n} \int_{0}^{\tau}\left[A_{i}-\hat{w}'X_{i}-\left\{\frac{\sum_{i=1}^nR_i(t)A_ie^{\alpha A_i+\hat{\beta}' X_i}}{\sum_{i=1}^nR_i(t)e^{\alpha A_i+\hat{\beta}' X_i}}-\hat{w}' \frac{\sum_{i=1}^nR_i(t)X_ie^{\alpha A_i+\hat{\beta}' X_i}}{\sum_{i=1}^nR_i(t)e^{\alpha A_i+\hat{\beta}' X_i}}
\right\}\right] \mathrm{d} N_{i}(t).
\]
To test the null hypothesis $\alpha=0$, they standardize $\hat{U}(0, \hat{\beta})$ to construct the test statistic. For further details, we refer the reader to \cite{fang2017}.

\subsection*{Appendix B: Figures and Tables}
\begin{figure}[htp]
\centering
\begin{subfigure}[b]{0.32\textwidth}
        \includegraphics[width=\textwidth]{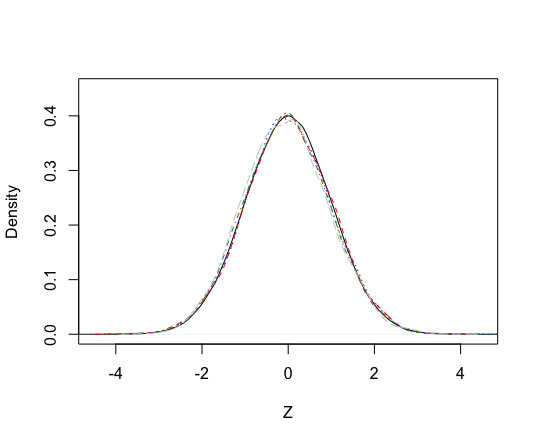}
        \caption{$\beta=0.2$, $\gamma_2=0.2$}
\end{subfigure}
\begin{subfigure}[b]{0.32\textwidth}
        \includegraphics[width=\textwidth]{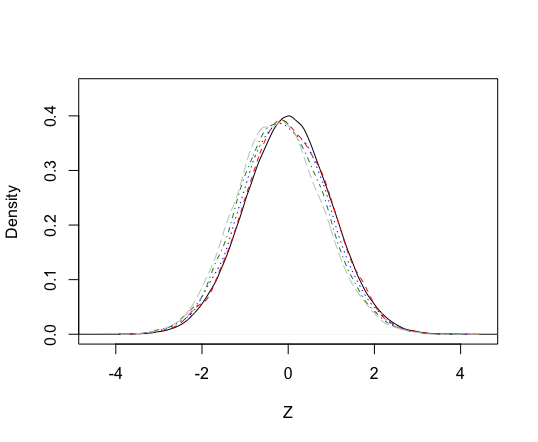}
        \caption{$\beta=2.2$, $\gamma_2=0.2$}
\end{subfigure}
\begin{subfigure}[b]{0.32\textwidth}
        \includegraphics[width=\textwidth]{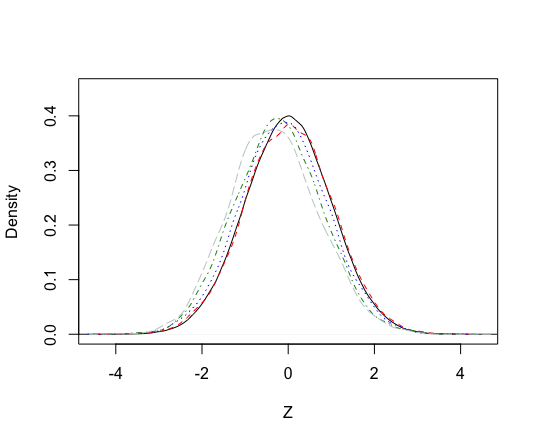}
        \caption{$\beta=0.2$, $\gamma_2=2.2$}
\end{subfigure}
\medskip

\begin{subfigure}[b]{0.32\textwidth}
        \includegraphics[width=\textwidth]{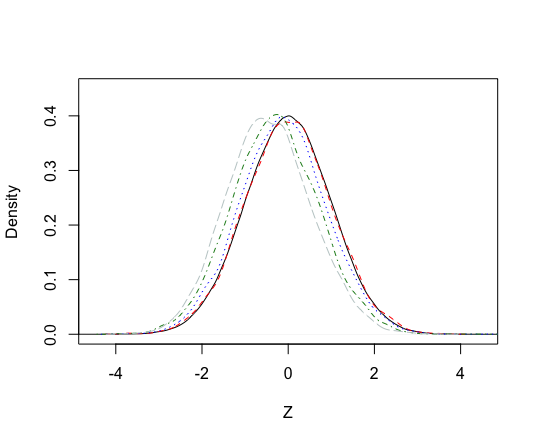}
        \caption{$\beta=0.5$, $\gamma_2=0.5$}
\end{subfigure}
\begin{subfigure}[b]{0.32\textwidth}
        \includegraphics[width=\textwidth]{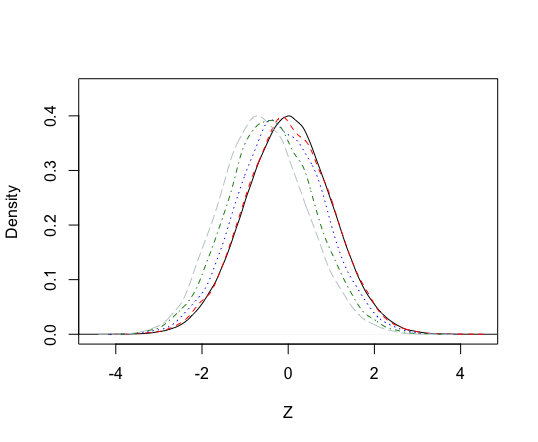}
        \caption{$\beta=2.2$, $\gamma_2=0.5$}
\end{subfigure}
\begin{subfigure}[b]{0.32\textwidth}
        \includegraphics[width=\textwidth]{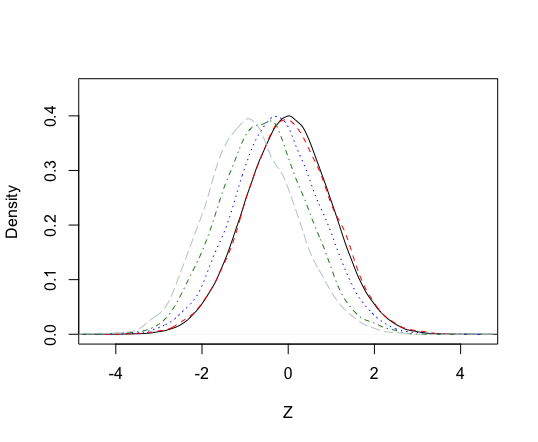}
        \caption{$\beta=0.5$, $\gamma_2=2.2$}
\end{subfigure}

\medskip

\begin{subfigure}[b]{0.32\textwidth}
        \includegraphics[width=\textwidth]{j6_k6.png}
        \caption{$\beta=0$, $\gamma_2=2.2$}
\end{subfigure}
\begin{subfigure}[b]{0.32\textwidth}
        \includegraphics[width=\textwidth]{j23_k6.png}
        \caption{$\beta=2.2$, $\gamma_2=0$}
\end{subfigure}
\begin{subfigure}[b]{0.32\textwidth}
        \includegraphics[width=\textwidth]{j6_k23.png}
        \caption{$\beta=2.2$, $\gamma_2=2.2$}
\end{subfigure}

\caption{Distribution of the logrank test statistic compared to the standard normal distribution (black, solid) for different values of $\beta$, $\gamma_2$ and $\gamma_1$. Red, dashed : $\gamma_1=0$; Blue, dotted: $\gamma_1=1$; Green, dotted-dashed: $\gamma_1=2$; Gray, long dashed: $\gamma_1=3$. Results are based on $10,000$ simulations, $n=100$, $X\sim N(0,1)$, $\lambda_0(t)=e$ and $\lambda_0^C(t)=e^{-1}$.}
\label{fig:LRT_bias}
\end{figure}

\begin{figure}[htp]
\centering

        \includegraphics[width=0.32\textwidth]{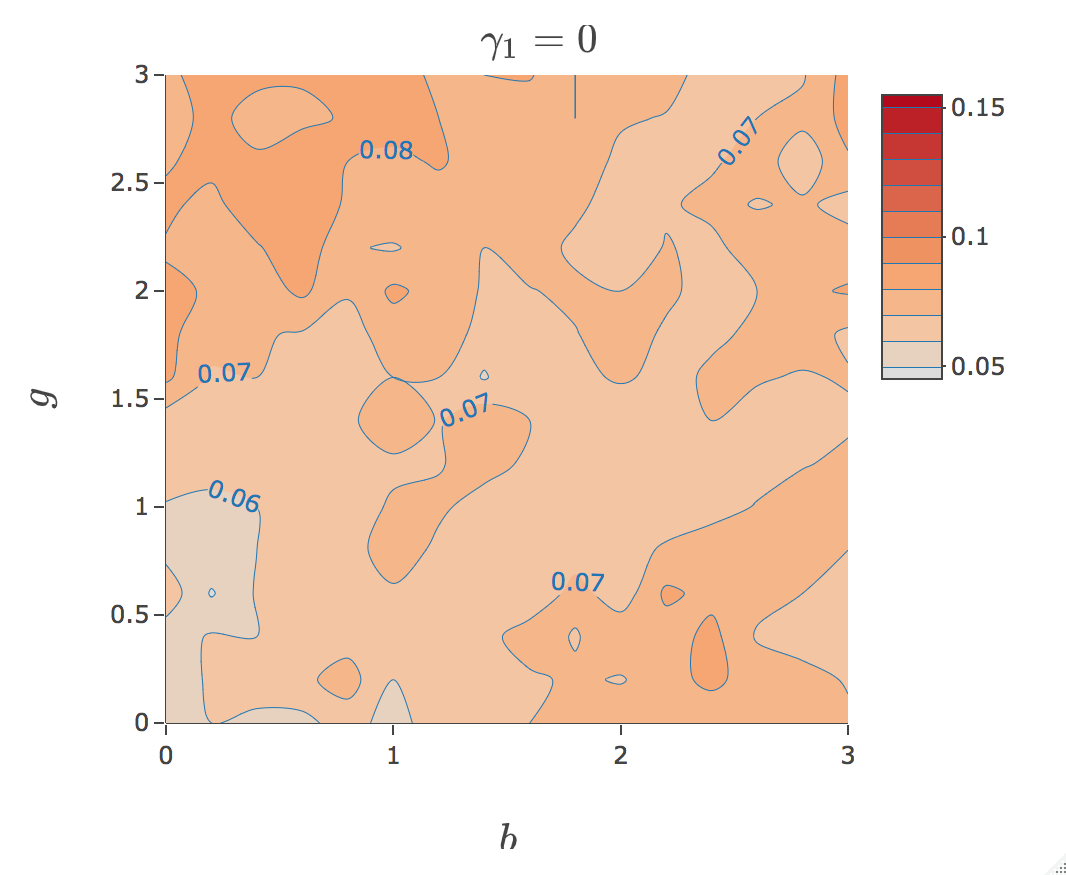}
        \includegraphics[width=0.32\textwidth]{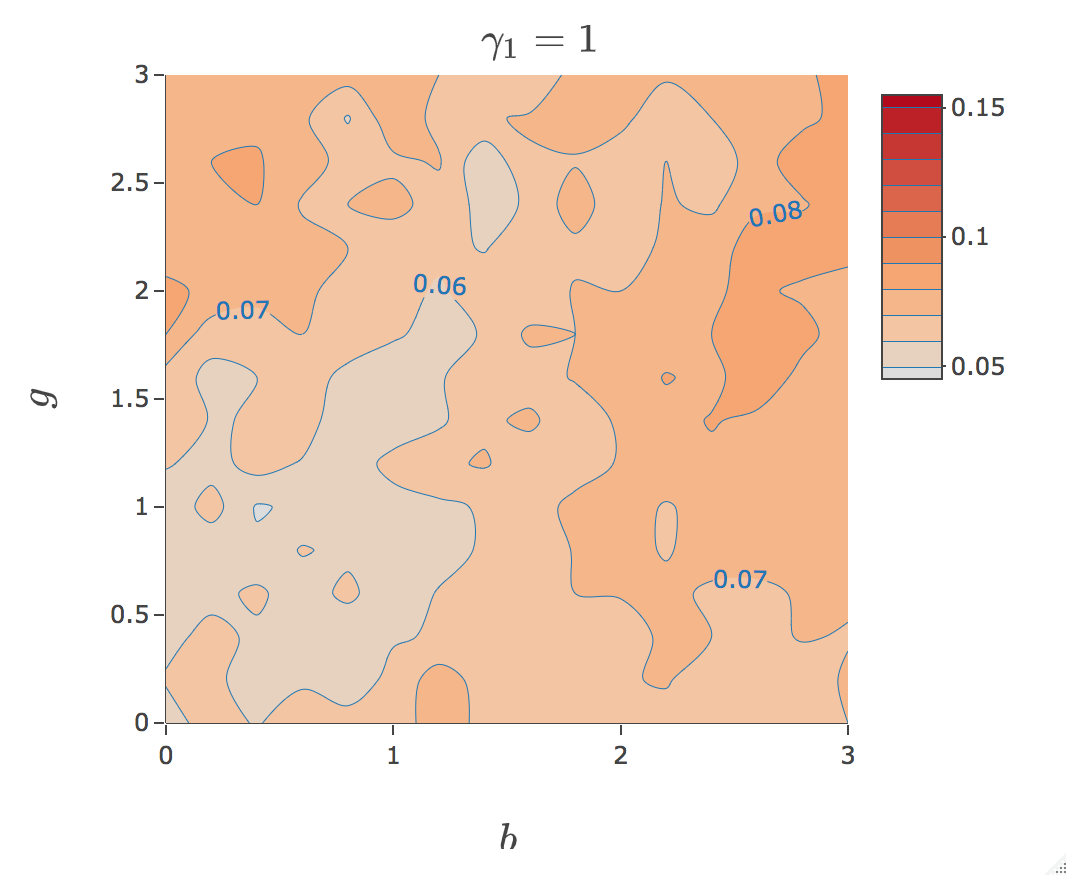}
        \includegraphics[width=0.32\textwidth]{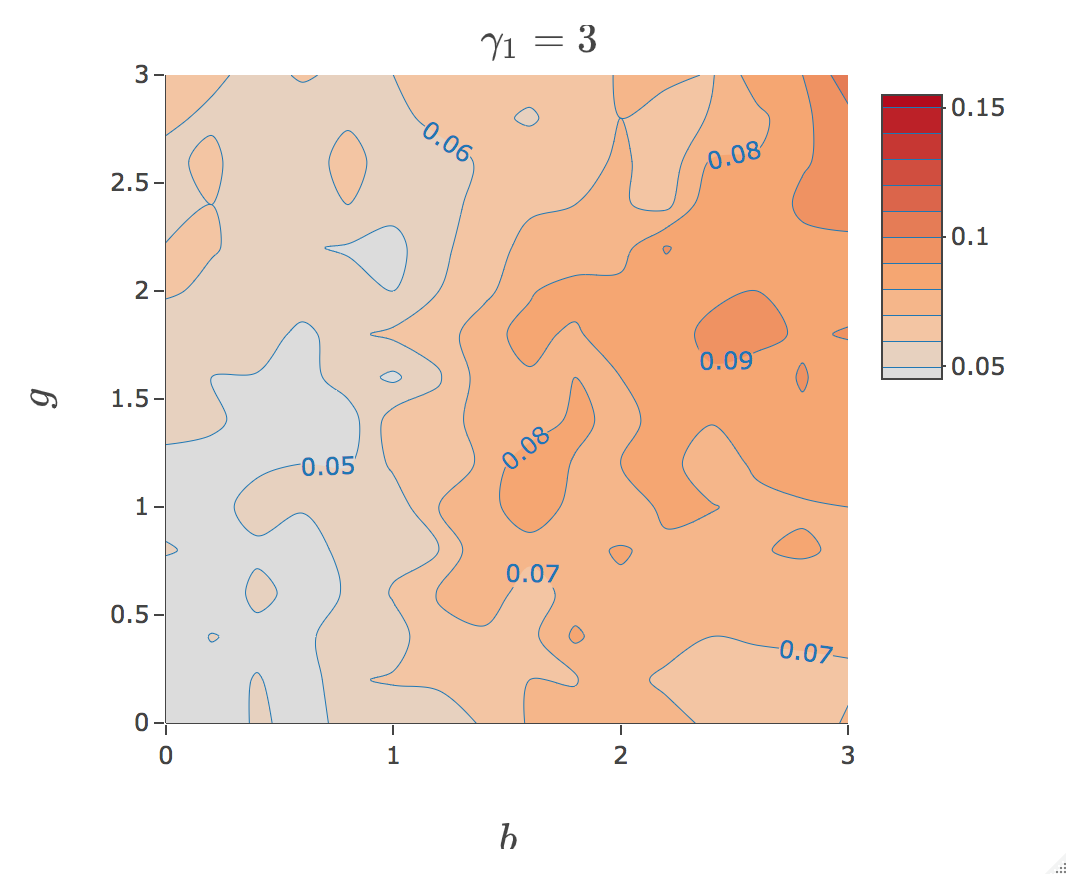}
\medskip

        \includegraphics[width=0.32\textwidth]{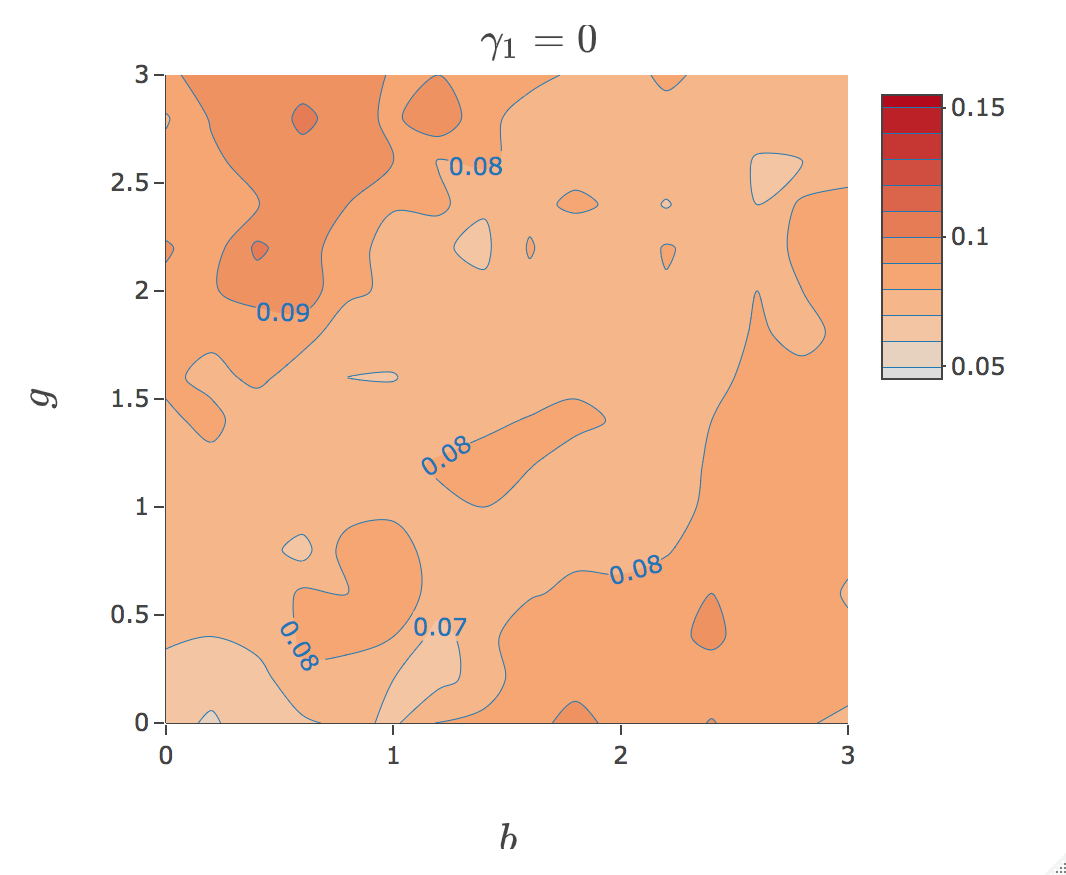}
        \includegraphics[width=0.32\textwidth]{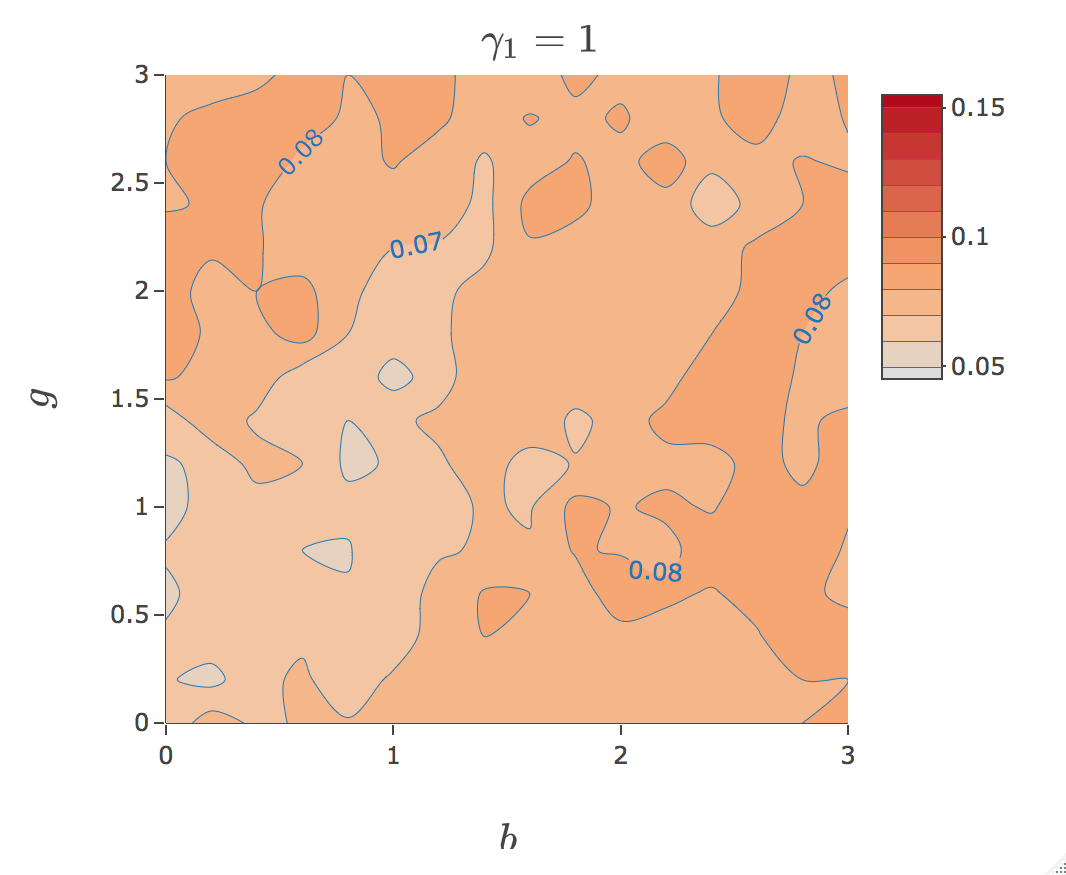}
        \includegraphics[width=0.32\textwidth]{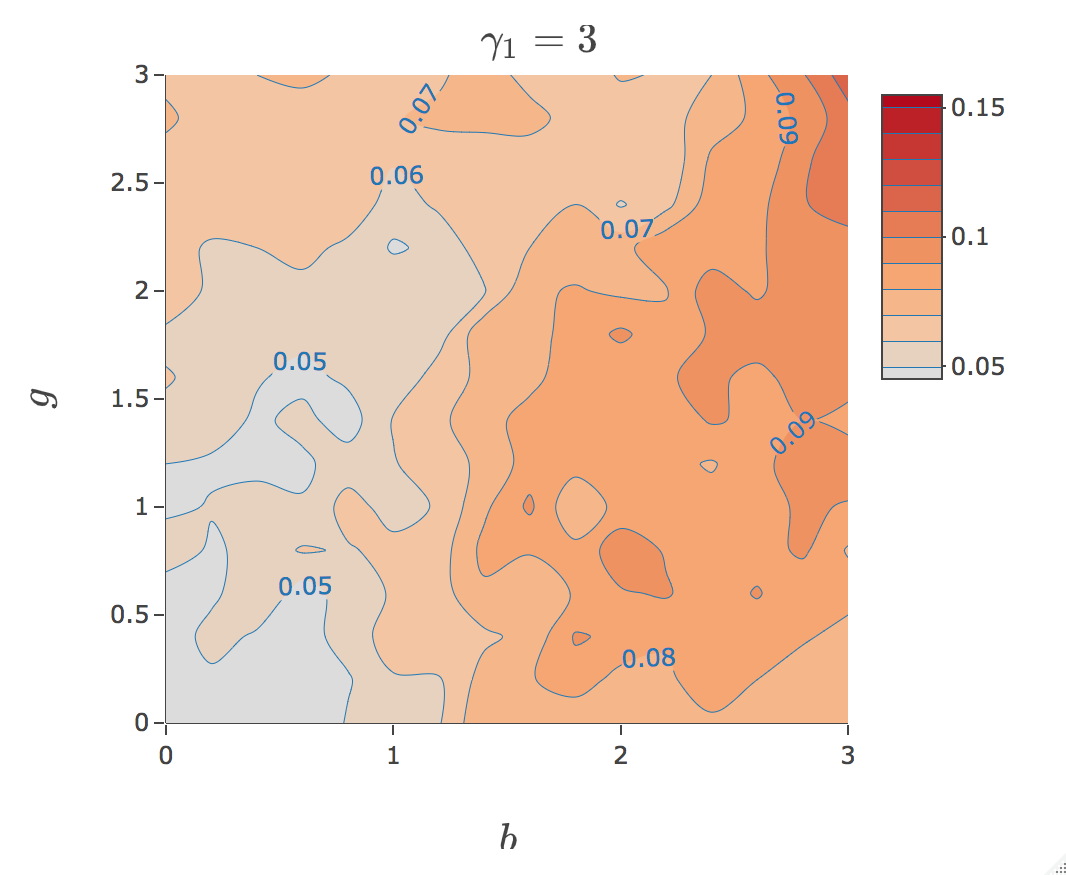}
\medskip

        \includegraphics[width=0.32\textwidth]{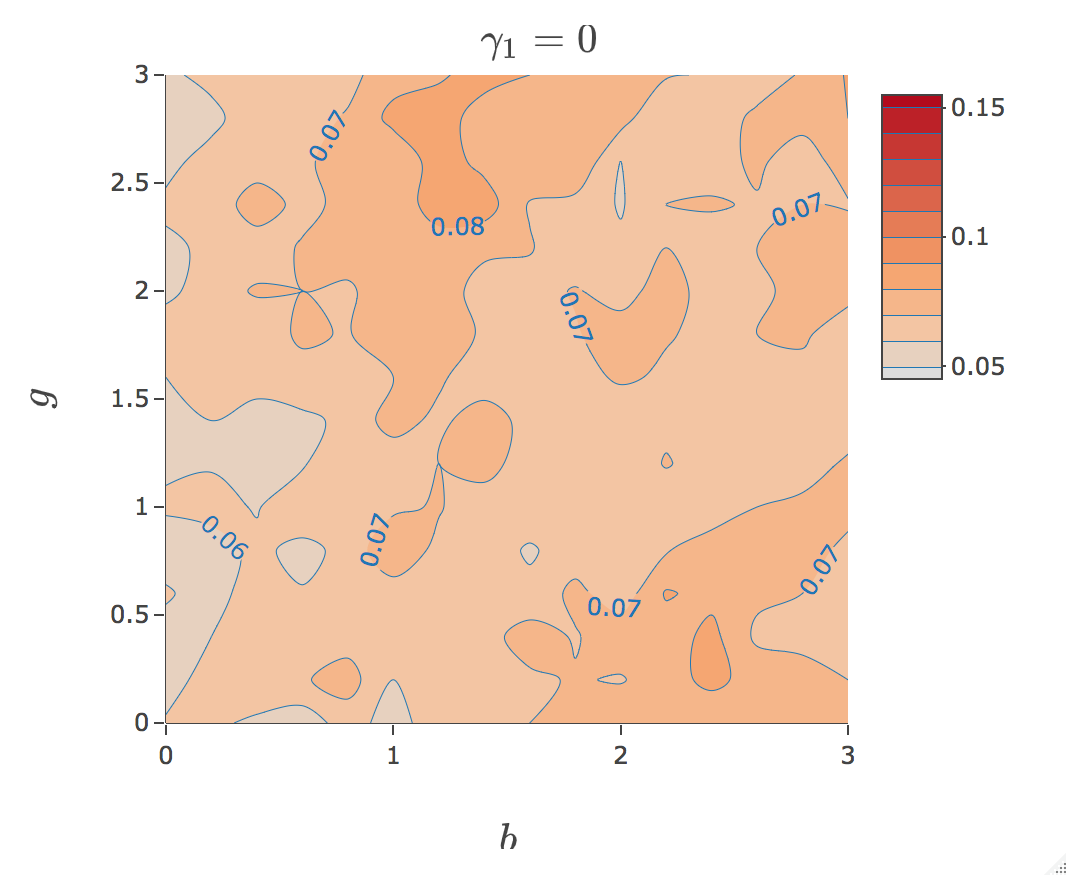}
        \includegraphics[width=0.32\textwidth]{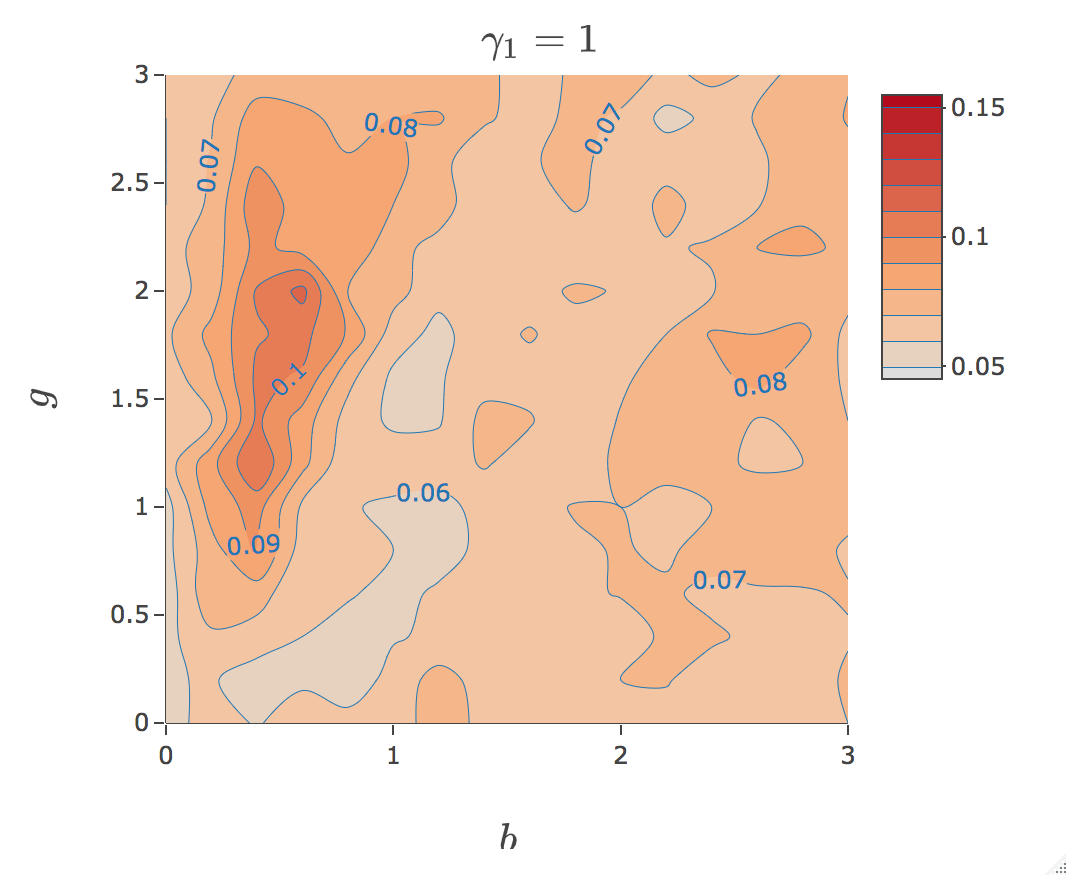}
        \includegraphics[width=0.32\textwidth]{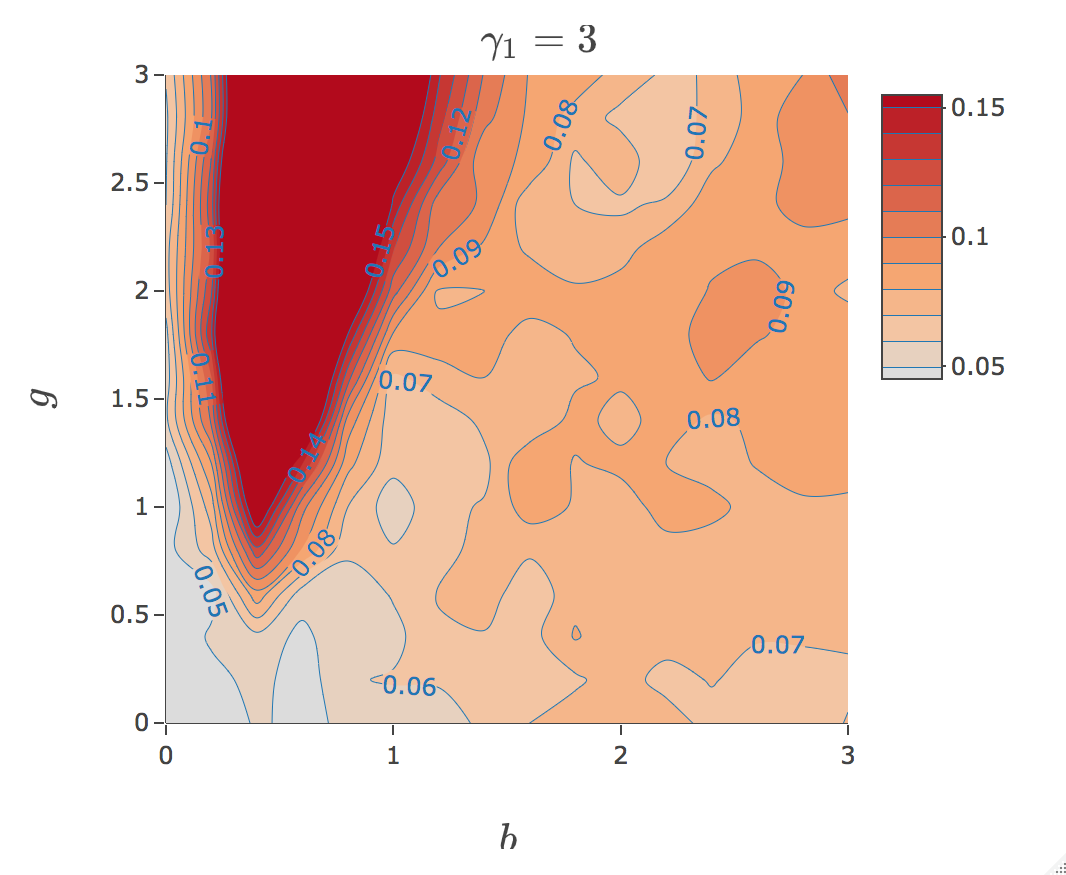}
\medskip

        \includegraphics[width=0.32\textwidth]{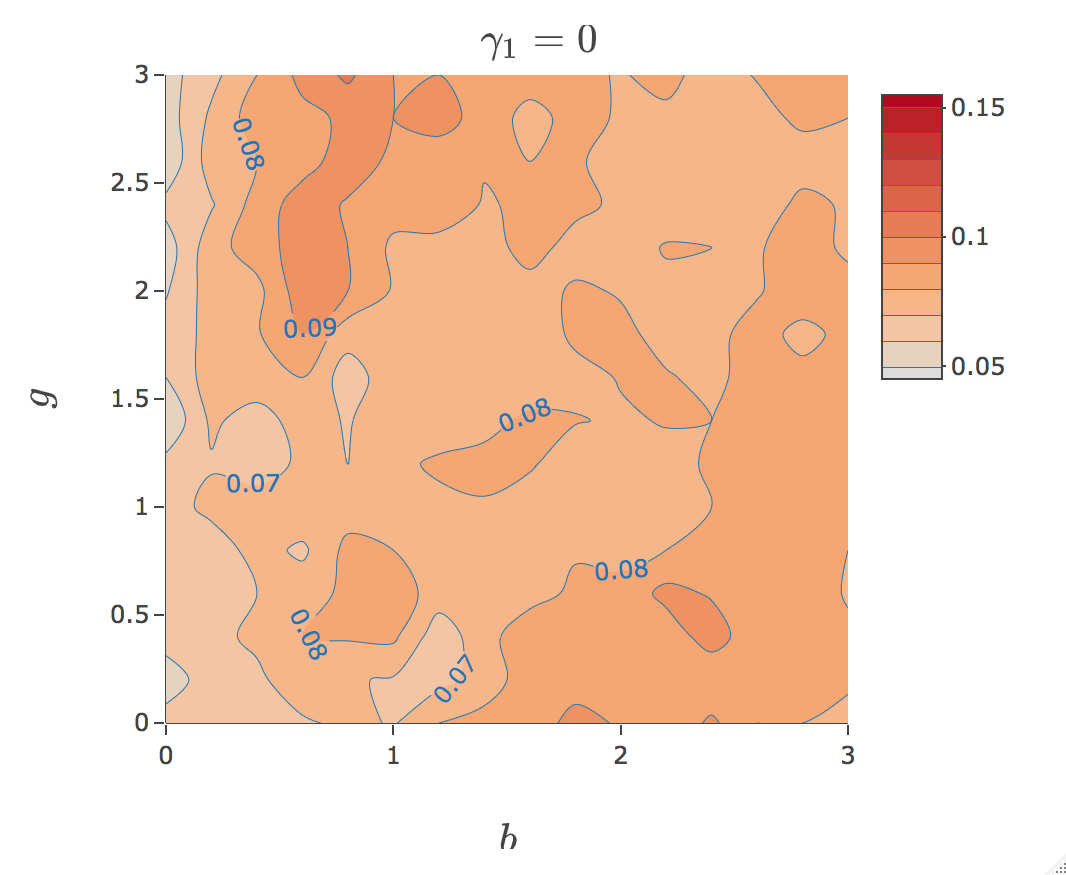}
        \includegraphics[width=0.32\textwidth]{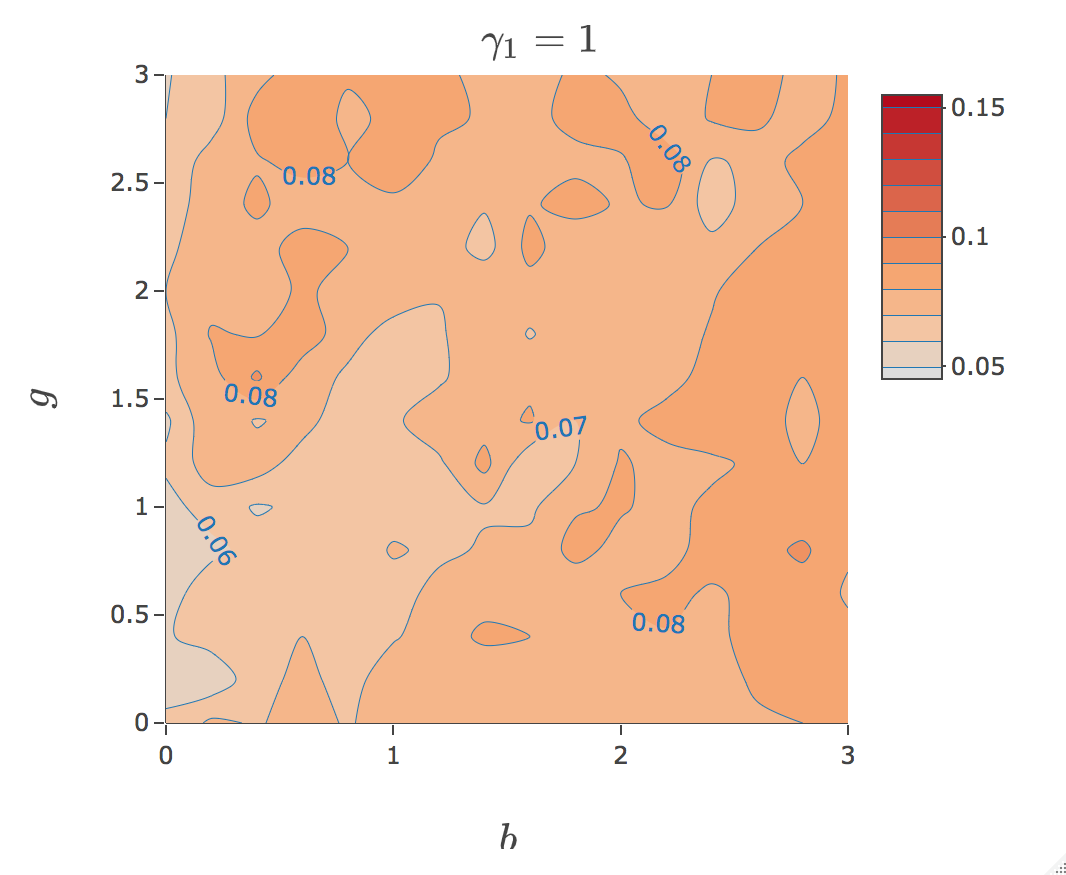}
        \includegraphics[width=0.32\textwidth]{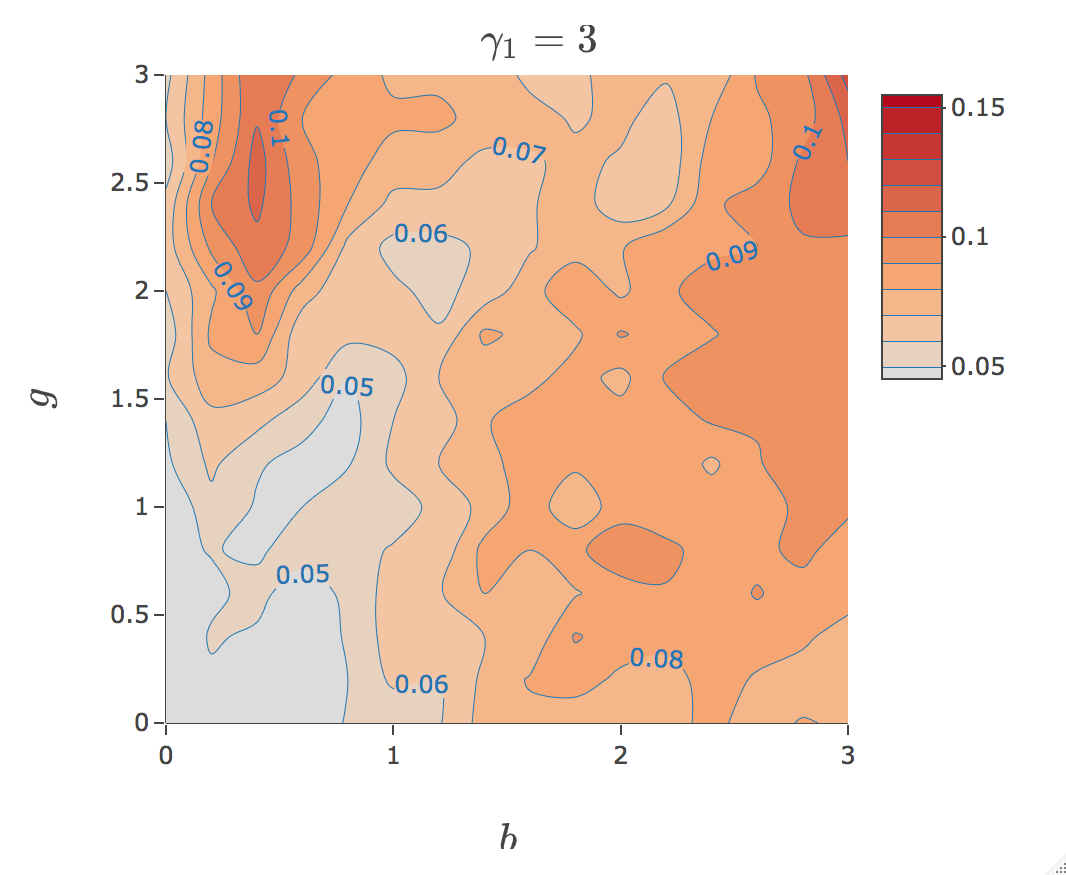}

\caption{Empirical Type I error rate at the $5\%$ significance level of the Lasso and the proposed double selection method under Setting $2$. Row $1$: double selection method with penalty parameter $\lambda_{1se}$; Row $2$:  double selection method with penalty parameter $\lambda_{min}$; Row $3$: Post-Lasso with penalty parameter $\lambda_{1se}$ ; Row $4$: Post-Lasso with penalty parameter $\lambda_{min}$. Left: $\gamma_1=0$; Middle: $\gamma_1=1$; Right: $\gamma_1=3$.
Results are based on $1,000$ simulations, $\lambda_0(t)=1$ ($\beta_0=0$) and $\lambda_0^C(t)=1$ ($\gamma_0=0$).
}
\label{fig:Sim_results3}
\end{figure}

\begin{figure}[htp]
\centering

        \includegraphics[width=0.32\textwidth]{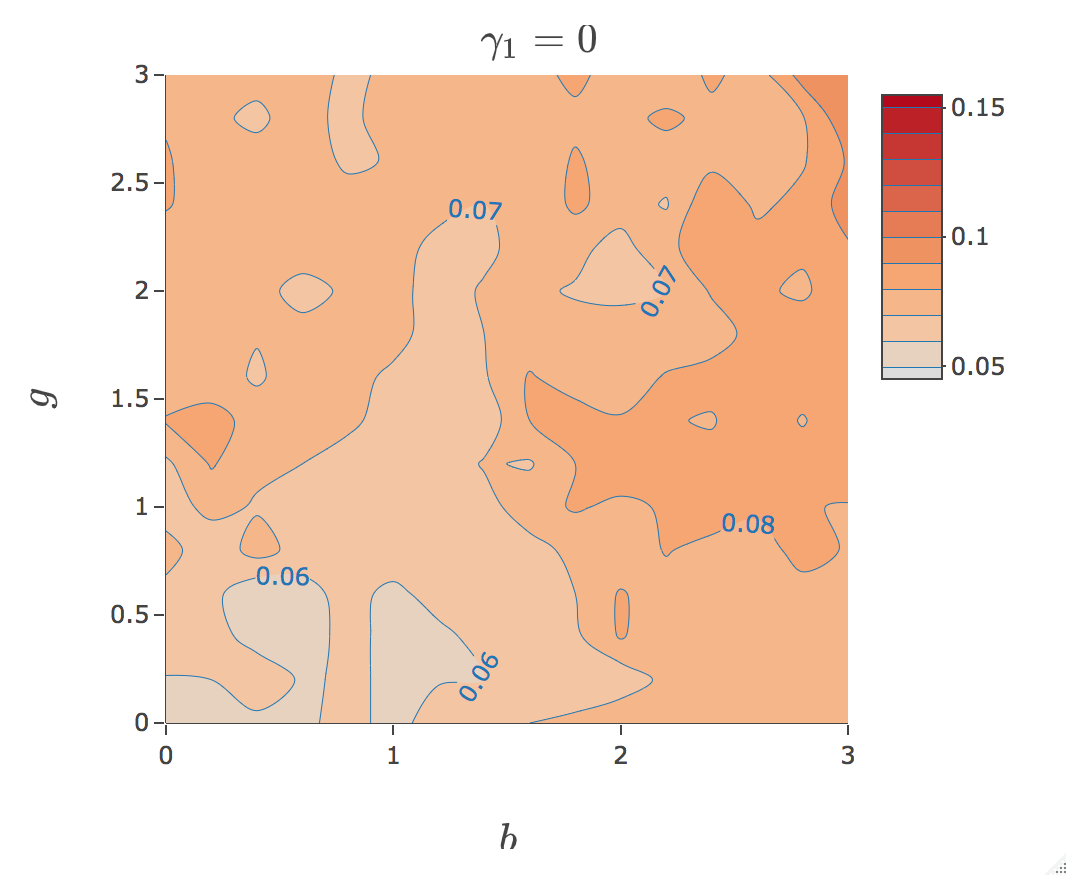}
        \includegraphics[width=0.32\textwidth]{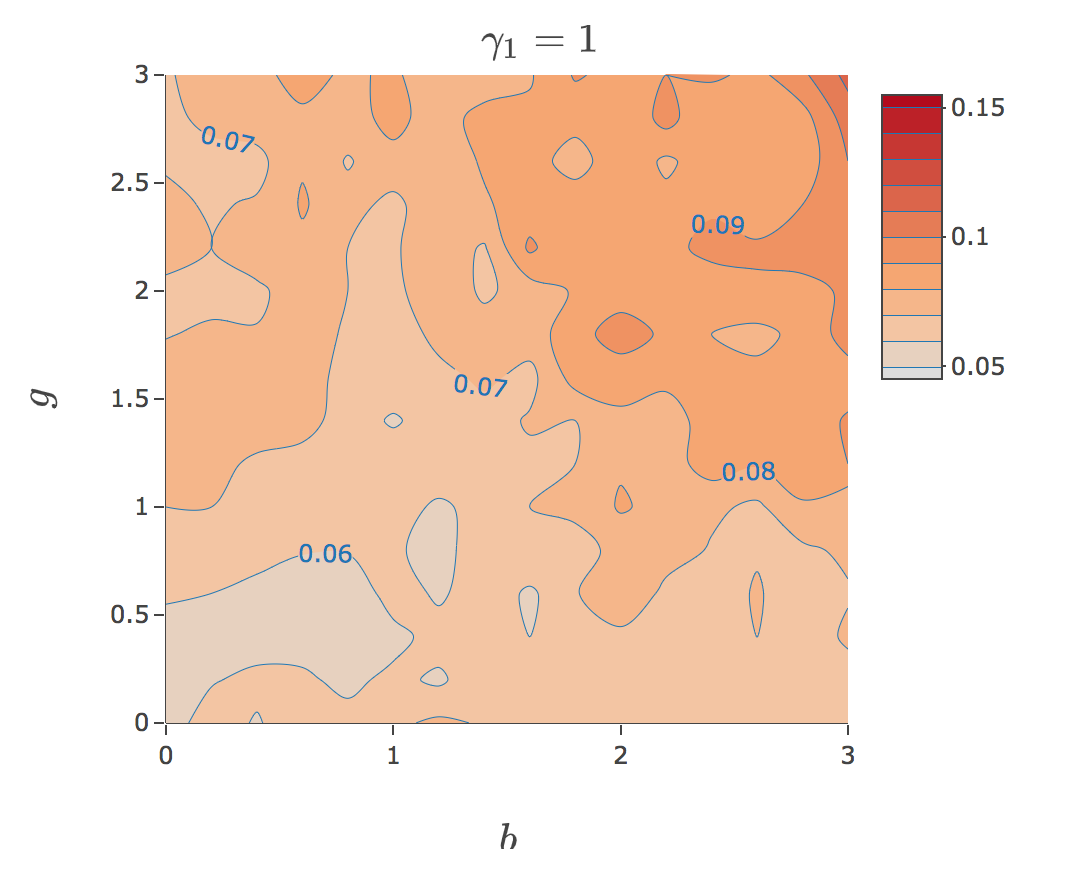}
        \includegraphics[width=0.32\textwidth]{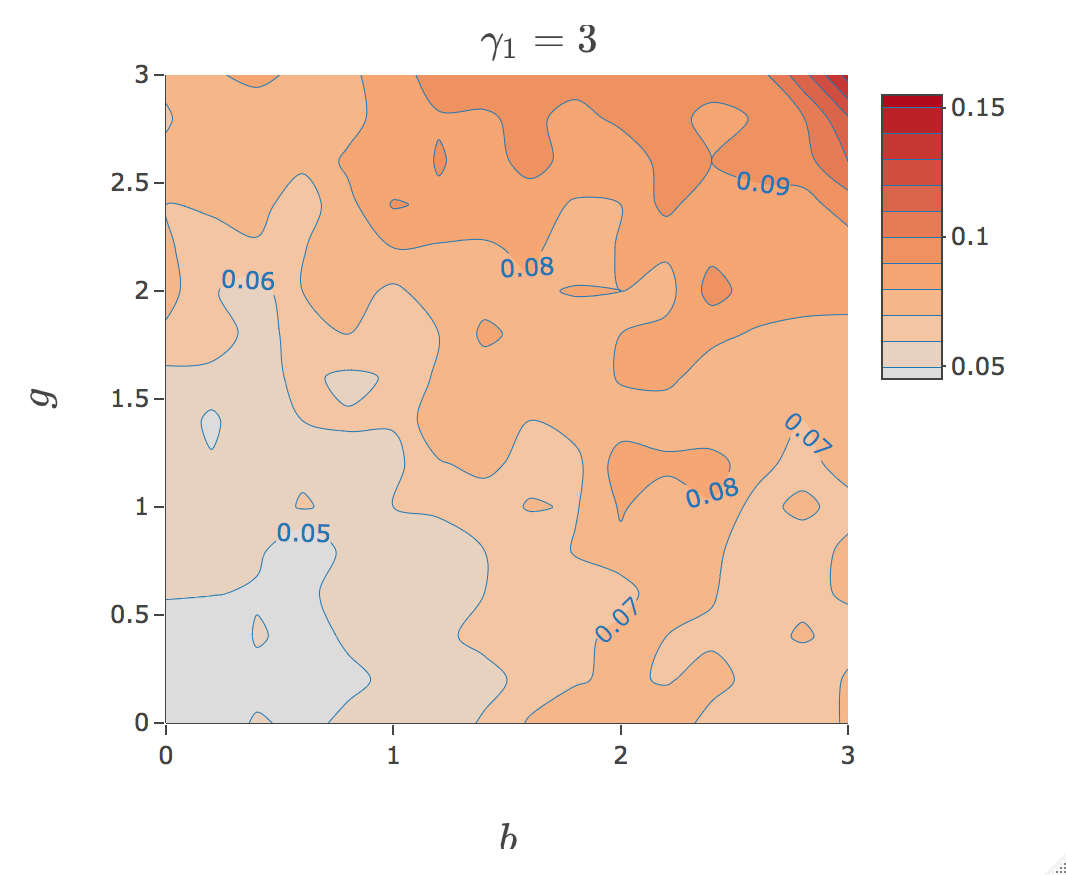}
\medskip

        \includegraphics[width=0.32\textwidth]{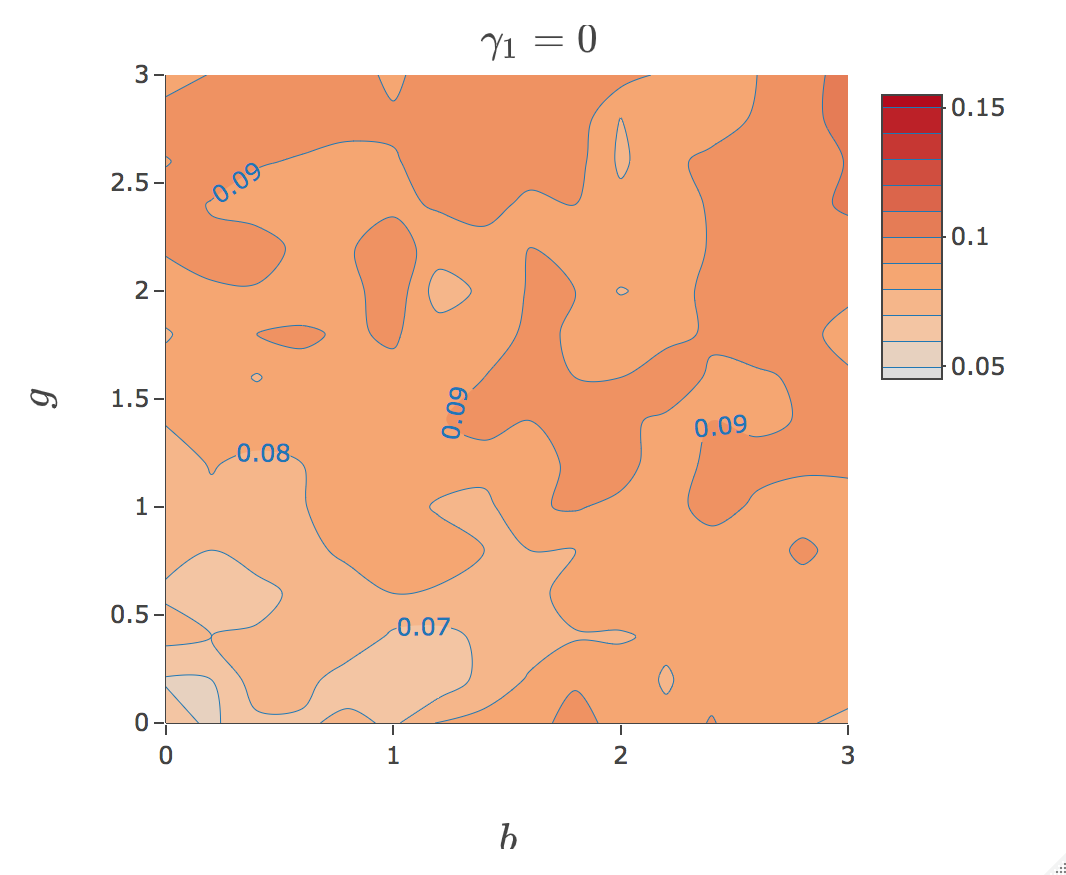}
        \includegraphics[width=0.32\textwidth]{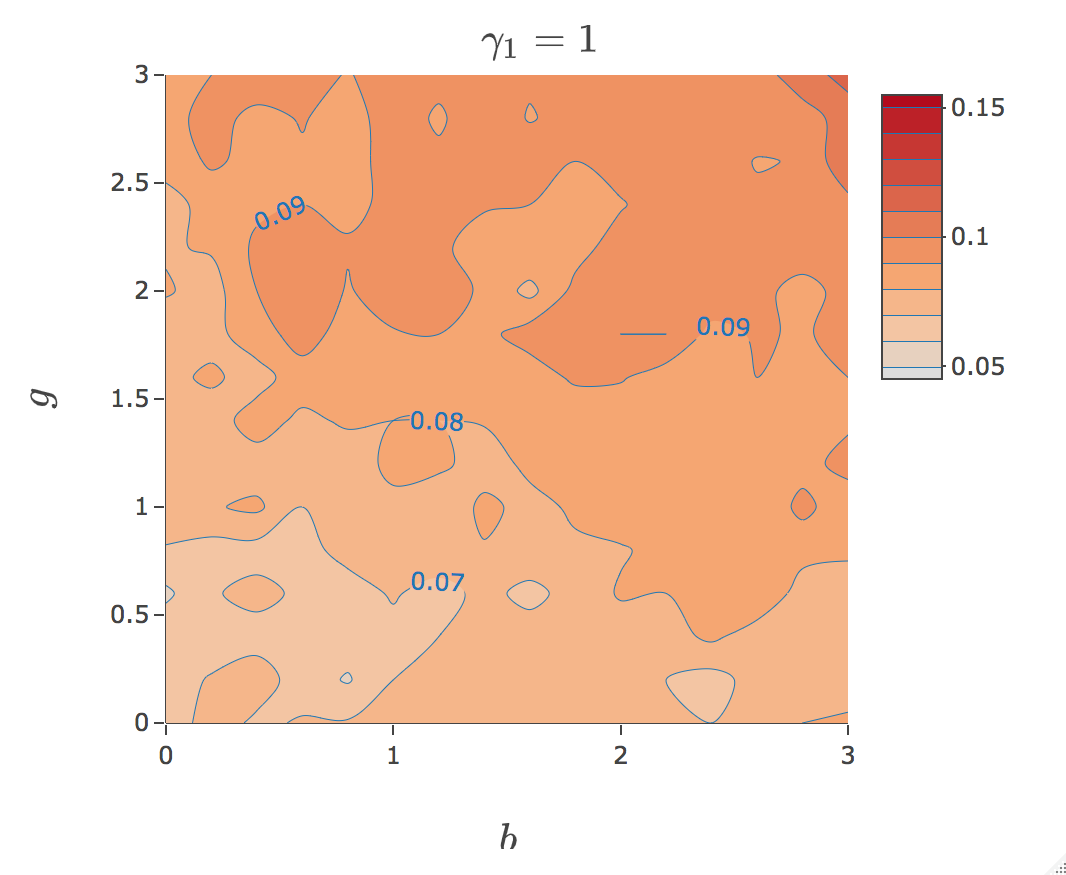}
        \includegraphics[width=0.32\textwidth]{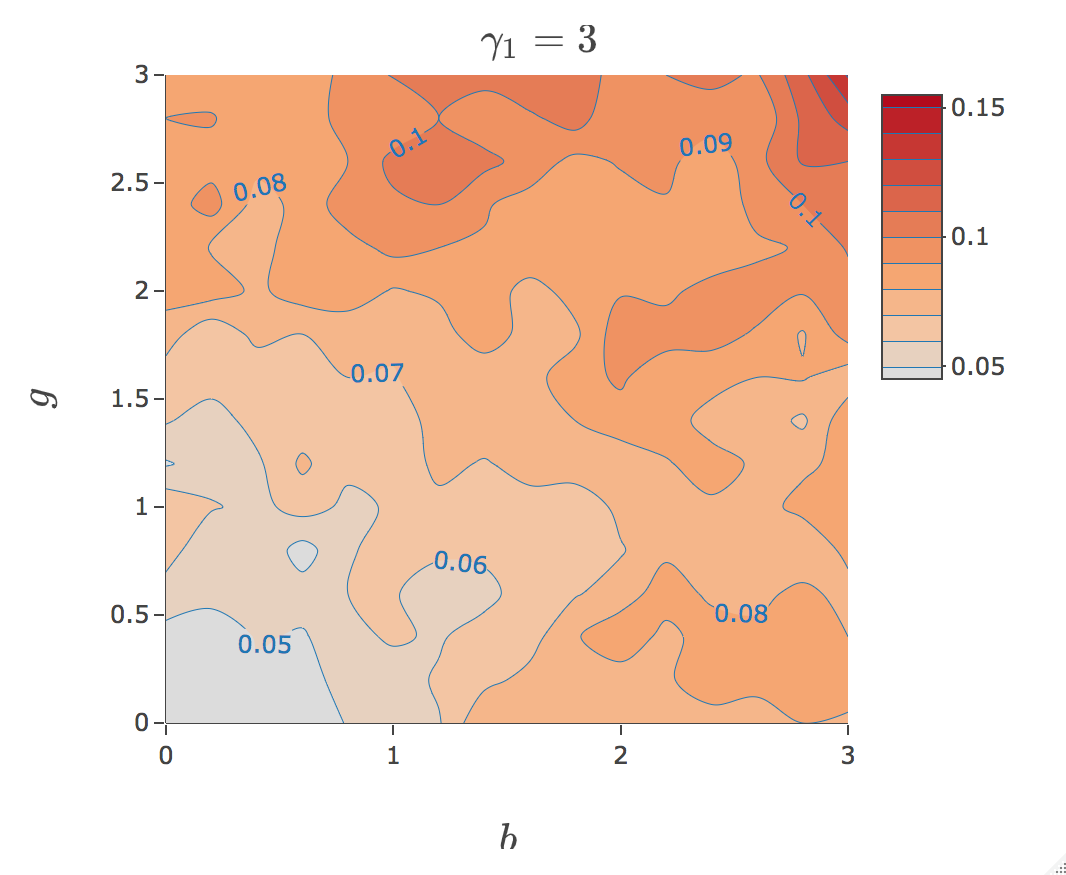}
\medskip

        \includegraphics[width=0.32\textwidth]{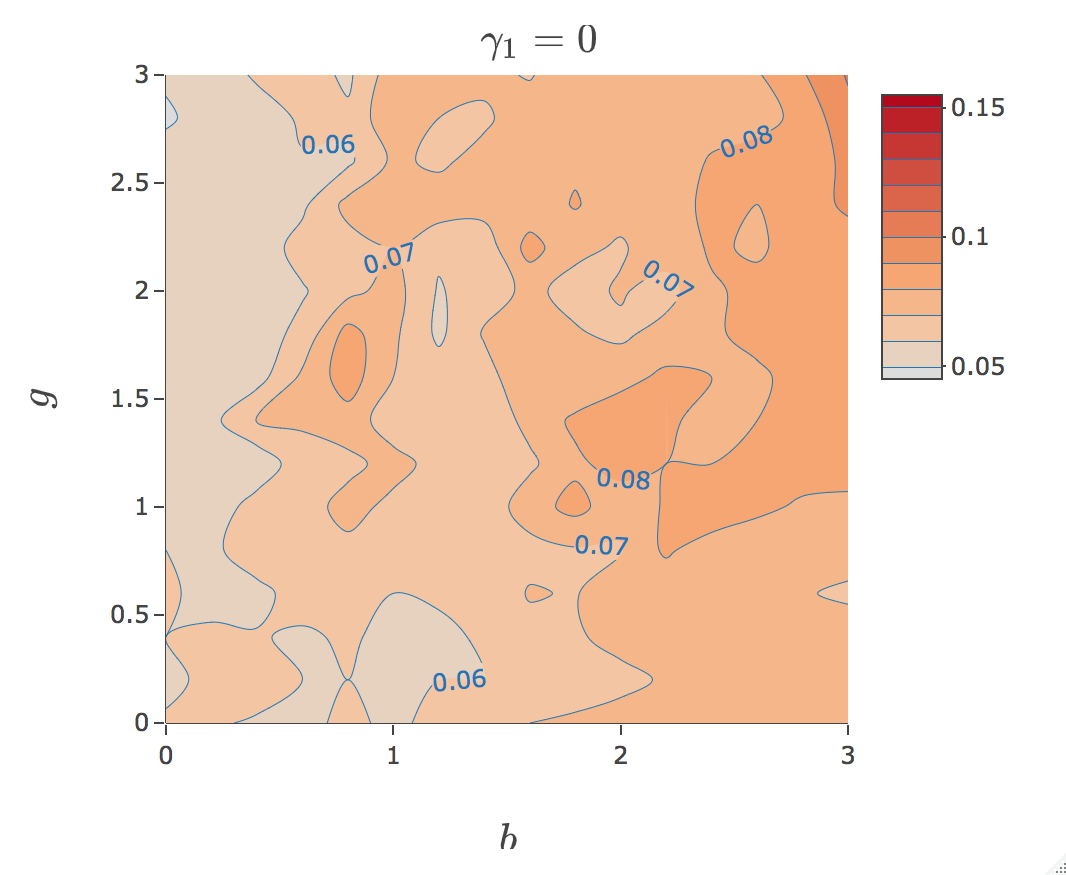}
        \includegraphics[width=0.32\textwidth]{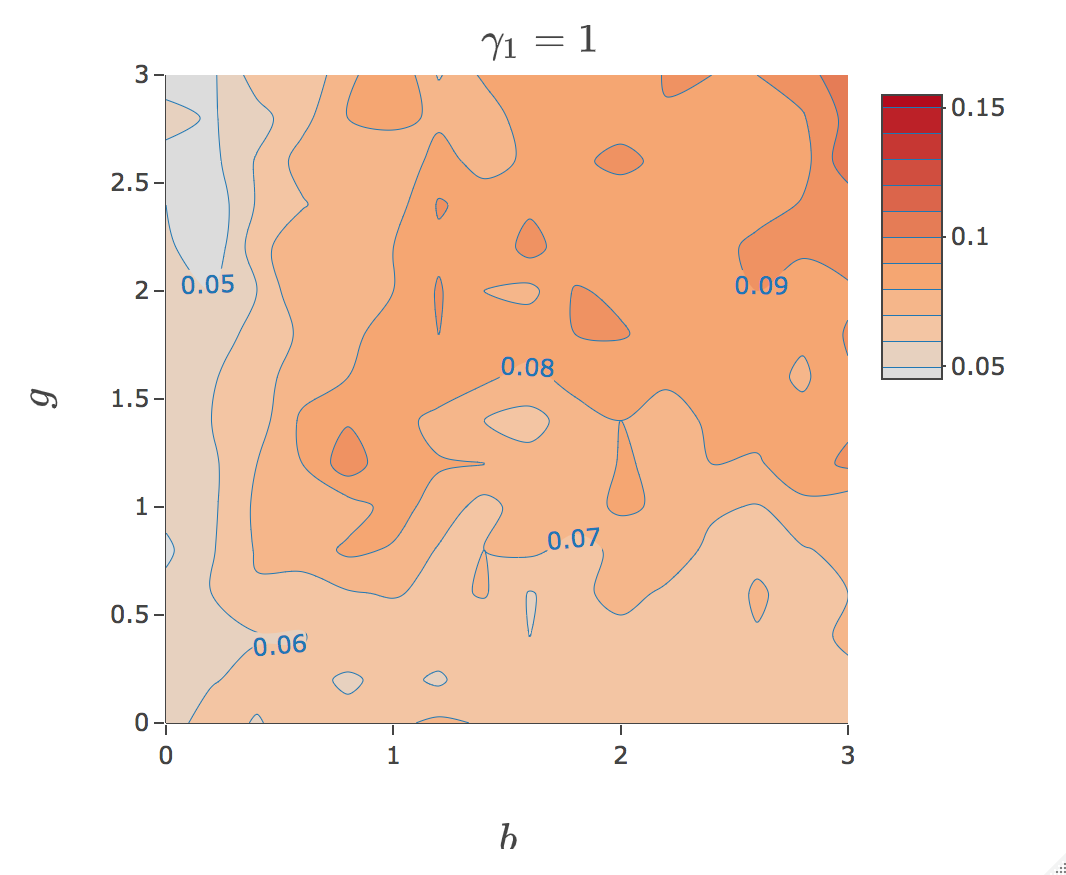}
        \includegraphics[width=0.32\textwidth]{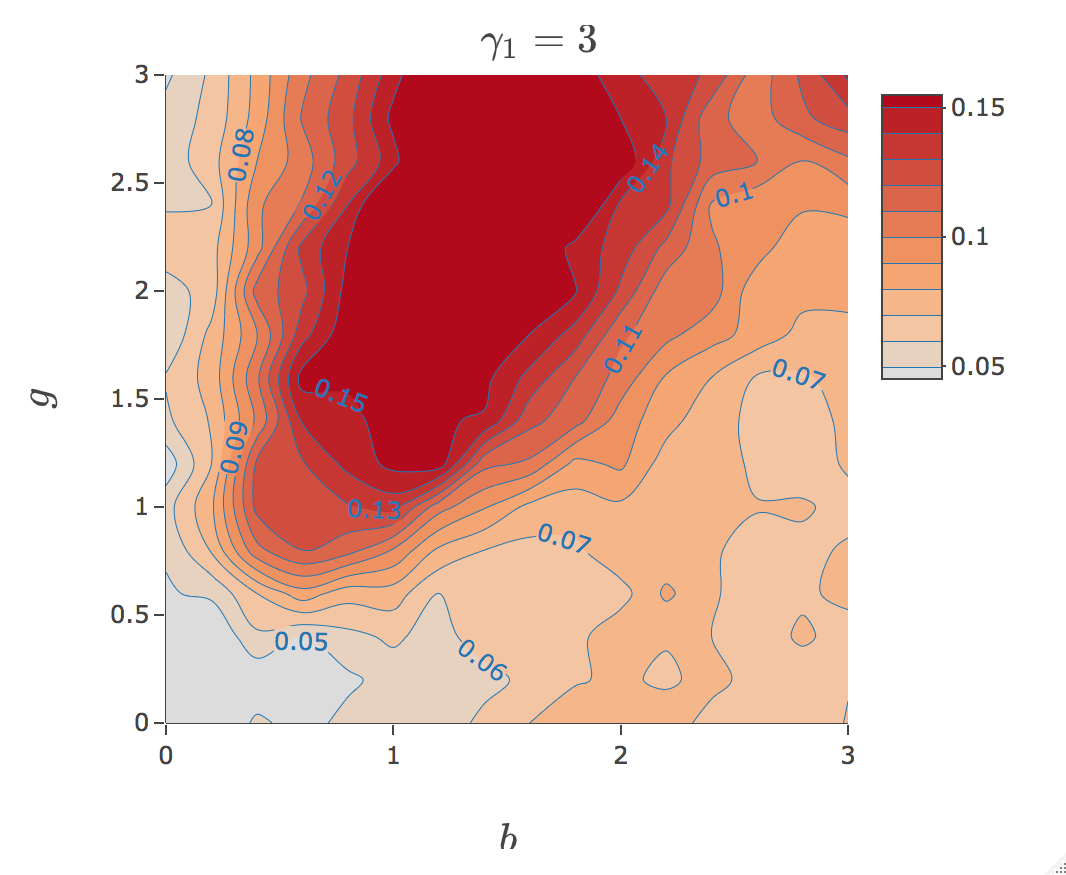}
\medskip

        \includegraphics[width=0.32\textwidth]{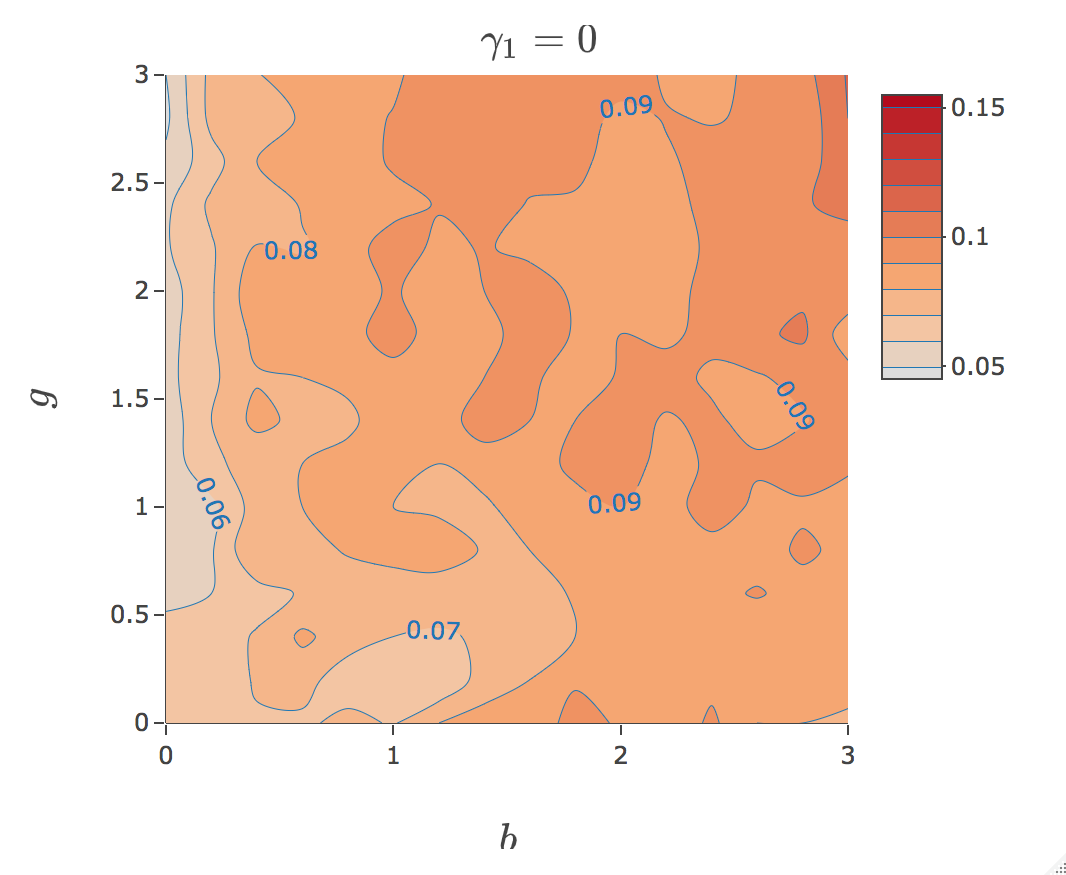}
        \includegraphics[width=0.32\textwidth]{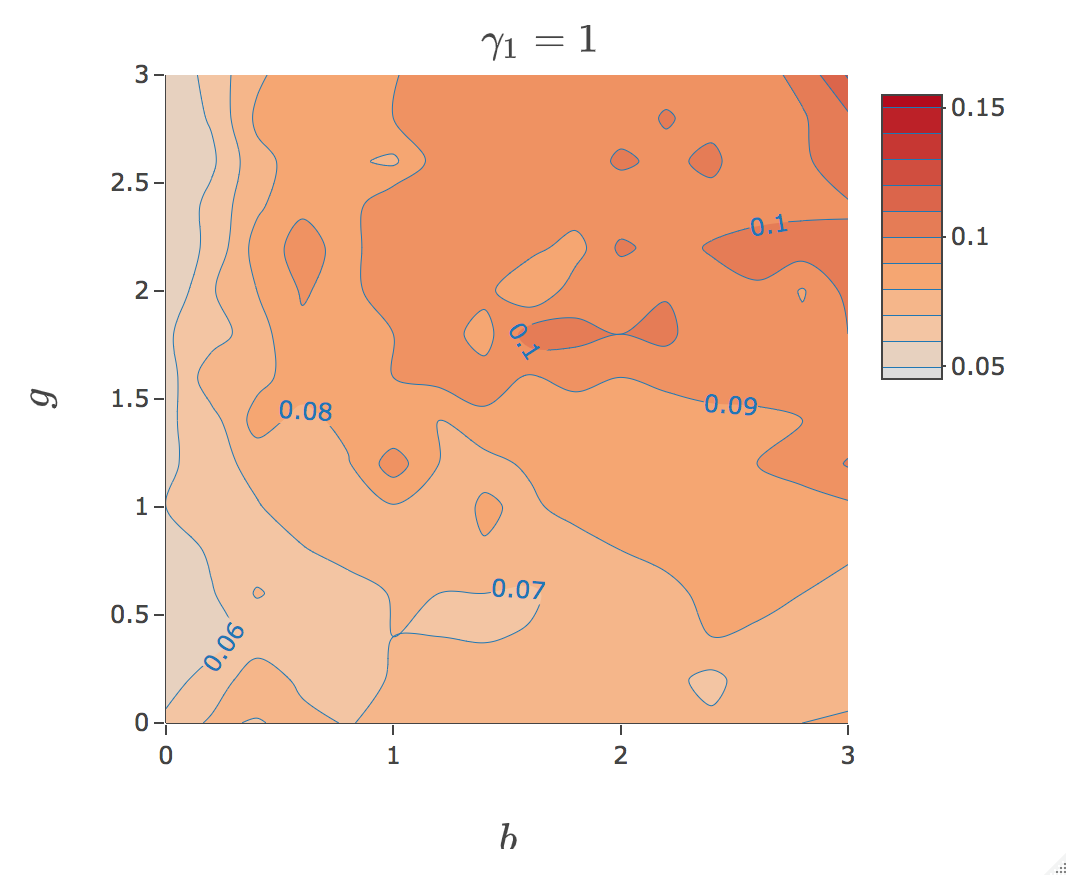}
        \includegraphics[width=0.32\textwidth]{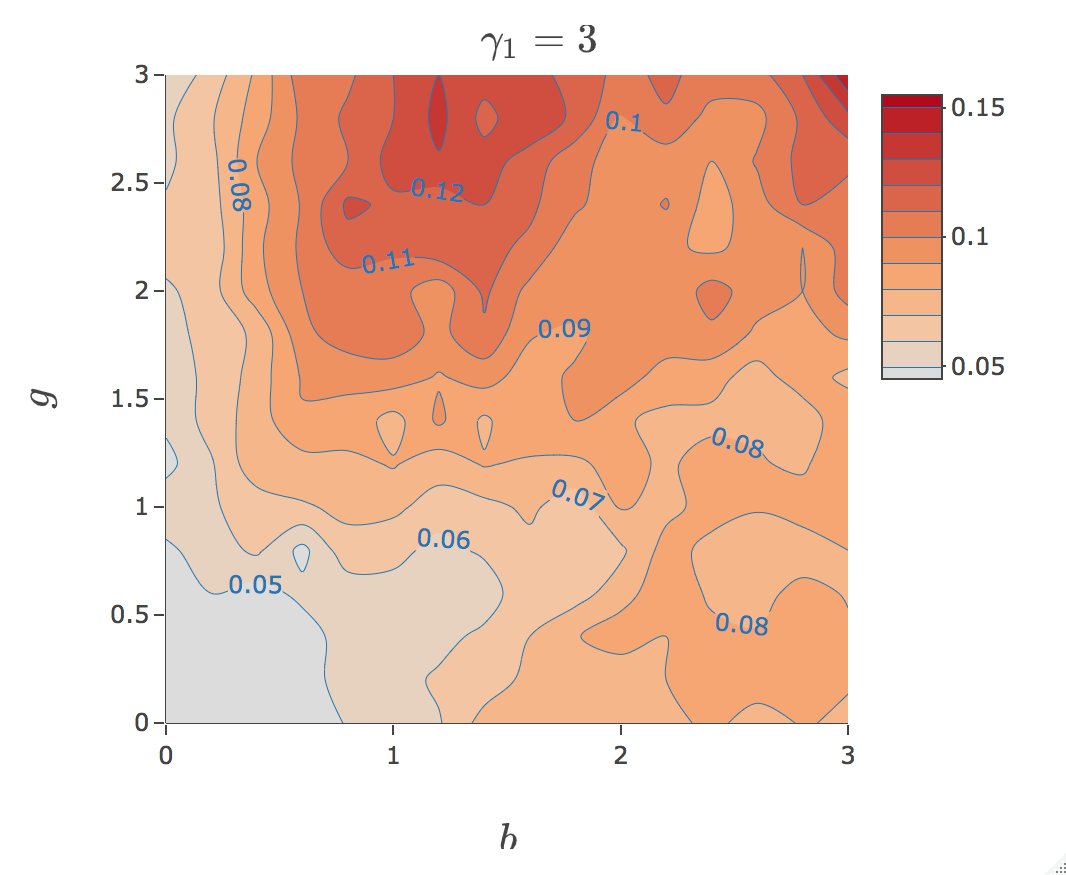}

\caption{Empirical Type I error rate at the $5\%$ significance level of the Lasso and the proposed double selection method under Setting $3$. Row $1$: double selection method with penalty parameter $\lambda_{1se}$; Row $2$:  double selection method with penalty parameter $\lambda_{min}$; Row $3$: Post-Lasso with penalty parameter $\lambda_{1se}$ ; Row $4$: Post-Lasso with penalty parameter $\lambda_{min}$. Left: $\gamma_1=0$; Middle: $\gamma_1=1$; Right: $\gamma_1=3$.
Results are based on $1,000$ simulations, $\lambda_0(t)=1$ ($\beta_0=0$) and $\lambda_0^C(t)=1$ ($\gamma_0=0$).
}
\label{fig:Sim_results5}
\end{figure}

\begin{figure}[h!]
    \centering
    \begin{subfigure}[b]{0.45\textwidth}
        \includegraphics[width=\textwidth]{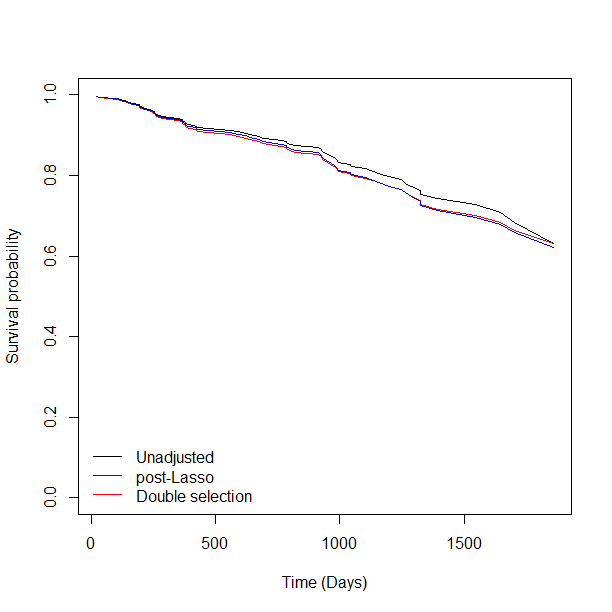}
        \caption{All deaths (transplants censored); under placebo.}
    \end{subfigure}
    \begin{subfigure}[b]{0.45\textwidth}
        \includegraphics[width=\textwidth]{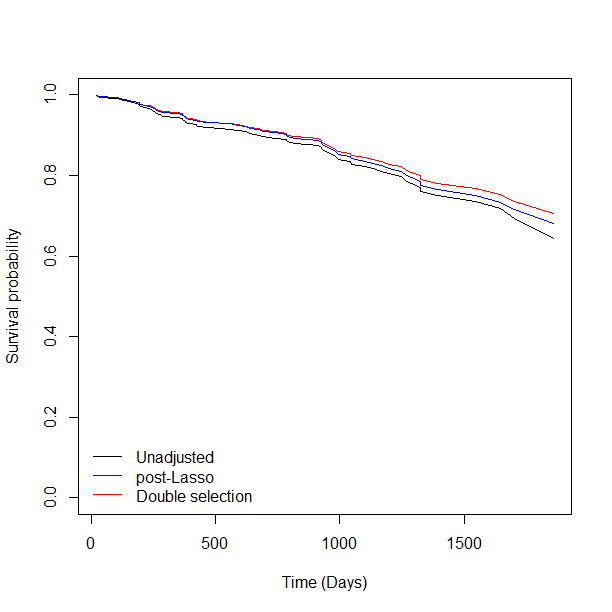}
        \caption{All deaths (transplants censored); under CyA.}
    \end{subfigure}
        \begin{subfigure}[b]{0.45\textwidth}
        \includegraphics[width=\textwidth]{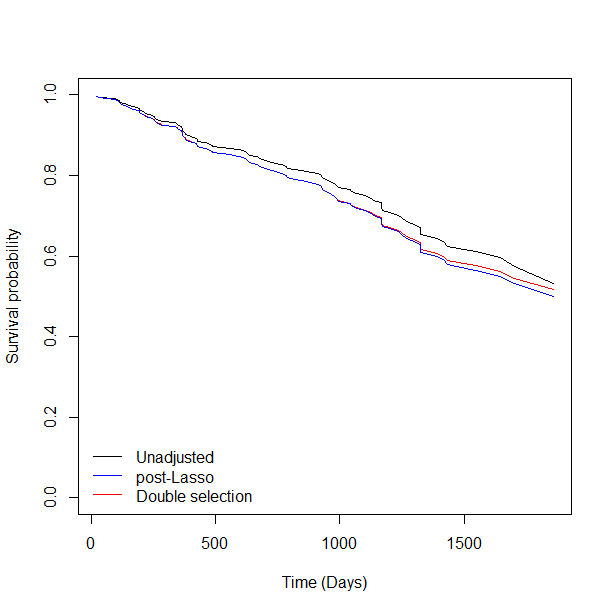}
        \caption{Progression to death or transplant; under placebo.}
    \end{subfigure}
    \begin{subfigure}[b]{0.45\textwidth}
        \includegraphics[width=\textwidth]{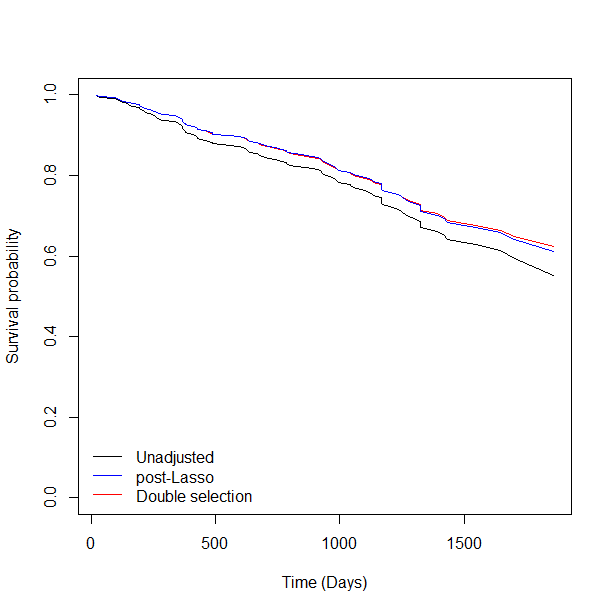}
        \caption{Progression to death or transplant; under CyA.}
    \end{subfigure}
    \caption{Unadjusted survival curves and adjusted/standardized survival curves based on the post-Lasso and double selection models for the PBC-$3$ dataset.
    Unadjusted: unadjusted survival curves; Double Selection: standardized survival curves based on double selection model with penalty parameter $\lambda_{1se}$;  Lasso: standardized survival curves based on Post-Lasso model penalty parameter $\lambda_{1se}$ .
    }
\label{fig:DataAnalysis}
\end{figure}

\begin{figure}[htp]
\centering

    \begin{subfigure}[b]{0.45\textwidth}
        \includegraphics[width=\textwidth]{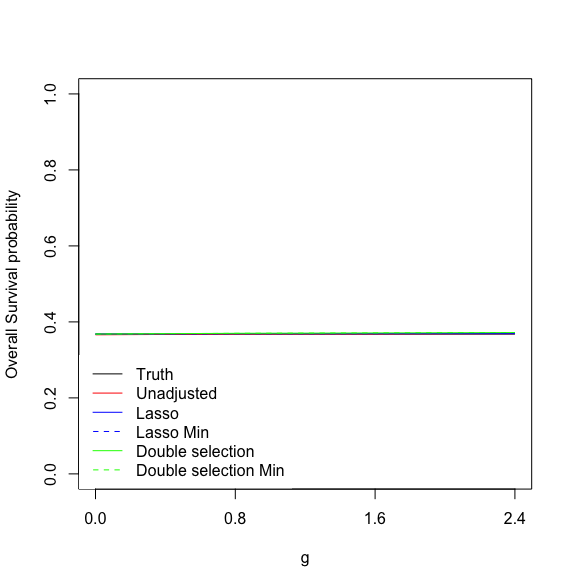}
        \caption{$b=0$}
    \end{subfigure}
        \begin{subfigure}[b]{0.45\textwidth}
        \includegraphics[width=\textwidth]{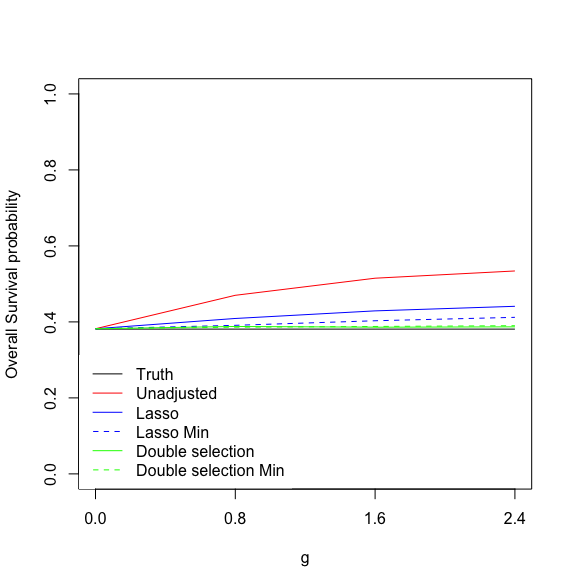}
        \caption{$b=0.8$}
    \end{subfigure}

\medskip
        \begin{subfigure}[b]{0.45\textwidth}
        \includegraphics[width=\textwidth]{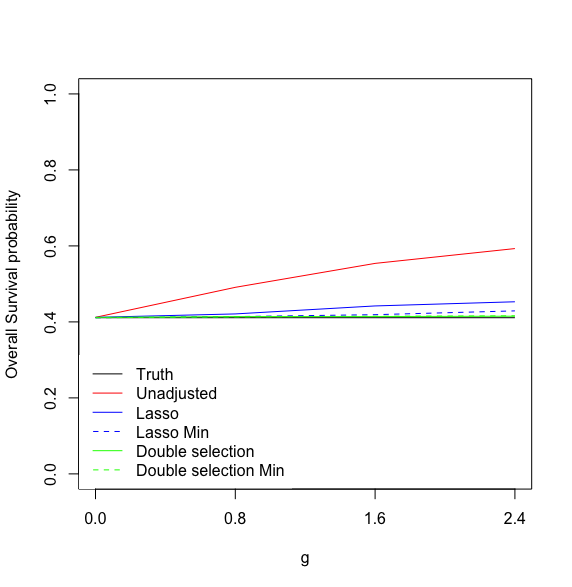}
        \caption{$b=1.6$}
    \end{subfigure}
        \begin{subfigure}[b]{0.45\textwidth}
        \includegraphics[width=\textwidth]{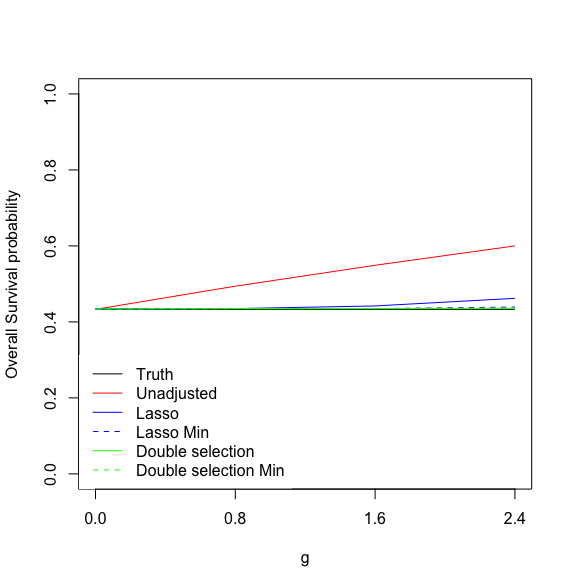}
        \caption{$b=2.4$}
    \end{subfigure}

\caption{True survival probabilities, unadjusted (Kaplan-Meier based) survival probabilities and adjusted/standardized survival probabilities based on the post-Lasso and double selection models evaluated at $1$ year.
    Results are shown for $A=1$ based on $1,000$ simulations under Setting 1 (see Section \ref{sec:montecarlo}), $n=400$, $p=30$, $\gamma_1=0$, $\lambda_0(t)=1$ ($\beta_0=0$) and $\lambda_0^C(t)=1$ ($\gamma_0=0$). 
    Truth: True survival probabilities; Unadjusted: unadjusted/Kaplan-Meier survival probabilities; Double Selection: standardized survival probabilities based on double selection model with penalty parameter $\lambda_{1se}$; Double Selection Min: standardized survival probabilities based on double selection model with penalty parameter $\lambda_{min}$; Lasso: standardized survival probabilities based on Post-Lasso model penalty parameter $\lambda_{1se}$; Lasso Min: standardized survival probabilities based on Post-Lasso model with penalty parameter $\lambda_{min}$. \label{fig:curves0}}
\end{figure}

\begin{figure}[htp]
\centering

    \begin{subfigure}[b]{0.45\textwidth}
        \includegraphics[width=\textwidth]{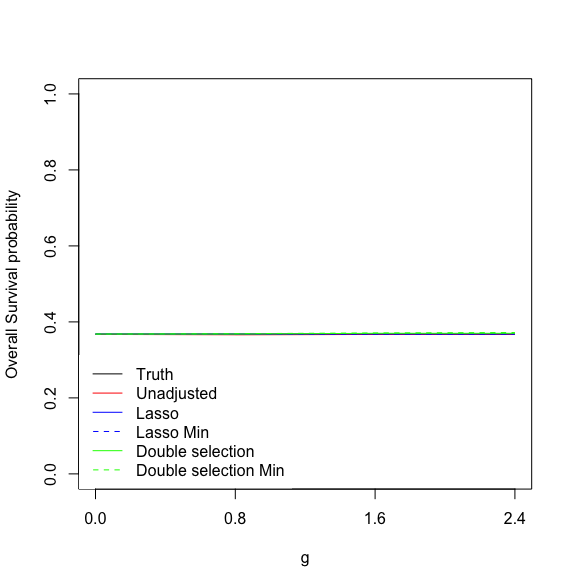}
        \caption{$b=0$}
    \end{subfigure}
        \begin{subfigure}[b]{0.45\textwidth}
        \includegraphics[width=\textwidth]{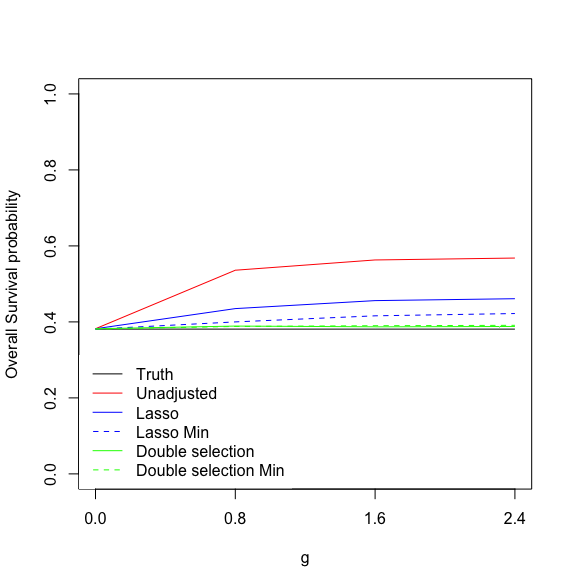}
        \caption{$b=0.8$}
    \end{subfigure}

\medskip
        \begin{subfigure}[b]{0.45\textwidth}
        \includegraphics[width=\textwidth]{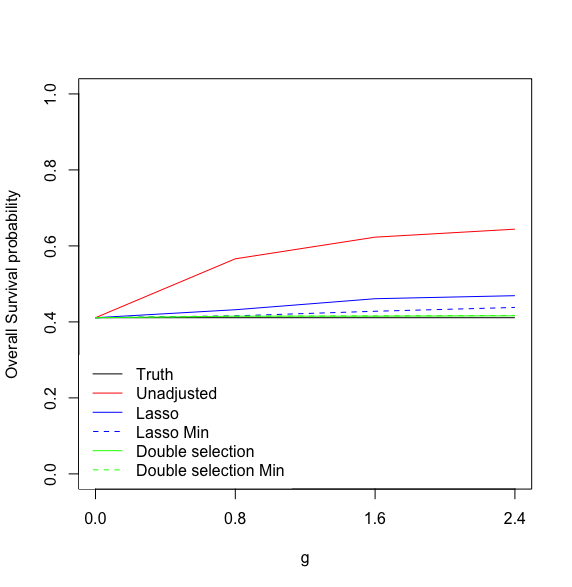}
        \caption{$b=1.6$}
    \end{subfigure}
        \begin{subfigure}[b]{0.45\textwidth}
        \includegraphics[width=\textwidth]{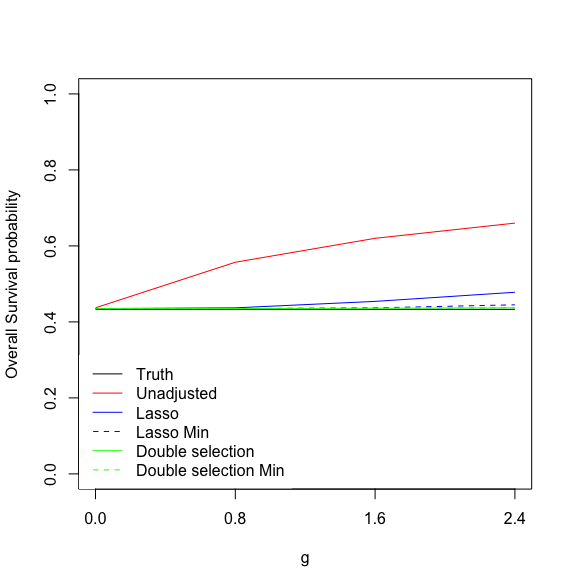}
        \caption{$b=2.4$}
    \end{subfigure}

\caption{True survival probabilities, unadjusted (Kaplan-Meier based) survival probabilities and adjusted/standardized survival probabilities based on the post-Lasso and double selection models evaluated at $1$ year.
    Results are shown for $A=1$ based on $1,000$ simulations under Setting 1 (see Section \ref{sec:montecarlo}), $n=400$, $p=30$, $\gamma_1=1$, $\lambda_0(t)=1$ ($\beta_0=0$) and $\lambda_0^C(t)=1$ ($\gamma_0=0$). 
    Truth: True survival probabilities; Unadjusted: unadjusted/Kaplan-Meier survival probabilities; Double Selection: standardized survival probabilities based on double selection model with penalty parameter $\lambda_{1se}$; Double Selection Min: standardized survival probabilities based on double selection model with penalty parameter $\lambda_{min}$; Lasso: standardized survival probabilities based on Post-Lasso model penalty parameter $\lambda_{1se}$; Lasso Min: standardized survival probabilities based on Post-Lasso model with penalty parameter $\lambda_{min}$. \label{fig:curves1}}
\end{figure}

\newpage
\bibliographystyle{apalike}
\bibliography{biblio2}
\end{document}